\documentclass[12pt]{iopart}

\usepackage[T1]{fontenc} % if needed

\usepackage{graphicx}% Include figure files
\usepackage{dcolumn}% Align table columns on decimal point
\usepackage{bm}% bold math
\usepackage[numbers]{natbib}
\usepackage{slashed}
\usepackage{contour}
\usepackage{lmodern}
\usepackage{ulem}
\usepackage{amsmath}
\usepackage{mathrsfs}
\usepackage{multirow}
\usepackage{xcolor}
\usepackage{booktabs}
\usepackage{graphicx}
\usepackage[caption=false]{subfig}
\usepackage{feynmf}
\usepackage{tcolorbox}
\usepackage{upgreek}
\usepackage{float}
\usepackage{placeins}
\usepackage[switch*]{lineno}
\usepackage{xspace}
\usepackage{enumitem}
\usepackage{xfrac}
\usepackage{doi}
\usepackage{pifont}
\usepackage{lineno}
\usepackage{nicematrix}

\newcommand{\dAUC}{$\Delta\mathrm{AUC}$\xspace}
\newcommand{\mbns}[1]{MB$#1$S}
\newcommand{\xmark}{\ding{55}}

\usepackage{iopams}  
\begin{document}

% \section*{Summary of Changes}
% \begin{itemize}
%     \item Paper style has been updated to conform to IOP style guideline
%     \item Added reference to : arxiv:2211.09912 (citation 46) and arxiv:1904.05948 (citation 80). The latter is the paper the reviewer asked to cite
%     \item Updated Ref. 44 to the official ICHEP proceedings
%     \item Updated Figures 16e and 16f to show correlation among disentangled principal components of latent space and the predicted class probability
%     \item Added the PFIN model in table A1

% \end{itemize}

\pagebreak
\title[A Detailed Study of Interpretability of Deep Neural Network based Top Taggers]{A Detailed Study of Interpretability of Deep Neural Network based Top Taggers}

\author{Ayush Khot$^1$, Mark S. Neubauer$^1$, and Avik Roy$^1$\footnote{corresponding author}}

\address{$^1$ Department of Physics \& National Center for Supercomputing Applications (NCSA), University of Illinois at Urbana-Champaign, Urbana, IL 61801}
\eads{avroy@illinois.edu}

\vspace{10pt}
\begin{indented}
\item[]June 2023
\end{indented}

\begin{abstract}
 Recent developments in the methods of explainable AI (XAI) allow researchers to explore the inner workings of deep neural networks (DNNs), revealing crucial information about input-output relationships and realizing how data connects with machine learning models. In this paper we explore interpretability of DNN models designed to identify jets coming from top quark decay in high energy proton-proton collisions at the Large Hadron Collider (LHC). We review a subset of existing top tagger models and explore different quantitative methods to identify which features play the most important roles in identifying the top jets. We also investigate how and why feature importance varies across different XAI metrics, how correlations among features impact their explainability, and how latent space representations encode information as well as correlate with physically meaningful quantities. Our studies uncover some major pitfalls of existing XAI methods and illustrate how they can be overcome to obtain consistent and meaningful interpretation of these models. 
  We additionally illustrate the activity of hidden layers as Neural Activation Pattern (NAP) diagrams and demonstrate how they can be used to understand how DNNs relay information across the layers and how this understanding can help to make such models significantly simpler by allowing effective model reoptimization and hyperparameter tuning. 
 These studies not only facilitate a methodological approach to interpreting models but also unveil new insights about what these models learn. Incorporating these  observations into augmented model design, we propose the Particle Flow Interaction Network (PFIN) model and demonstrate how interpretability-inspired model augmentation can improve top tagging performance.
\end{abstract}

\vspace{2pc}
\maketitle

\section{Introduction}

Machine learning (ML) models are ubiquitous in experimental High Energy Physics (HEP). With an ever increasing volume of data coupled with complex detector phenomenology, these models are useful to find meaningful information from these large datasets. Over time, machine learning models have grown in complexity and simpler regression and classification models have been replaced by intricate and deep neural networks. Owing to their intractably large number of trainable parameters and arbitrarily complex non-linear nature, deep neural networks (DNNs) have often been treated as \textit{black boxes}. It has always been challenging to understand how different input features contribute to the network's computational process and how the inter-connected neural pathways convey information. In recent years, advances in \textit{explainable} Artificial Intelligence (XAI)~\cite{MILLER20191} have made it possible to build intelligible relationship between an AI model's inputs, architecture, and predictions~\cite{xAI-intro,xAI-review,xAI-review-2}. While some methods remain model agnostic, a substantial subset of these methods have been developed to infer interpretability of computer vision models where an intuitive reasoning can be extracted from human-annotated datasets to validate XAI techniques. However, in other data structures such as large tabular data or relational data constructs like graphs, use of XAI methods are still quite novel~\cite{sahakyan2021explainable, xAI-GNN-survey}. In recent times, XAI has been successful in learning the underlying physics of a number of problems in high energy detectors~\cite{9302535}, including parton showers at the Large Hadron Collider (LHC)~\cite{LAI2022137055} and jet reconstruction using particle flow algorithms~\cite{mokhtar2021explaining}. 

One of the major applications of ML in the field of HEP is classification of jets, which is referred to as \textit{jet tagging}. Jets represent hadronic showers observed as conical spray of particles originating from quarks and gluons produced in the high energy collisions at a collider experiment like the LHC. Identifying jets that originate from decay products of a particle such as the top quark ($t$) and being able to separate them from other jet categories, such as jets originating from the quantum chromodynamics (QCD) background, is an important challenge in many physics analyses. Traditional top tagging algorithms based on kinematic features of jets and clustering of jet constituents (see Refs.~\cite{kaplan2008top, almeida2009top, almeida2010template, plehn2012top} for example) have been used in particle phenomenology research as well as by the ATLAS and CMS experiments and their predecessors. 
In Run 1 physics analyses, these top tagging algorithms along with low-complexity statistical models like decision trees took the center stage in dealing with top tagging~\cite{aad2016identification, CMS-PAS-JME-09-001, CMS-PAS-JME-13-007}. 
%\todo{AR}{Add references to simple top tagger usages.}
However, owing to their superior performance, models based on DNNs started becoming  popular in Run 2 at a higher center-of-mass energy of 13~TeV~\cite{atlastopodnn, cms2020identification}. 

For top quarks produced with large momenta, the decay products can be packed close to one another and be reconstructed as a single jet. For such \textit{boosted} jets, top tagging can be particularly challenging and require a better analysis of \textit{jet substructures}, a collection of constituents and their derivative properties that can offer better discrimination between jet classes. DNNs have proven to be useful to exploit the jet substructure properties in performing jet classification. A wide variety of deep learning models have been developed to optimize top tagging~\cite{pearkes2017jet, liam2019reports, datta2017how, louppe2019qcd, butter2018deep, komiske2019energy, qu2020jet, macaluso2018pulling, erdmann2019lorentz,egan2017long, bogatskiy2020lorentz, moreno2020jedi, gong2022efficient,bogatskiy2022pelican,qu2022particle}. A comprehensive review and comparison of many of these models is given in Ref.~\cite{kasieczka2019machine}. Some of these models have exploited DNN's capacity to approximate arbitrary non-linear functions~\cite{hornik1989multilayer} and their huge success with problems in the field of computer vision, other models have been inspired by the underlying physics information like jet clustering history~\cite{louppe2019qcd}, physical symmetries~\cite{butter2018deep} and physics-inspired feature engineering~\cite{erdmann2019lorentz}. These efforts have inspired novel model architectures and feature engineering by creating or augmenting input feature spaces with physically meaningful quantities~\cite{erdmann2019lorentz,chakraborty2019interpretable,agarwal2021explainable}. 

The rich history of physics-inspired model development makes the problem of top tagging an excellent playground to better understand the modern XAI tools themselves. This allows us to traverse a rare two-way bridge in exploring the relationship between data and models- our physics knowledge will allow us to better understand the inner workings of modern XAI tools and perfect them while those improved tools would allow us to take a deeper look at the models- paving ways for analyzing and reoptimizing them.
%
%Nevertheless, exploring \textit{post-hoc} interpretability of jet tagger models in light of modern XAI tools is still quite novel in literature. 
%Development of ML models for top tagging has been mostly interpretation-agnostic-- exploration of tractable input-output relationship or information propagation through neural architecture has been largely ignored. 
%Attempts to develop interpretable, physics-inspired architectures has mostly been concentrated within clever feature engineering, in creating or augmenting input feature spaces with physically meaningful quantities~\cite{erdmann2019lorentz,chakraborty2019interpretable,agarwal2021explainable}.
%Identifying the relative importance of different input features in a trained model to establish tractable input-output relationships or standardize model training paradigm has been only minimally addressed. 
As it has been pointed out in Ref.~\cite{shanahan2022snowmass}, such insights into explainability of DNN-based models are important to validate them, to make them reliable and reusable.
Additionally, the broader scope of uncertainty quantification in association with ML models relies on developing robust explanations~\cite{seuss2021bridging} and in the field of HEP for problems like top tagging will require dedicated understanding of how robust as well as interpretable these models are~\cite{grojean2022lessons}.

Yet another remarkable application of interpretability is to understand how the model conveys information and in doing so, which parts of a DNN most actively engage in forward propagation of information. Such studies could be useful to understand and reoptimize model complexity. Given DNNs have shown remarkable success in jet and event classification, recent work has placed emphasis on developing DNN-enabled FPGAs for trigger-level applications at the LHC~\cite{duarte2018fast, iiyama2021distance, heintz2020accelerated}. As resource consumption and latency of FPGAs directly depend on the size of the network to be implemented, it is definitely easier to embed simpler networks on these devices. Hence, methods that  allow interpreting a network's response patterns as well as provide critical insights about model optimization without compromising its performance can greatly benefit these new budding fields of ML applications, especially for online event selection and jet tagging at current and future high energy colliders.

Application of state-of-the-art explainability techniques for interpreting jet tagger models is receiving more attention recently~\cite{agarwal2021explainable,roy2022interpretability,neubauer2022explainable,Mokhtar:2022pwm} and has been demonstrated to be successful in identifying feature importance for models like the Interaction Network~\cite{IN}. 
In this paper, we study the interpretability of a subset of existing ML-based top tagging models. The models we have chosen use multi-layer perceptrons (MLPs) as underlying neural architecture.
Choosing simpler neural architecture allows us to elucidate the applicability and limitations of existing XAI methods and develop new tools to examine them without convoluting these efforts with the complexity of larger models or unorthodox data structures.
To compare our results for different models as well as with existing benchmarks in published literature, we use the dataset developed by the authors of Ref.~\cite{butter2018deep} and later used in the top tagger model review in Ref.~\cite{kasieczka2019machine}. The models explored in this paper along with the dataset have been reviewed in section~\ref{sec:review-tagger}. The model hyperparameters explained in this section will constitute the \textit{baseline} model in each category. Variants of each model are studied to better understand their interpretability where the underlying architecture remains the same but model hyperparameters, input features, or data preprocessing might be changed.  Section~\ref{sec:review-interp} reviews modern XAI methods that we will use in investigating the explainability of top tagger models. In section~\ref{sec:interp}, we analyze the results of applying XAI methods on different top tagger models. Section~\ref{sec:conclusion} summarizes our findings and illustrates new dimensions to explore in the conjunction of XAI and HEP. %Finally, inspired by these models' inner workings, we achieve state-of-the-art top tagging performance in section~\ref{sec:new-tagger} by interpretation-inspired augmentation of existing models. 
\section{Review of Top Tagging Dataset and Models}
\label{sec:review-tagger}

\begin{figure}[!h]
\centering
\subfloat[]{
\includegraphics[width=0.33\textwidth]{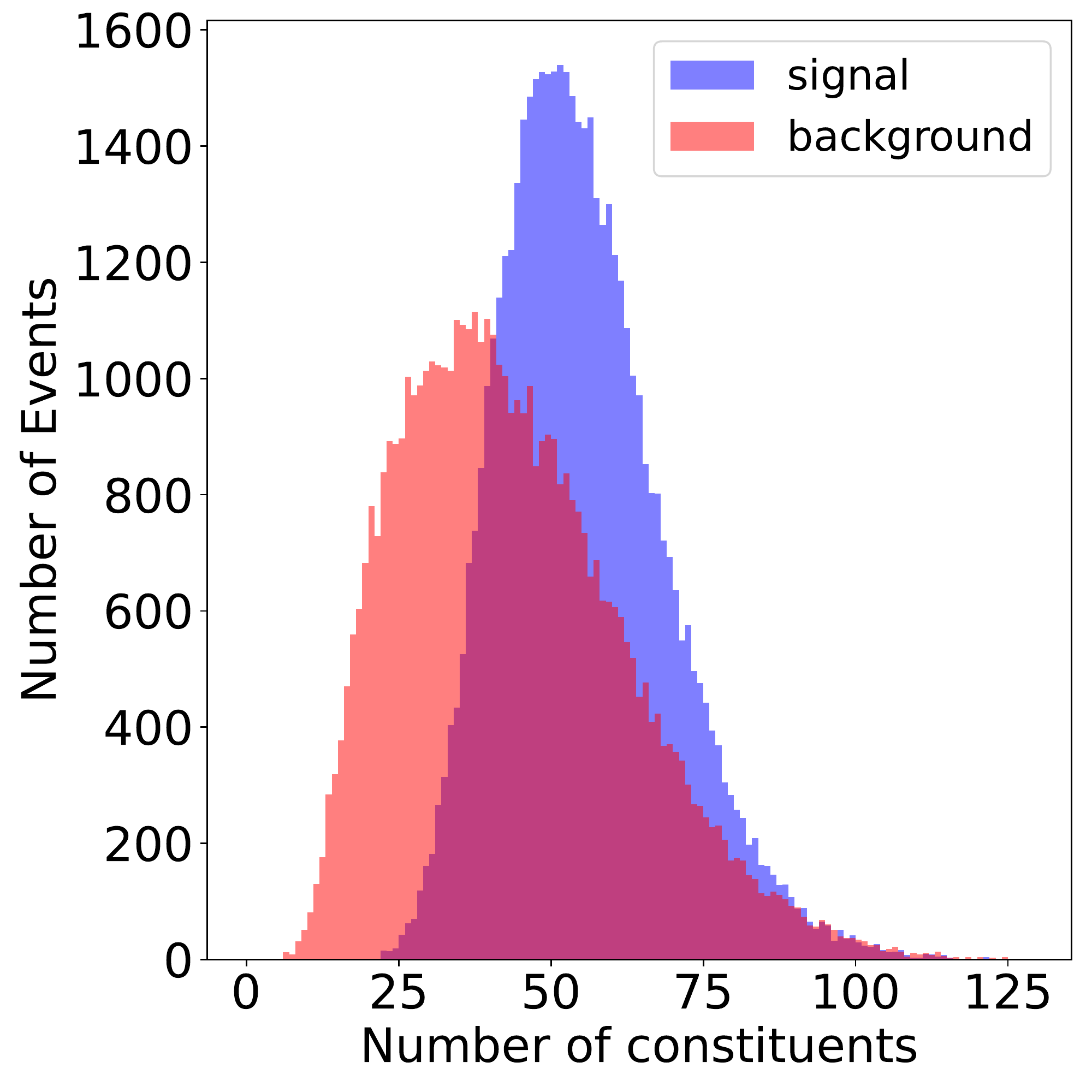}
\label{fig:Nconst}            
}
\subfloat[]{
\includegraphics[width=0.33\textwidth]{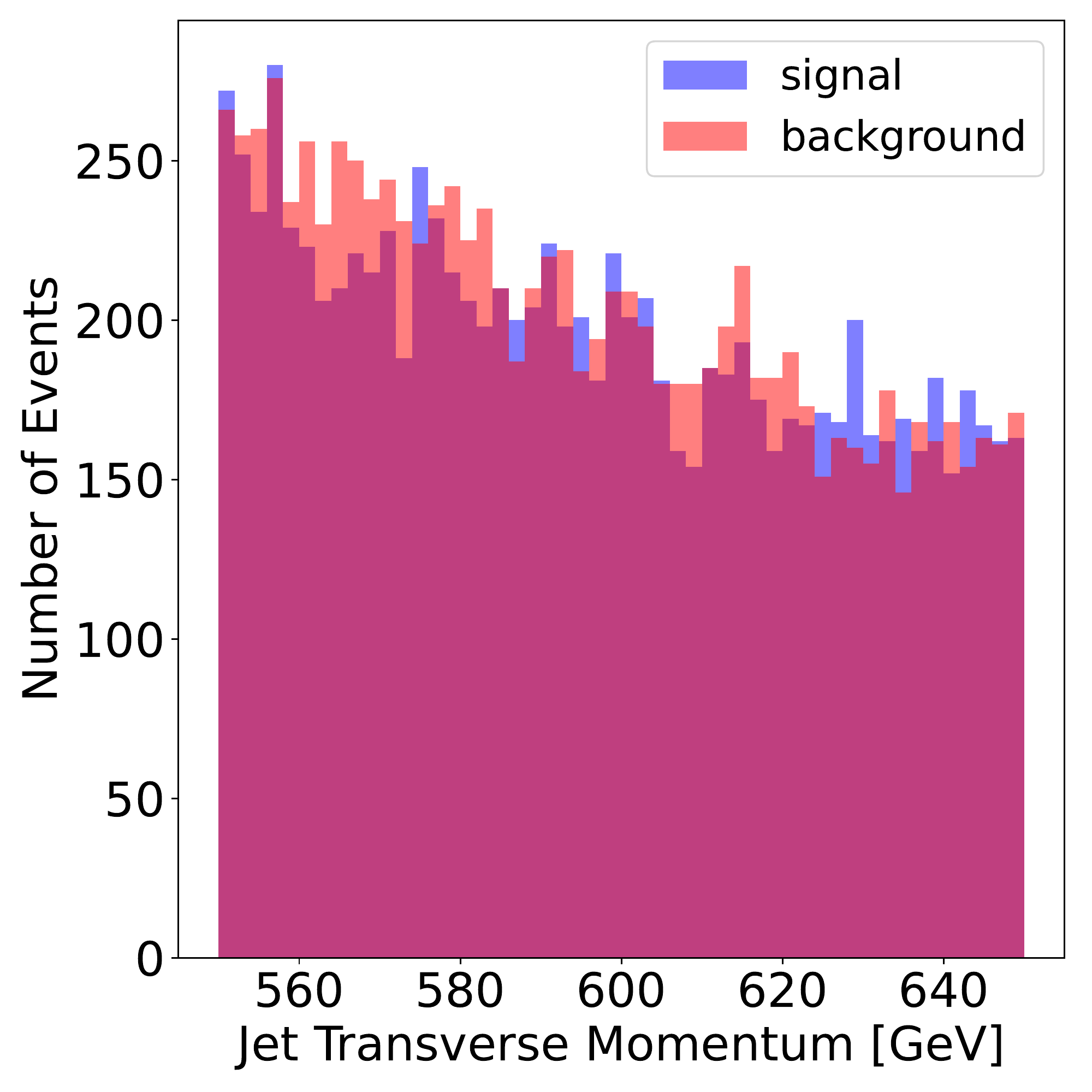}
\label{fig:jet-pt}            
}
\subfloat[]{
\includegraphics[width=0.33\textwidth]{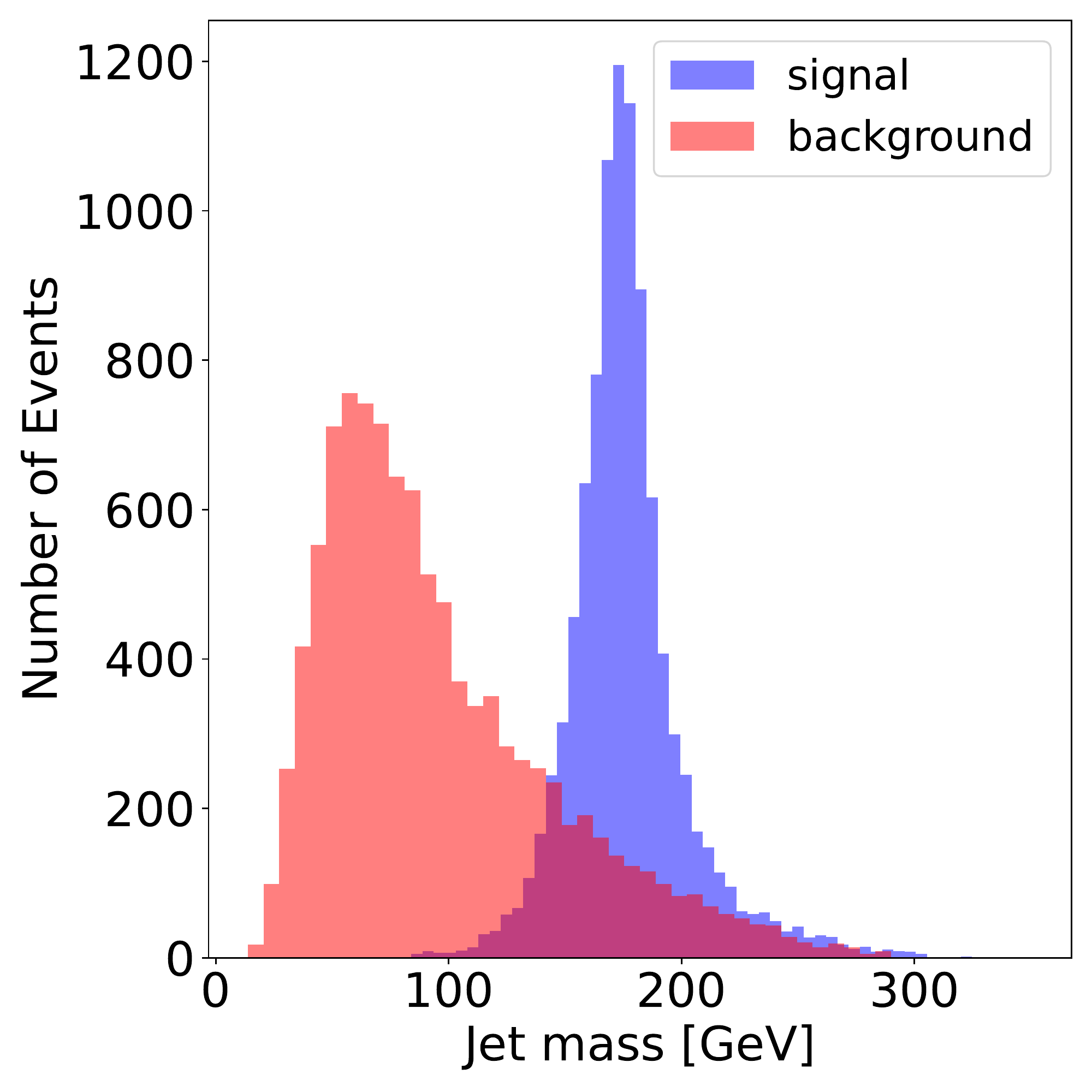}
\label{fig:jet-m}            
}
\caption{Distribution of \protect\subref{fig:Nconst} number of constituent particles,  \protect\subref{fig:jet-pt} jet transverse momentum ($p_{T,J}$), and  \protect\subref{fig:jet-m} jet mass ($m_J$) for background (QCD) and signal (top) jets}
\label{fig:jet-feats}
\end{figure}

The dataset used in this paper has been used to perform model benchmarking studies in Ref.~\cite{kasieczka2019machine} and publicly available at Ref.~\cite{topdata}. This dataset consists of 1 million top (signal) jets and 1 million QCD (background) jets genetated with \textsc{Pythia8}~\cite{pythia8} with its default tune at 14~TeV center of mass energy for proton-proton collisions. The detector simulation was performed with \textsc{Delphes}~\cite{delphes} and jets were reconstructed using the $anti-k_t$ algorithm~\cite{cacciari2008anti} with a jet radius of $R = 0.8$ using \textsc{FastJet}~\cite{cacciari2012fastjet}. Only jets with transverse momenta within the range of $550$ and $650$~GeV are considered. For each jet, the dataset contains the four momenta of up to 200 constituents with zero-padded entries for missing constituents. The dataset is divided into traning, validation, and testing sets with a 6:2:2 split. Some characteristic jet features from a random subsample of the training data are shown in Figure~\ref{fig:jet-feats}.

In this paper we consider three different NN-based models for top tagging. Given the tagger distinguishes between two jet classes, minimizing the  standard binary cross-entropy (BCE) loss has been used as the training objective for all models. The training is done using the \textsc{Adam} optimizer with minibatches. All networks showed comparable performance with different batchsizes. The architecture, hyperparameters, and data preprocessing for each of the baseline models is summarized below-

\begin{itemize}[leftmargin=*, label={}]
    \item \textbf{TopoDNN}~\cite{pearkes2017jet, atlastopodnn}: The simplest top tagging model we consider is a fully connected multi-layer perceptron (MLP) trained with transverse momentum ($p_T$), azimuthal angle ($\phi$), and pseudorapidity ($\eta$) of the 30 most energetic particles. Usually referred to as TopoDNN, this model represents a quintessential MLP network. Although TopoDNN is outperformed by many other ML-models for top tagging, its simple architecture allows us to explore different XAI metrics, their limitations, and the best practices to overcome them. Since MLPs are still widely used in HEP for a wide variety of applications, our studies of modern XAI for this model will also illustrate the best practices to interpret the input-output relations for such models. 
    
   TopoDNN is trained on preprocessed data where (i) the jet is rotated on the $\eta-\phi$ plane to have the most energetic component aligned along the central coordinate (0,0), (ii) the second most energetic component falls along the negative-$\phi$ axis, and (iii) all momenta are scaled with an arbitrarily chosen factor of 1/1700. The transformations (i) and (ii) take advantage of the underlying Lorentz invariance of collider physics and (iii) converts the momenta into unitless quantities and scales them down to a numerical range comparable to those of $\eta, \phi$ quantities. The baseline model is constructed with 4 hidden layers with  300, 102, 12, and 6 nodes respectively and \textsc{ReLU} activation function. The output layer consists of a single node which is converted by the sigmoid function to represent the probability of the jet being classified as a signal jet. 
    \item \textbf{Multi-body $N$-subjettiness (\mbns{N})}~\cite{liam2019reports,datta2017how}: Top-tagging with $N$-subjettiness variables uses an MLP as the underlying trainable architecture. However, the input to the network is different from the usual kinematic variables. It uses the multi-body $N$-subjettiness variables~\cite{thaler2011identifying}, defined as 
    
    \begin{equation}
        \tau_n^{(\beta)} = \frac{1}{p_{T,J}}\sum_i p_{T,i}\min\left\{\Delta R_{1i}^\beta, \Delta R_{2i}^\beta, ..., \Delta R_{ni}^\beta\right\}
        \label{eqn:subjettiness}
    \end{equation}
    where $p_{T,J}$ and $p_{T,i}$ represent the transverse momenta of the jet and its $i$-th constituent and $\Delta R_{ki}$ is the distance between the $k$-th jet axis and the $i$-th particle constituent. The $n$ jet axes chosen for claculating $\tau_n^{(\beta)}$ are obtained using the $k_t$ algorithm~\cite{ellis1993successive} with $E$-scheme recombination~\cite{blazeya2000run}. Figure~\ref{fig:jet-taus} shows the distribution of some of the $\tau$ variables for QCD and top jets. The input to MB$N$S tagger is the set of subjettiness variables
    \begin{equation}
        \left\{ 
        \tau_1^{(0.5)}, \tau_1^{(1)}, \tau_1^{(2)},
        \tau_2^{(0.5)}, \tau_2^{(1)}, \tau_2^{(2)},
        ...,
        \tau_{N-2}^{(0.5)}, \tau_{N-2}^{(1)}, \tau_{N-2}^{(2)},
        \tau_{N-1}^{(1)}, \tau_{N-1}^{(2)}
        \right\} \bigcup \left\{p_{T,J}, m_J\right\}
        \label{eqn:tauinputs}
    \end{equation}
    where, besides the subjettiness variables, the jet $p_T$ and jet mass $(m_J)$ variables are used as inputs to provide a kinematic scale for the jet event. However, the latter inputs are scaled by a factor of 1/1000 to mitigate the several orders of magnitude gap between their numerical range and those of the $\tau$s. 
    \begin{figure}[!h]
\centering
\subfloat[]{
\includegraphics[width=0.25\textwidth]{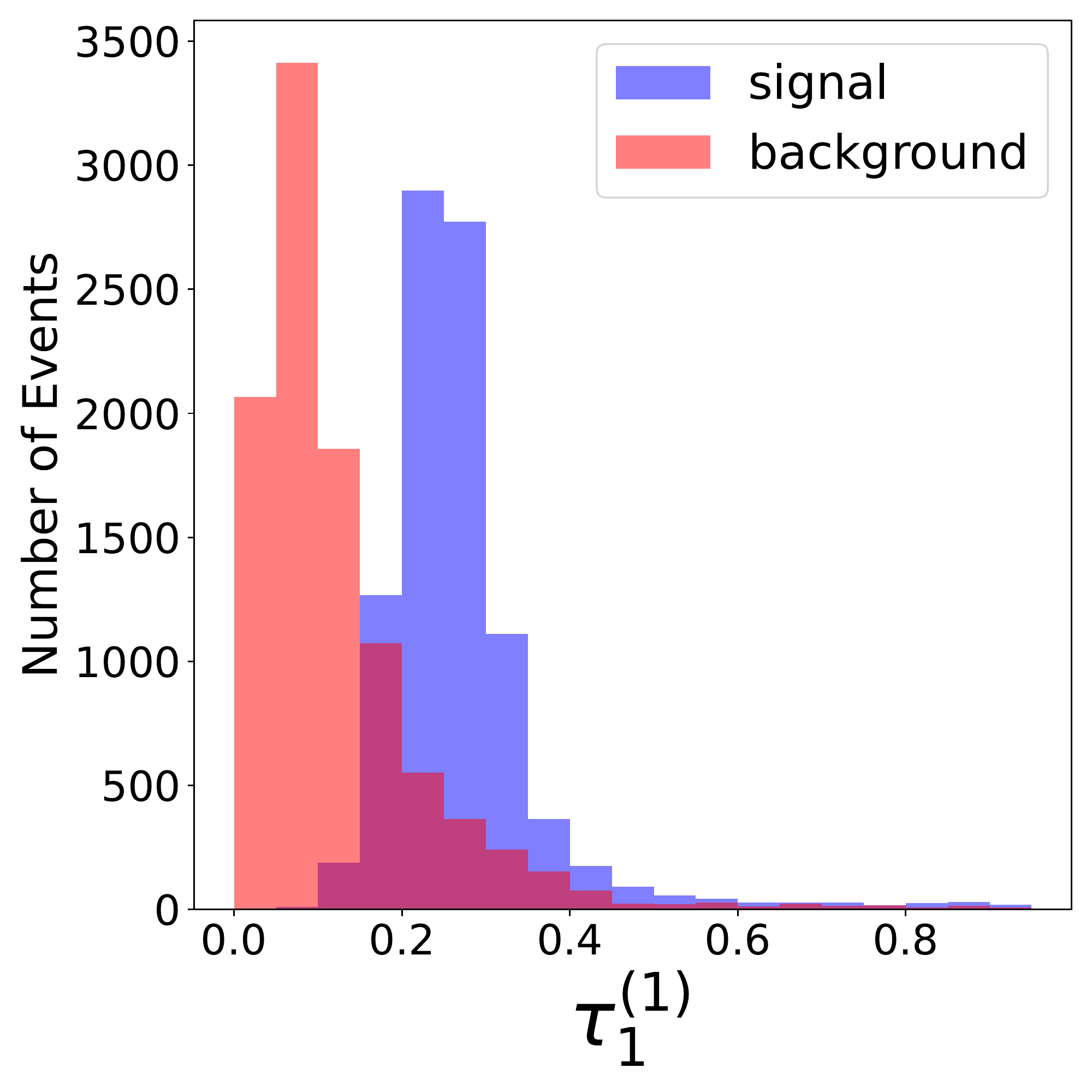}
\label{fig:tau_1_10}            
}
\subfloat[]{
\includegraphics[width=0.25\textwidth]{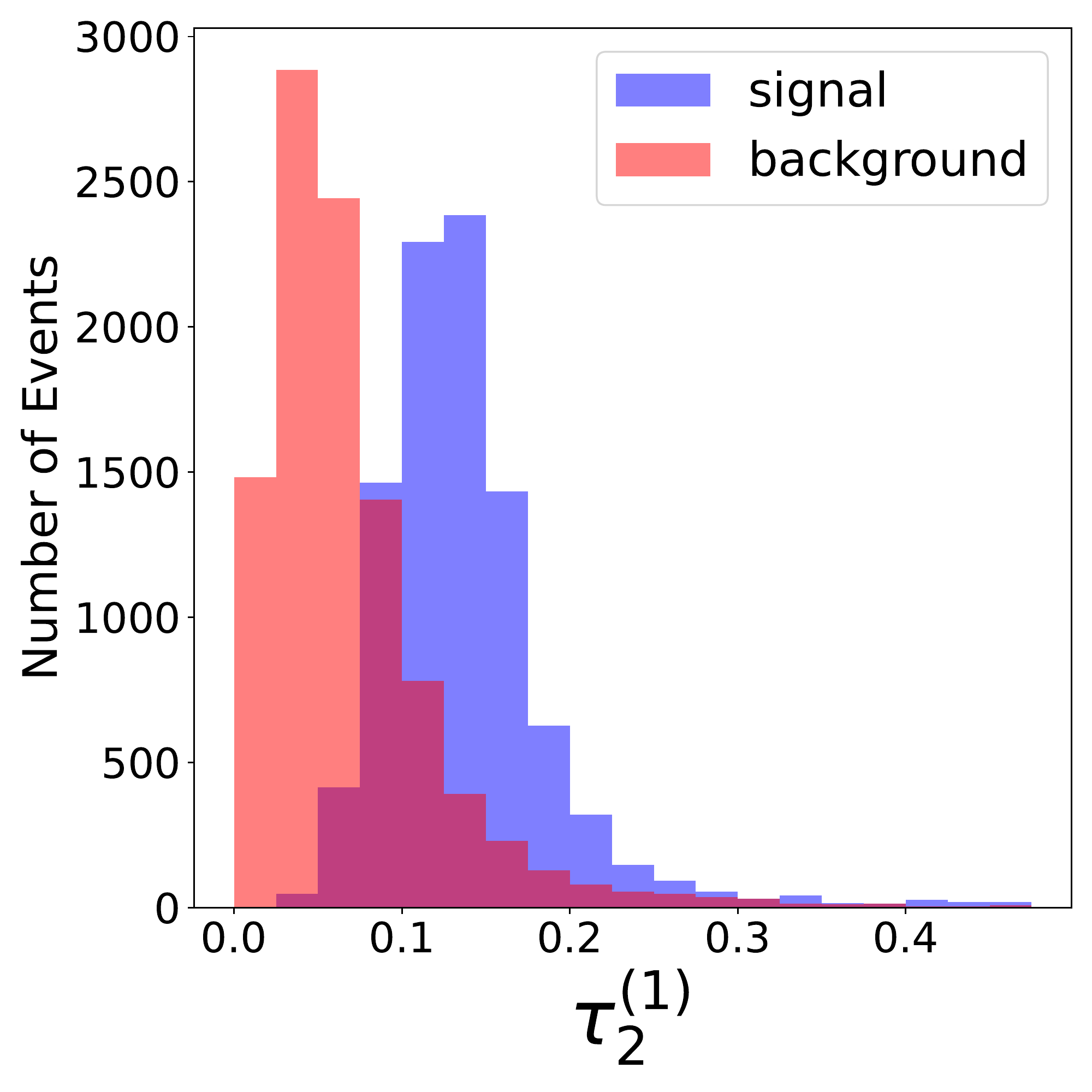}
\label{fig:tau_2_10}            
}
\subfloat[]{
\includegraphics[width=0.25\textwidth]{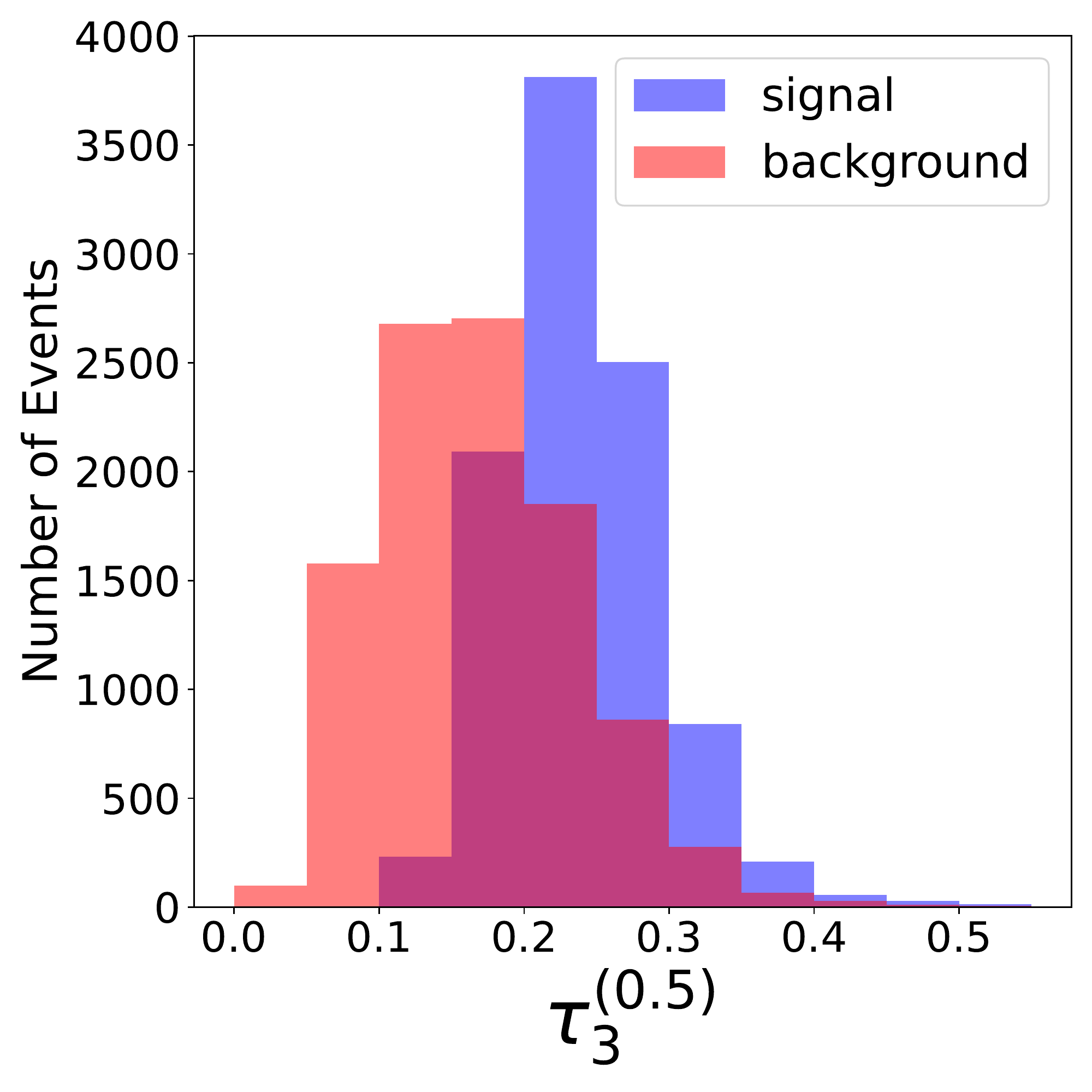}
\label{fig:tau_3_05}            
}
\subfloat[]{
\includegraphics[width=0.25\textwidth]{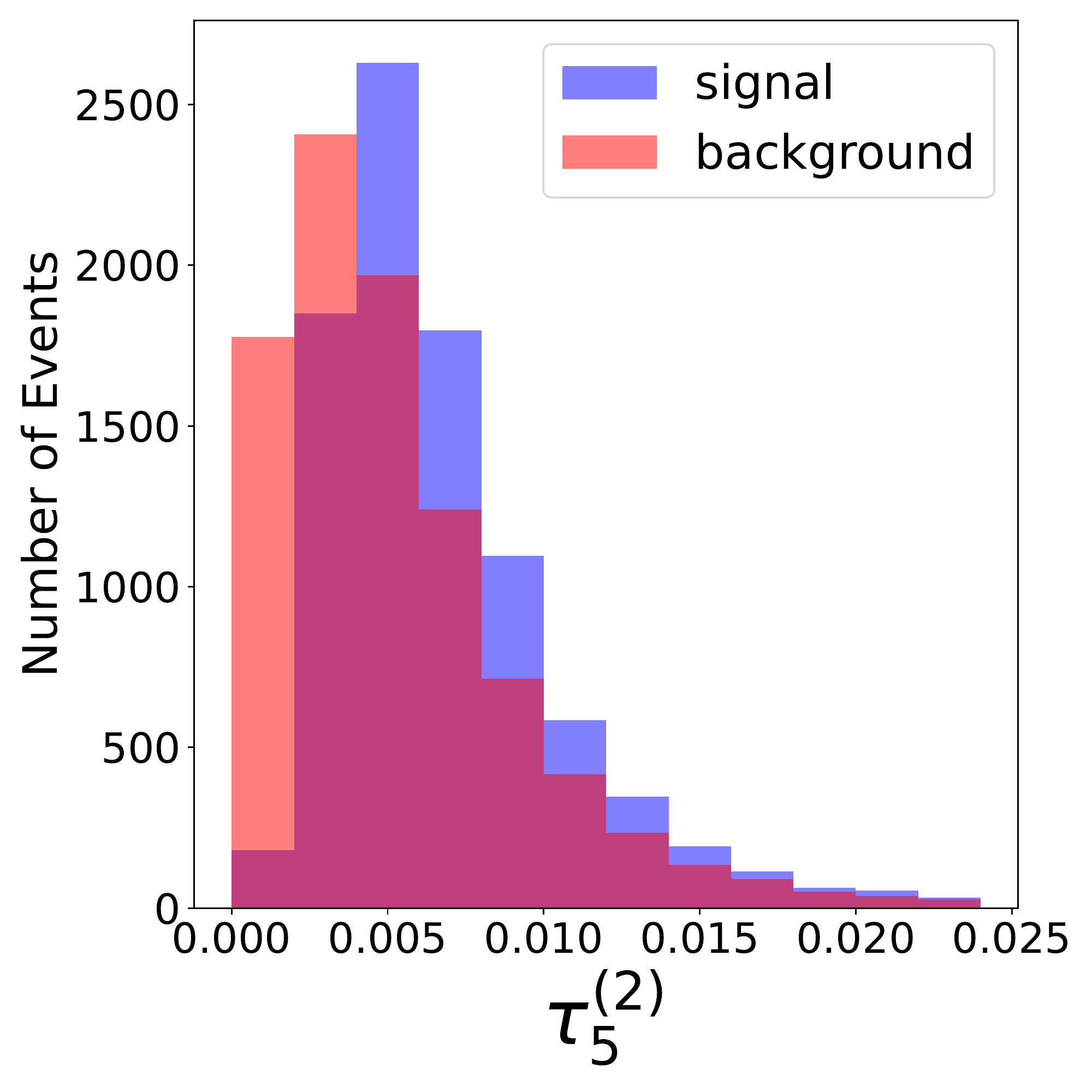}
\label{fig:tau_5_20}            
}
\caption{Distribution of \protect\subref{fig:tau_1_10} $\tau_1^{(1)}$, \protect\subref{fig:tau_2_10} $\tau_2^{(1)}$,  \protect\subref{fig:tau_3_05} $\tau_3^{(0.5)}$, and  \protect\subref{fig:tau_5_20} $\tau_5^{(2)}$ for background and signal jets}
\label{fig:jet-taus}
\end{figure}
    In our work we consider the \mbns{8} model- an MLP consisting of 4 hidden layers with (200, 200, 50, 50) nodes respectively. The \textsc{ReLU} activation function is used for the hidden layers. The output layer consists of 2 nodes transformed by the \textsc{SoftMax} function to respectively represent the probabilities of the jet being classified as a background or signal jet.
    
    \item \textbf{Particle Flow Network (PFN)}~\cite{komiske2019energy}: PFNs are built
    %as infrared- and colinear- (IRC) safe  networks
    following the deep set~\cite{zaheer2017deep} architecture,
    %. The deep set architecture 
    inherently making the network invariant under permutation of particle constituents. The model implements the following relation
    
    \begin{equation}
        \mathrm{PFN} = F\left( \sum_{i=0}^{N-1} \Phi\left(p_i \right) \right)
        \label{eqn:pfn}
    \end{equation}
    where $F$ and $\Phi$ represent non-linear functions implemented as trainable NNs, $N$ is the number of jet constituents and $p_i$ is the four momentum of the $i$-th jet constituent.
    The $\Phi$ network learns to create a latent embedding of each constituent particle. These latent embeddings are summed over for all constituent particles of a jet creating jet-level latent embeddings, making the network invariant under permutation of particle constituents. Finally, the jet-level embeddings are passed on to the $F$ network which performs the jet classification. 
    
    We train the network with the $p_T, \eta, \phi$ of jet constituents as input. As a part of data preprocessing, we standardized the constituents' $\eta$ and $\phi$ by subtracting the jet's  $\eta$ and  $\phi$. Also, the $p_T$ values of the jet constituents are scaled by a the inverse of sum of constituent $p_T$s, i.e. $\sfrac{1}{\sum_i p_{T,i}}$. The $\Phi$ network is implemented as an MLP with 3  layers of 100, 100, and 256 nodes respectively. Each layer is followed by a \textsc{ReLU} activation layer. The output layer of $\Phi$ represents a 256 dimensional latent space of jet representation. The $F$ network consists of 3 hidden layers with 100 nodes per layer with \textsc{ReLU} activations. The output consists of two nodes transformed by the \textsc{SoftMax} operation to represent the probabilities corresponding to each jet class. 
    
\end{itemize}

\section{Interpretability of Machine Learning Models: Tools and Methods}
\label{sec:review-interp}
While a number of XAI techniques have been developed in ML literature, how any one these methods actually \textit{explain} an ML model can actually be quite different from the others~\cite{wang2009feature, van2021evaluating, jesus2021can}. Often we find the XAI techniques producing diverging explanations, making it challenging to rely on these methods. We investigate a number of these techniques and compare the corresponding results and try to understand what may contribute to their divergence. 
%Our investigation looks into which input features are considered as the most important ones by a model and how these feature importance metrics vary across different xAI methods.
Here,  we summarize the methods that we are going to use to explore the interpretability of top tagger models.

\begin{itemize}[leftmargin=*, label={}]
    \item \textbf{The \dAUC method:} Identifying feature importance has been an important part of studying classification models~\cite{tang2014feature}. In standard feature selection tasks, a reasonable subset of the features that excels in some model performance metric is chosen. Although it is conceptually different from feature ranking in \textit{post-hoc} model interpretation, many interpretation metrics also rely on identifying feature importance with a simpler surrogate model which is trained to minimize a model's performance loss~\cite{ribeiro2016should}. One of the most useful model analysis tools for binary classification is the  Region Operator Characteristic (ROC) curve, and the Area Under the Curve (AUC) serves as a scalar metric for evaluating model performance. ROC-AUC based feature ranking has been widely promoted in ML literature~\cite{chen2008fast,wang2009feature,serrano2010feature}. We adapt those same principles for our model interpretation studies. One straightforward way of evaluating a feature's contribution in making predictions is to investigate the model's performance when a particular feature is masked from the input- by replacing it with a population-wide average value or a zero value, whichever is contextually relevant to the model's relationship with the training dataset.
    
    \item \textbf{Shapely Additive Explanations (SHAP)~\cite{lundberg2017unified}:} The SHAP scores represent a game theoretic approach in identifying the importance of difference features. For each instance of the dataset, the input features are assigned an additive score that determines to what extent a particular feature contributes to the classifier prediction. An average model prediction is determined by replacing each feature by its population average and then individual features are added back to the model to find their impact on leading the prediction towards the optimal value. For each feature, the SHAP score is determined by evaluating the average contribution of adding the feature over all possible feature subsets defined without that feature. Given that evaluating exact SHAP scores require iterating over $2^n$ sets of feature combinations for each data instance with $n$ features, several simplifying assumptions are made to reduce the computational complexity. In our work, we use the kernel SHAP method- a model agnostic approach to obtain local explanations similar to the LIME framework~\cite{ribeiro2016model} and obtained by generating random samples around the data point and performing a mean-squared-error-minimizing linear regression over the samples to evaluate the SHAP scores. In order to avoid any overfitting in obtaining the SHAP score, the number of samples in each model were chosen to be at least twice as many as the number of input features
    
    \item \textbf{Layerwise Relevance Propagation (LRP)~\cite{LRP-NN, LRP-overview}:} The LRP technique propagates the classification score predicted by the network backwards through the layers of the network and attributes a partial relevance score to each input. The backpropagation of LRP scores in an MLP network is obtained by the following relation-
    \begin{equation}
        r_j^{(n)} =\sum_k \frac{a_j^{(n)}w_{jk:n}}{\sum_m a_m^{(n)}w_{mk:n}}r_k^{(n+1)}
        \label{eqn:LRP-0}
    \end{equation}
    where $a_j^{(n)}$ and $r_j^{(n)}$ are the activation and relevance scores of the $j$-th node in $n$-th layer and $w_{jk:n}$ is the weight that determines the contribution of the $j$-th activation in the $n$-th layer to the $k$-th node in layer $n+1$. The inputs to the network are identified as the $0$-th layer and the relevance scores assigned to them are denoted as $r_j^{(0)}$. 
    The original LRP method has been developed for simple MLP networks. Variants of this method have been explored to propagate relevance across convolutional neural networks~\cite{LRP-pixel} and graph neural networks~\cite{schnake2021higher}. While the basic LRP rule in Eqn.~\ref{eqn:LRP-0} conserves the total relevance score i.e. the classifier network's output, based on the distribution of weights and activations, relevance scores can become unbounded when  $\sum_k a_j^{(n)}w_{jk:n} \to 0$. To overcome this, the LRP rule is modified to treat postive and negative weights asymmetrically. We use the so called LRP-$\gamma$ rule defined as-
    \begin{equation}
        r_j^{(n)} = \sum_k\frac{a_j^{(n)}\left(w_{jk:n} + \gamma w_{jk:n}^+\right)}{\sum_m a_m^{(n)}\left(w_{mk:n} + \gamma w_{mk:n}^+\right)}r_k^{(n+1)}
        \label{eqn:LRP-gamma}
    \end{equation}
    where $w^+ = w\cdot\Theta(w)$, $\Theta$ being the Heaviside step function and $\gamma$ is a regularization parameter. In our representation, we always present the normalized relevance scores so that $\sum_j r_j^{(0)} = 1$.
    
    \item \textbf{Neural Activation Pattern (NAP) Diagrams~\cite{roy2022interpretability, neubauer2022explainable}:}
    While the aforementioned methods help identify the importance of features for a trained NN model, the NAP diagrmas visualize the information propagation pathways through the network's architecture. The NAP diagram visualizes the Relative Neural Activation (RNA) score, defined as-
    \begin{equation}
    \mathrm{RNA}(j,k;\mathcal{S}) = 
    \frac{\sum_{i=1}^{N} a_{j,k} (s_i)}{\max_j\sum_{i=1}^{N} a_{j,k}(s_i)}
    \label{eqn:rna}
    \end{equation}
    where $\mathcal{S} =\{s_1, s_2 ... s_N\}$ represents a set of samples over which the $\mathrm{RNA}$ score is evaluated. The quantity $a_{j,k} (s_i)$ is the activation of $j$-th neuron in the $k$-th layer when the input to the network is $s_i$. When summed over all the samples in the evaluation set $\mathcal{S}$, this represents the cumulative neural response of a node, which is normalized with respect to the largest cumulative neural response in the same layer to obtain the $\mathrm{RNA}$ score. Hence, in each layer, there will be at least one node with an $\mathrm{RNA}$ score of 1. Since the neurons are activated with \textsc{ReLU} activation in the models we consider, the $\mathrm{RNA}$ score will be strictly non-negative, and $\leq 1$. In a qualitative way, we are trying to see which neurons most actively engage to obtain the predictions from our models. Since the MLPs in our models consist of only Dense layers, each layer takes all the activations from the previous layer as inputs. As all nodes within a given layer are subject to the same set of inputs, we can reliably estimate how strongly they perceive and transfer that information to the next layer by looking at their activation values. For the same reason, we normalize the cumulative activation of a node with respect to the largest aggregate in the same layer. 
    
    The NAP diagram is obtained by presenting the RNA scores for the different layers of the model as a two dimensional heatmap where along the horizontal axis  lies the different activation layers of the network and the vertical axis represents the different nodes in those activation layers. NAP diagrams illustrate the relative activity level of different nodes within each layer and hence can demonstrate the sparsity of the model's activity.

\end{itemize}

The methods that we have explored so far adopt widely different approaches to understanding different aspects of a NN. A summary of their properties is given in  Table~\ref{tab:xAI-comp}. In our analyses, we use models with $\mathcal{O}(10-100)$ inputs, so scalability is not a major bottleneck for application of these methods. On the other hand, although some of these methods allow exploring local explanations, i.e. explanations for individual data samples, we concern our  studies with global explanations alone.

\begin{table}[h]
  \centering
  \begin{tabular}{|c|c|c|c|c| }
  \hline
   & \dAUC & SHAP & LRP & RNA/NAP \\
   \hline 
  Scalability in input dimension &
  \xmark &  \xmark & \checkmark & \checkmark  \\
  \hline 
  Local explanation &
  \xmark &  \checkmark & \checkmark & \xmark  \\
  \hline 
  Global explanation &
  \checkmark &  \checkmark & \checkmark & \checkmark  \\
  \hline 
  Requires Forward Propagation &
  \checkmark &  \checkmark & \checkmark & \checkmark \\
  \hline 
  Requires Backward Propagation &
  \xmark & \xmark & \checkmark & \xmark \\
  \hline 
  Susceptible to spurious correlations &
  \checkmark & \checkmark & \checkmark & \xmark \\
  \hline
  Addresses Model Complexity &
  \xmark & \xmark & \xmark & \checkmark \\
  \hline 
  Requires Retraining &
  \xmark & \xmark & \xmark & \xmark \\
  \hline
  \end{tabular}
  \caption{Comparison of different XAI methods in terms of their implementation heuristics. We consider a method scalable if the complexity of its implementation grows at most by linear order. A local explanation refers to explanation metrics assigned to individual features for a given data point while a global explanation refers to explanation metrics assigned to individual features for the entire dataset.}
  \label{tab:xAI-comp}
\end{table}

\section{Model Interpretability for Top Taggers}
\label{sec:interp}
Ideally, we expect an XAI metric to correctly identify features that the NN consider most important. Hence, any \textit{post-hoc} feature ranking XAI method should ideally identify the same set of features though their relative rankings can moderately vary. However, there is no straightforward correlation among XAI methods introduced in the previous section. In the following subsections, we first investigate and validate these methods in the context of simpler TopoDNN and \mbns{8} models. Building on the insights obtained from these studies, we use these tools to interpret the PFN model and investigate its latent space representation.

\subsection{TopoDNN}
\label{sec:topodnn}
\begin{figure}[!h]
\centering
\subfloat[]{
\includegraphics[width=0.5\textwidth]{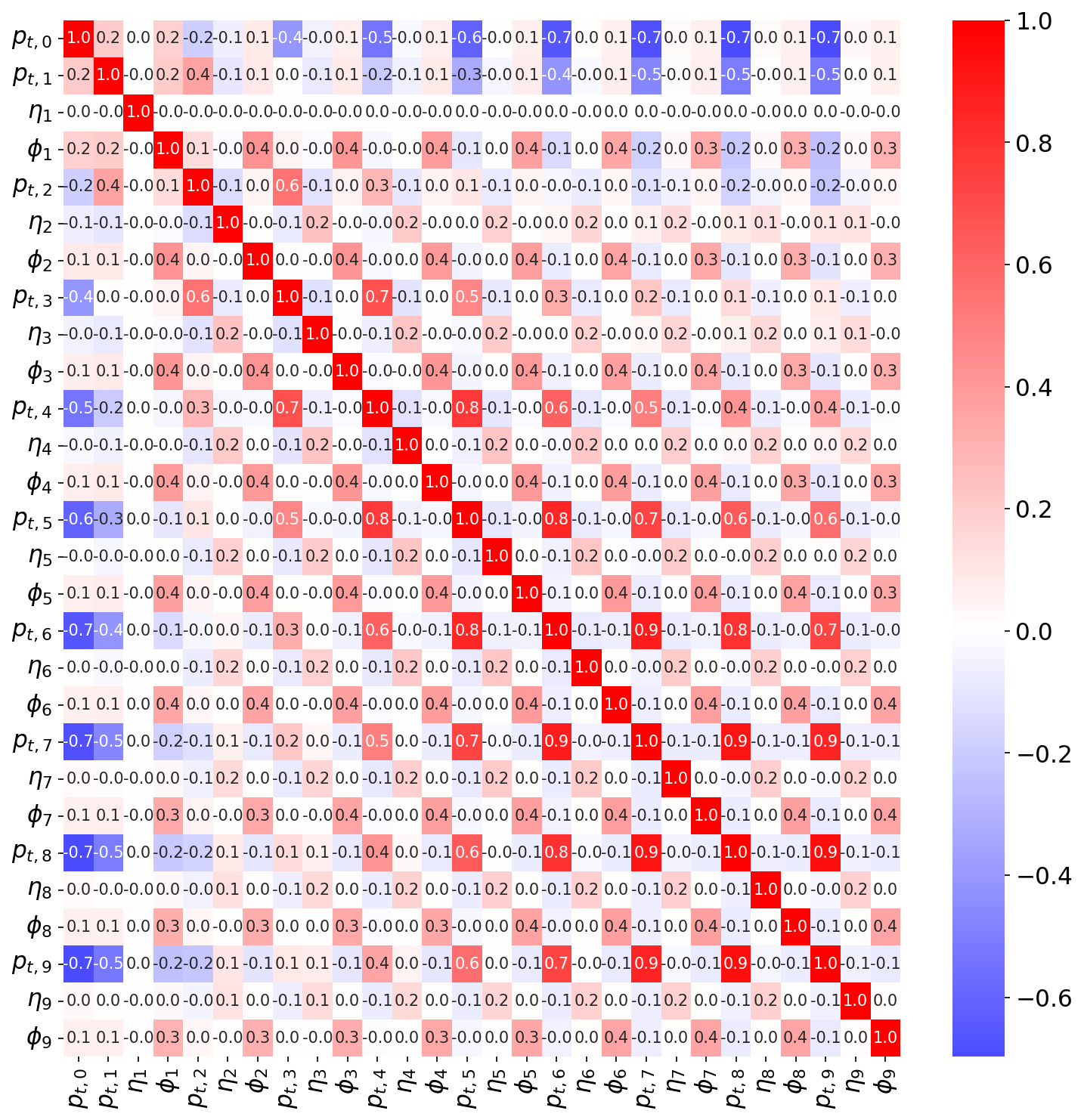}
\label{fig:featTopoDNNbkg}            
}
\subfloat[]{
\includegraphics[width=0.5\textwidth]{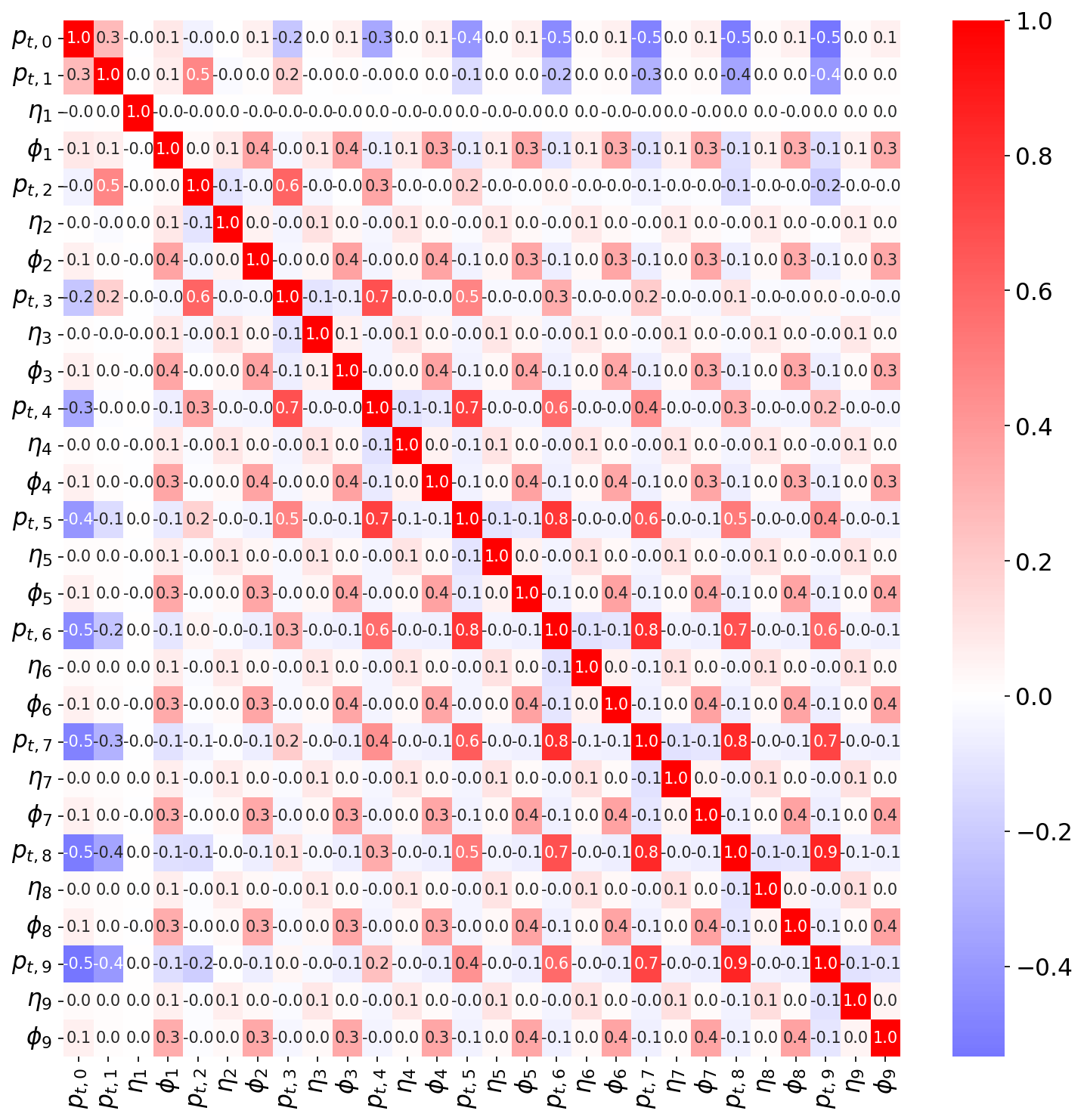}
\label{fig:featTopoDNNsig}            
}
\caption{Feature correlation matrix for $p_T, \eta,$ and $\phi$ of the 10 highest energy constituents of \protect\subref{fig:featTopoDNNbkg} background QCD jets and \protect\subref{fig:featTopoDNNbkg} signal top jets.}
\label{fig:TopoDNfeatcorr}
\end{figure}

\begin{figure}[!h]
\centering
\subfloat[]{
\includegraphics[width=0.33\textwidth]{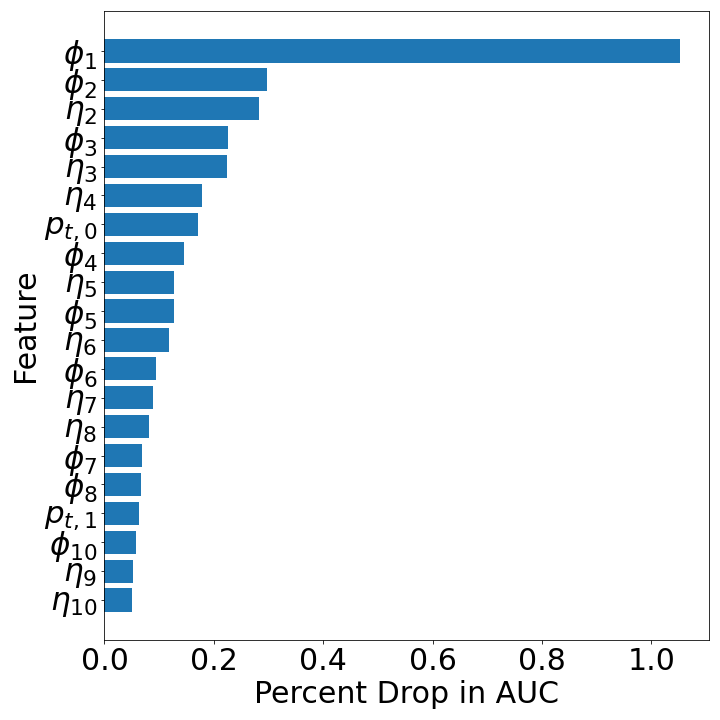}
\label{fig:dAUCTopoDNN}            
}
\subfloat[]{
\includegraphics[width=0.33\textwidth]{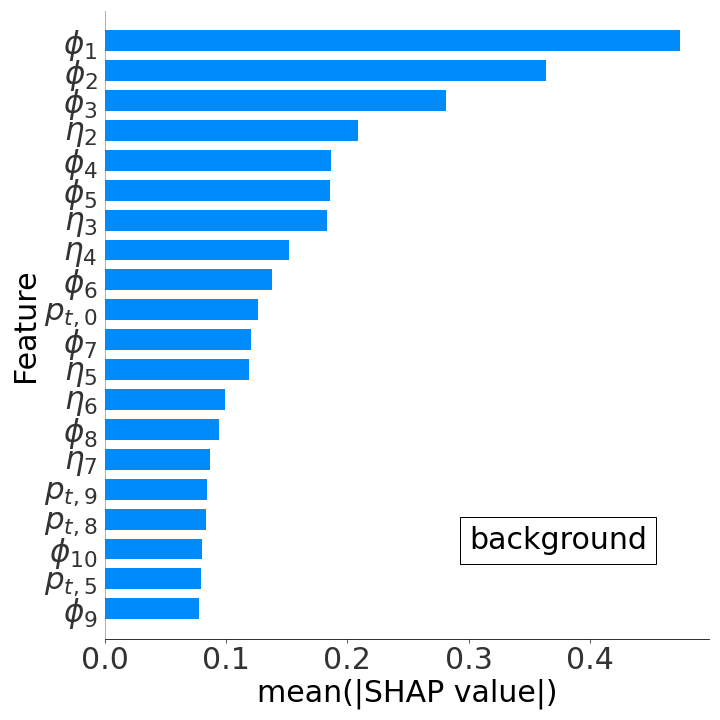}
\label{fig:bkgSHAPTopoDNN}            
}
\subfloat[]{
\includegraphics[width=0.33\textwidth]{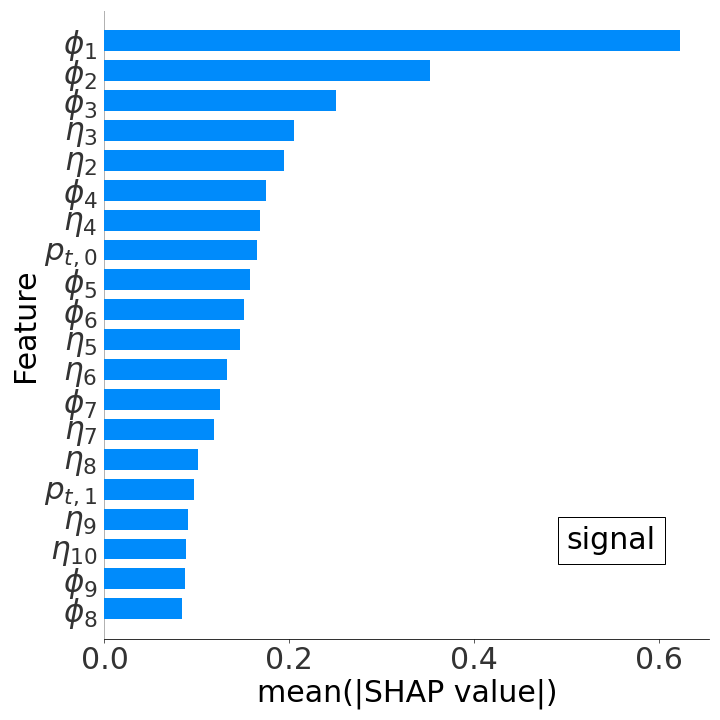}
\label{fig:sigSHAPTopoDNN}
}\\
\subfloat[]{
\includegraphics[width=0.25\textwidth]{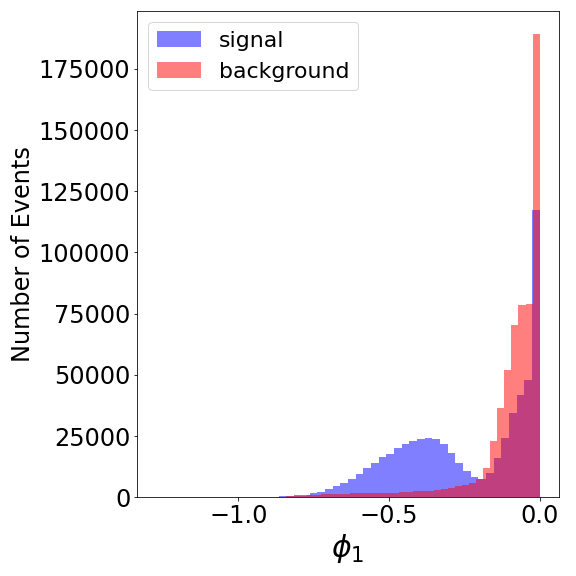}
\label{fig:phi1TopoDNN}            
}
\subfloat[]{
\includegraphics[width=0.25\textwidth]{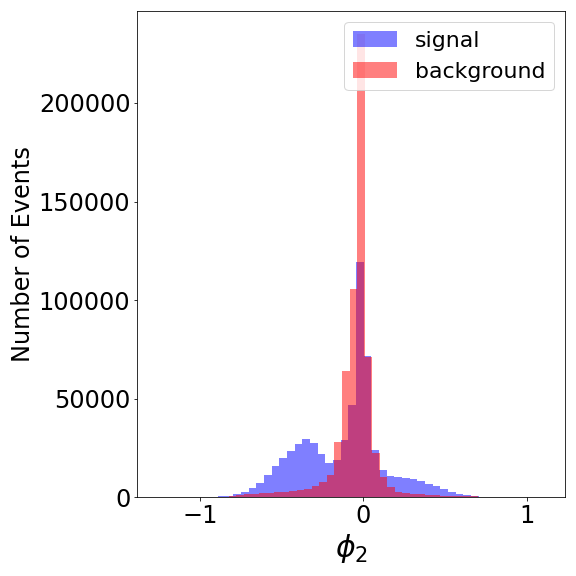}
\label{fig:phi2TopoDNN}            
}
\subfloat[]{
\includegraphics[width=0.25\textwidth]{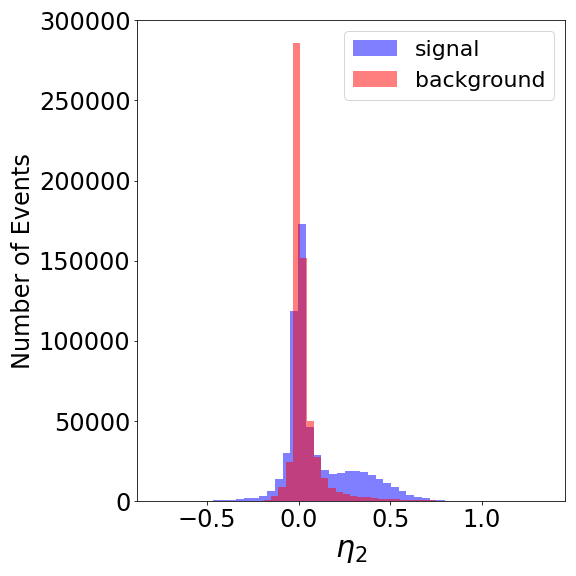}
\label{fig:eta2TopoDNN}
}
\subfloat[]{
\includegraphics[width=0.25\textwidth]{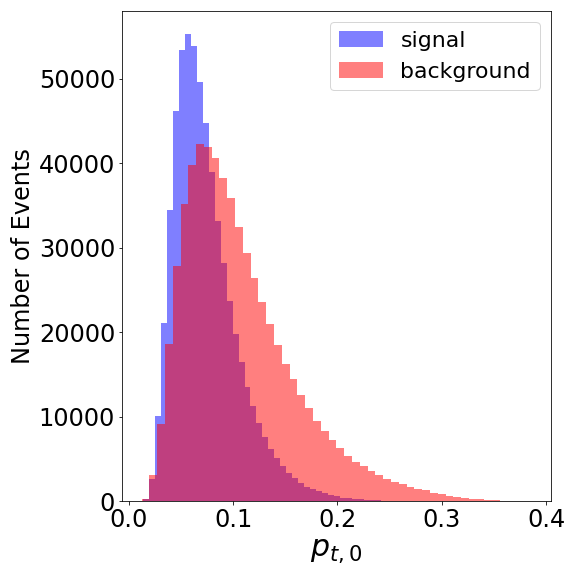}
\label{fig:pt0TopoDNN}
}
\caption{Feature ranking obtained from the \protect\subref{fig:dAUCTopoDNN} \dAUC scores, \protect\subref{fig:bkgSHAPTopoDNN} SHAP scores for background QCD jets, and \protect\subref{fig:sigSHAPTopoDNN} SHAP scores for signal top jets. Only the 20 highest-ranked features are shown. Figures \protect\subref{fig:phi1TopoDNN}, \protect\subref{fig:phi2TopoDNN}, \protect\subref{fig:eta2TopoDNN}, \protect\subref{fig:pt0TopoDNN} show the distributions of the inputs $\phi_1, \phi_2, \eta_2$, and ${p_{t,0}}$ for two jet categories after performing the preprocessing described in Section~\ref{sec:review-tagger}.}
\label{fig:featrankTOPODNN-1}
\end{figure}

Since TopoDNN is arguably the simplest NN-based model to perform top tagging, it is perhaps the most ideal model to investigate different aspects of XAI. Given that correlation among features has been demonstrated to be an important aspect of identifying feature rankings in classical machine learning~\cite{tolocsi2011classification} as well as modern XAI methods, we start by examining the pairwise Pearson correlation coefficient for a subset of the input features for background and signal jets in Figure~\ref{fig:TopoDNfeatcorr}. The correlation matrices for both jet categories are mostly sparse except for some large anti-correlations between $p_{t,0}$, the transverse momentum of the most  energetic jet constituent, and that of some of the low energy constituents. Given the dataset has been generated within a limited jet $p_T$ range, such anti-correlations are expected- the higher the energy of the most energetic constituent, the lower the energy of the remaining constituents. Given that the numerical range of the $p_t$ of lower energy constituents is typically much smaller than that of the highest energy constituent, we can expect the impact of their anti-correlations with $p_{t,0}$ on the NN's performance to be rather small.  We can indeed verify that in Figures~\ref{fig:dAUCTopoDNN}--\ref{fig:sigSHAPTopoDNN} where we identify the important features for the TopoDNN model using the \dAUC  (Figure~\ref{fig:dAUCTopoDNN}) and SHAP scores  (Figures~\ref{fig:bkgSHAPTopoDNN}~and~~\ref{fig:sigSHAPTopoDNN}). 

The \dAUC score cannot independently identify the features that contribute to identification of signal and background jets. However, by evaluating the SHAP scores for subsets of the dataset that only contain one kind of jets, one can identify the features that most dominantly contribute to identification of the corresponding jet class. 
However, as shown in Figures~\ref{fig:bkgSHAPTopoDNN}~and~~\ref{fig:sigSHAPTopoDNN}, there is a significant overlap between the features that are identified as the most important ones for both jet categories. Unlike computer vision models that deal with image or videos as input data and importance distribution for different images can vary based on which pixels carry the most relevant information, the same feature can contribute equally importantly for different classes in models with tabular data. Why the network treats the same set of variables as important becomes clearer upon inspecting the distribution of some of these preprocessed features as shown in Figures~\ref{fig:phi1TopoDNN}--\ref{fig:pt0TopoDNN}. $\phi_1, \phi_2, \eta_2$ all show strong classification characteristics, and given these variables are either loosely correlated or almost uncorrelated as shown in Figure~\ref{fig:TopoDNfeatcorr}, they all can independently contribute to the network's ability to tell apart the different jet classes. On the other hand, $p_{t,0}$  by itself is a modest discriminator and hence identified as having a modest impact on the model's performance. This has also been verified by training a variant of the TopoDNN model that excludes the $p_{t,0}$ variable and as shown in Table~\ref{tab:TopoDNN-perf}, performs almost equally as well as the baseline model.

\begin{figure}[!h]
\centering
\subfloat[]{
\includegraphics[width=0.5\textwidth]{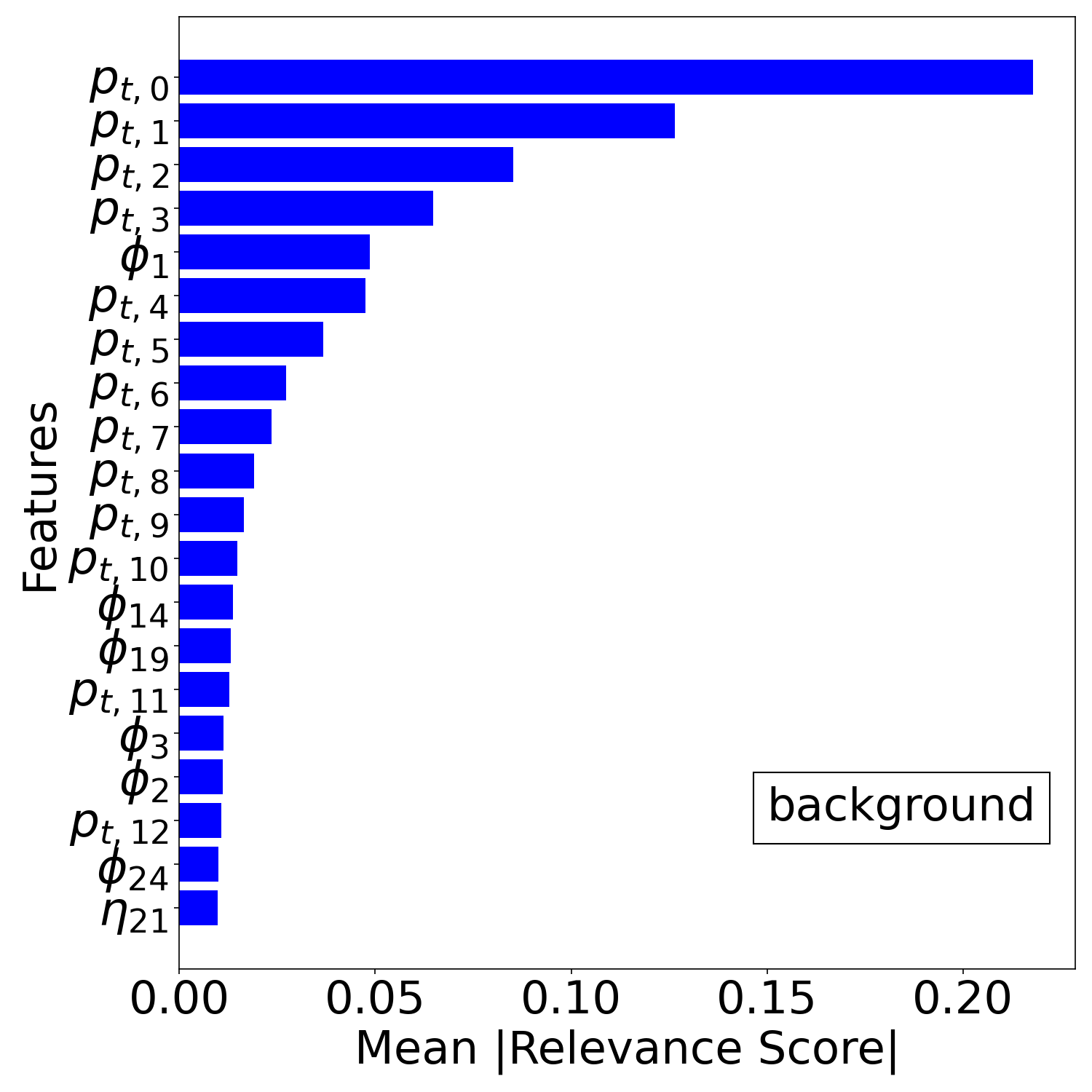}
\label{fig:bkgLRPTopoDNN}            
}
\subfloat[]{
\includegraphics[width=0.5\textwidth]{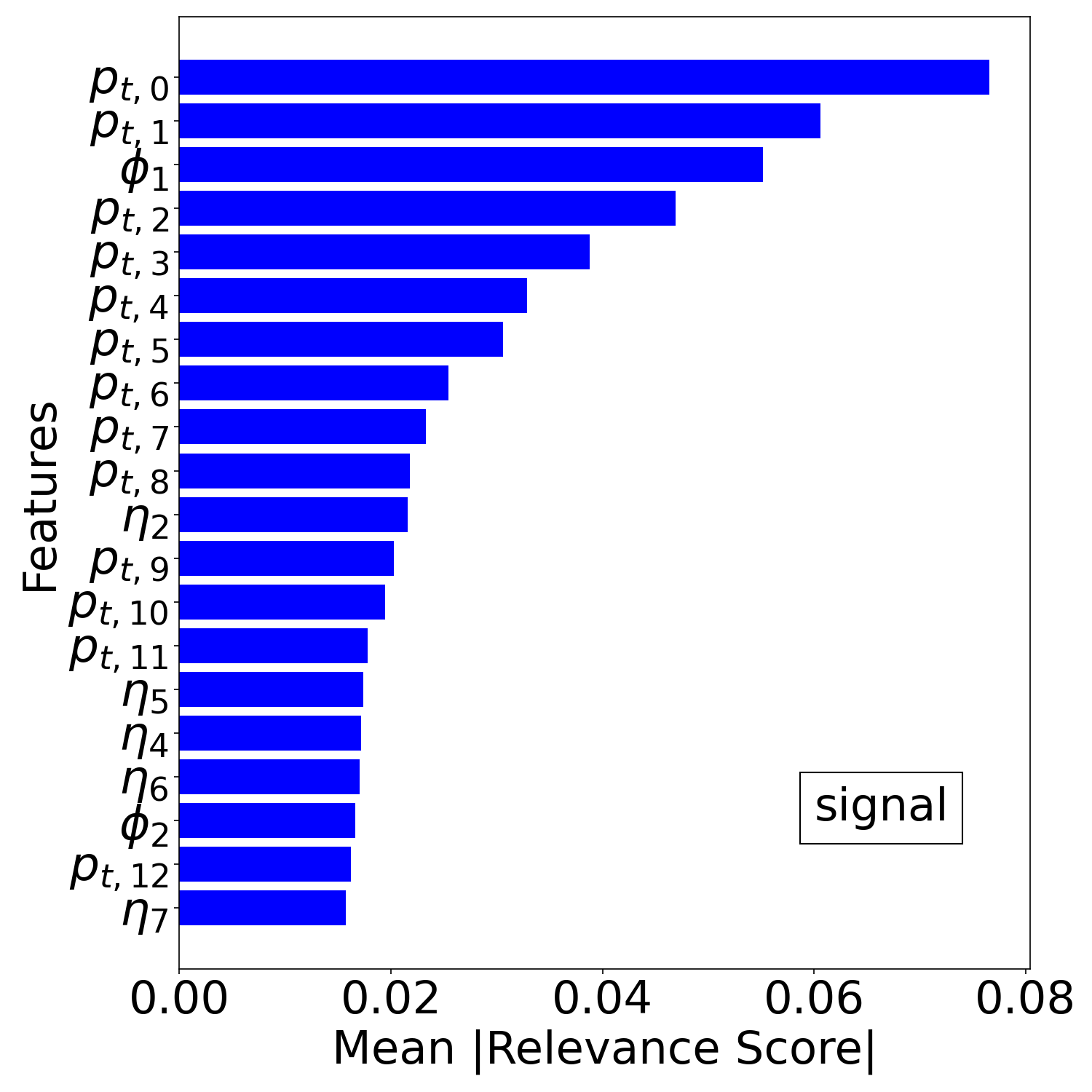}
\label{fig:sigLRPTopoDNN}            
}
\caption{Distribution of the relevance scores obtained from LRP for \protect\subref{fig:bkgLRPTopoDNN} background QCD jets and \protect\subref{fig:sigLRPTopoDNN} signal top jets. Only the 20 highest-ranked features are shown for each jet category.}
\label{fig:featrankTOPODNN-2}
\end{figure}

\begin{figure}[!h]
\centering
\subfloat[]{
\includegraphics[width=0.33\textwidth]{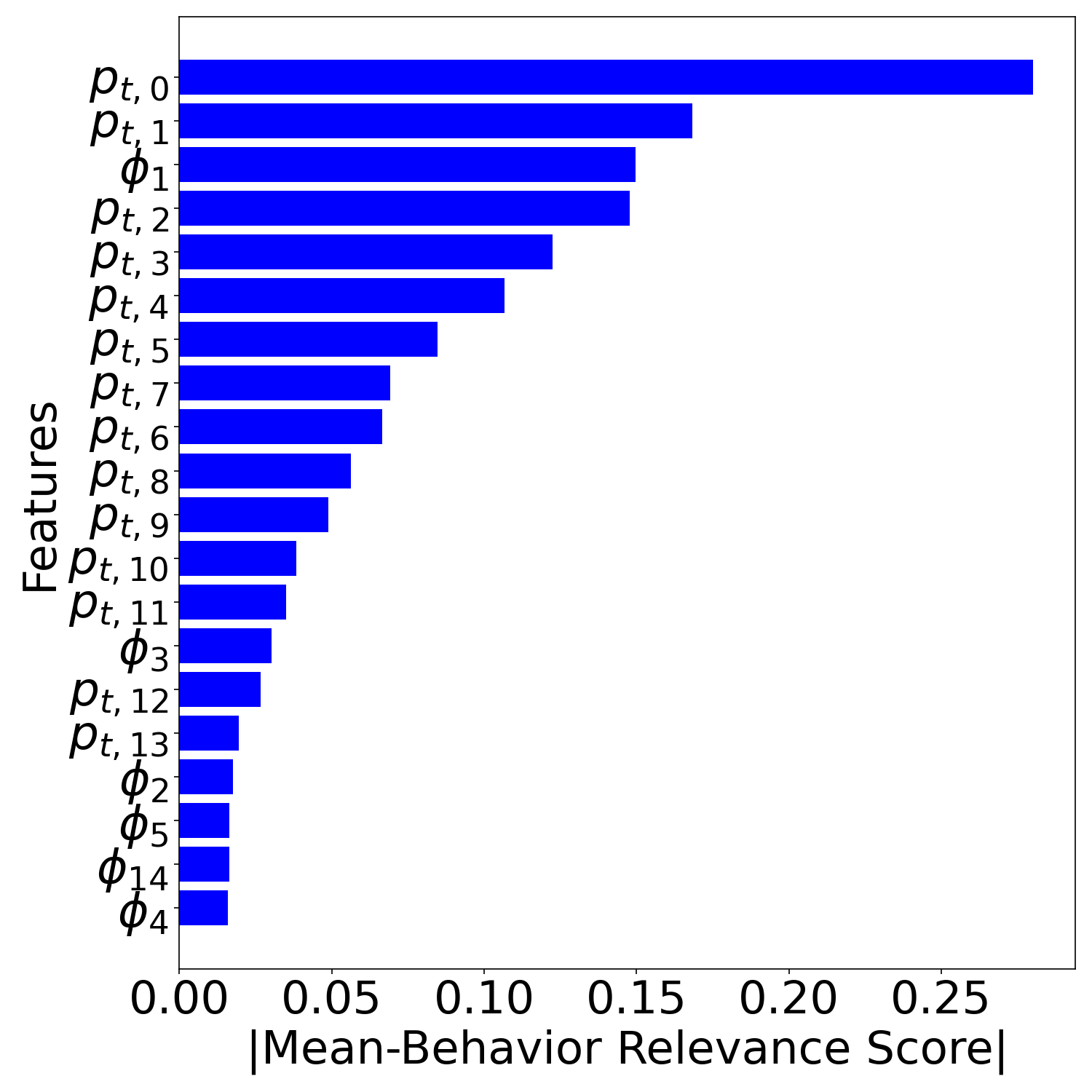}
\label{fig:avgLRPTopoDNN}            
}
\subfloat[]{
\includegraphics[width=0.33\textwidth]{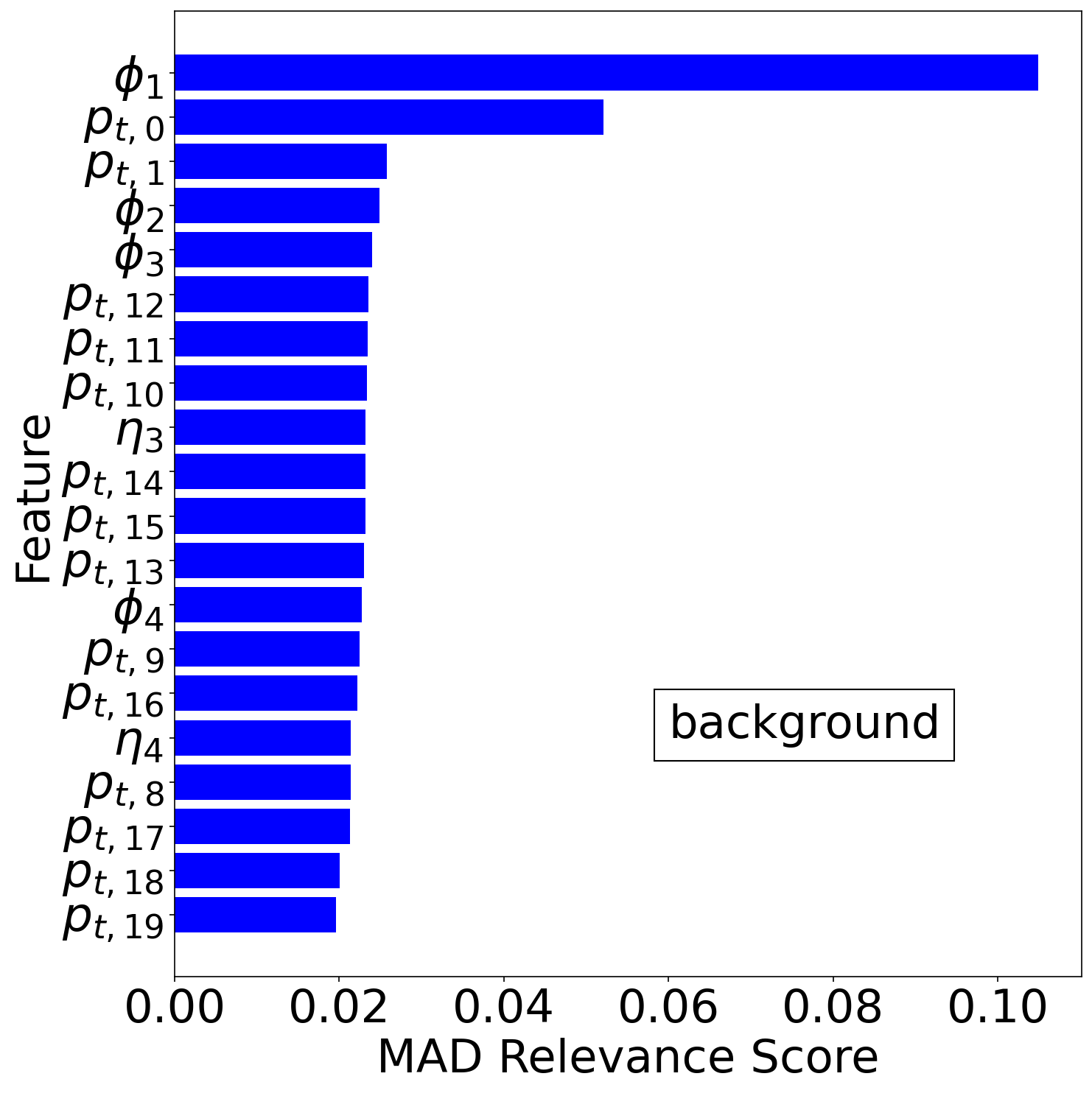}
\label{fig:bkgdLRPTopoDNN}            
}
\subfloat[]{
\includegraphics[width=0.33\textwidth]{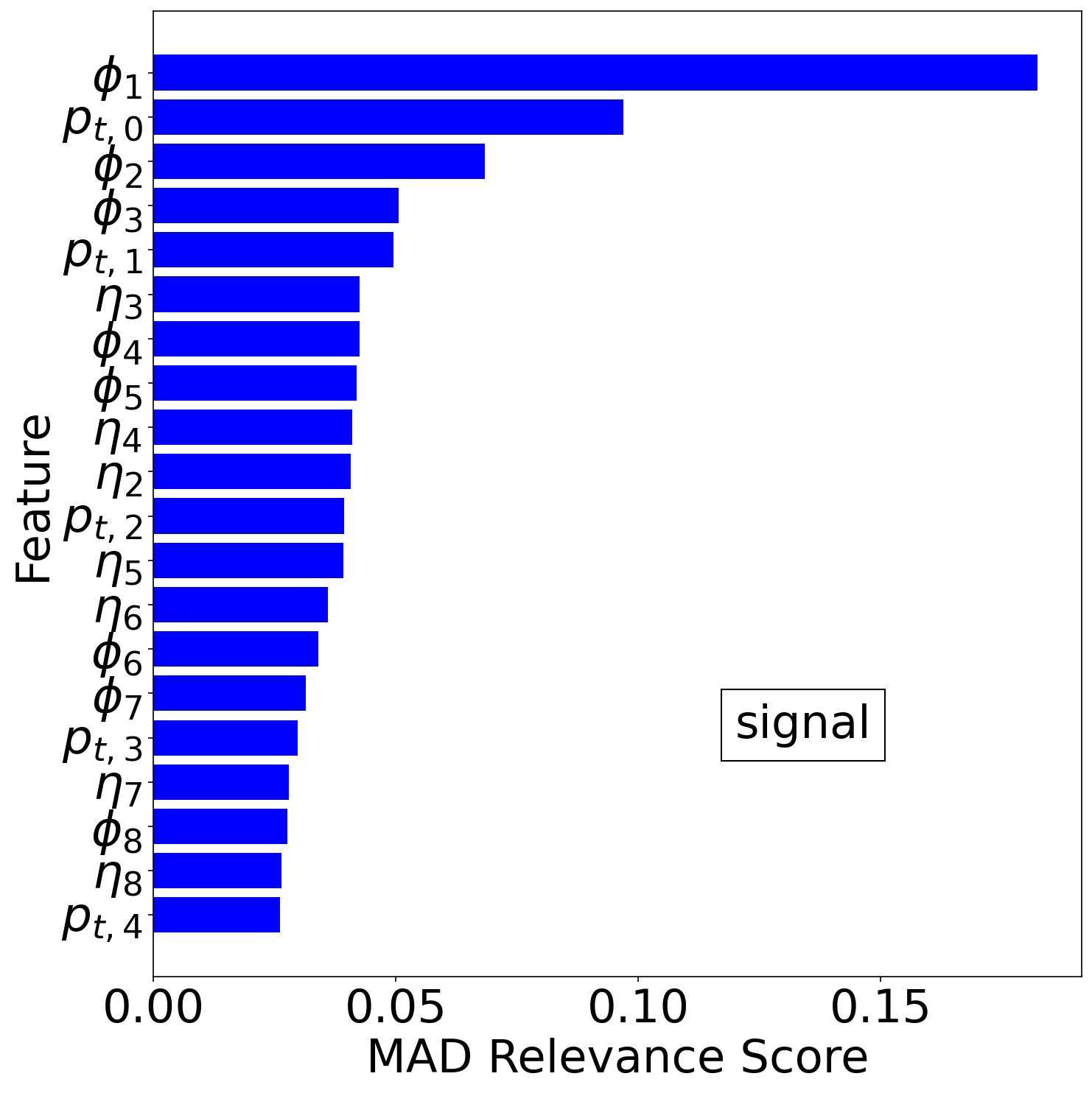}
\label{fig:sigdLRPTopoDNN}            
}
\caption{Distribution of the \protect\subref{fig:avgLRPTopoDNN} absolute mean-behavior relevance score, \protect\subref{fig:bkgdLRPTopoDNN} MAD relevance score for  background QCD jets, and \protect\subref{fig:sigdLRPTopoDNN} MAD relevance score signal top jets.  Only the 20 highest-ranked features are shown. }
\label{fig:LRPanaTOPODNN}
\end{figure}

However, we see a stark difference in the distribution of the relevance scores (Figure~\ref{fig:featrankTOPODNN-2}) among different features, obtained from the LRP method,  when compared to other feature ranking metrics we have considered so far. Unlike SHAP or \dAUC scores, a subset of the $p_t$ variables have the largest relevance scores for both jet categories. While the  most highly ranked features from the previous two methods show strong discriminating characterisitcs, some of the highly ranked features from the LRP method show very little discriminating capacity. This difference can be understood from the nature of these ranking methods. Both \dAUC and SHAP calculate the model's deviation from the \textit{mean behavior}, i.e. qualitatively, they both represent how much information is obtained from inclusion of the true value of a feature instead of using the population mean as a feature mask. On the other hand, LRP calculates the feature's cumulative relevance, which additively includes the relevance scores attributed to each feature's mean behavior. Assuming $\vec{x} = \{x_i\}$ be a sample jet event taken from the set of events $X$ and $\vec{{x}}_{\backslash k} = \vec{x} \backslash \{x_k\} \bigcup  \{\mathbf{E}\left(X_k\right)\}$ be the 
event set where we mask the $k$-th input feature by replacing it with its mean value,
%global mean of the event set, 
the linear order behavior of the NN can be approximated using the Deep Taylor Decomposition formalism~\cite{montavon2017explaining}-

% \begin{equation}
%     f(\vec{x}) \approx f(\vec{\bar{x}}) + \sum_i \frac{\partial f}{\partial x_i}\left(x_i - \bar{x}_i \right)
%     \label{eqn:DTD}
% \end{equation}
\begin{equation}
    f(\vec{x}) \approx f(\vec{{x}}_{\backslash k}) + 
    \frac{\partial f}{\partial x_k}\left(x_k - \bar{x}_k \right)
    \label{eqn:DTD}
\end{equation}

where $f(\vec{x})$ represents the output of the NN before the final \textsc{Sigmoid} activation. 
%While we recognize that Eqn.~\ref{eqn:DTD} is an approximation, especially given that a random sample $\vec{x}$ can be quite far away from $\vec{{x}}_{\backslash k}$ in the Euclidean vector space spanned by $X$ so that nonlinear effects might become important, we only intend to use this relation as a diagnostic tool to inspect the linear order effect observed on relevance scores when the model's inputs are varied away from the population mean. 
Noting that the relevance scores additively distribute the functional output among the different inputs, i.e. $f(\vec{x}) = \sum_i r(x_i)$ and $f(\vec{{x}}_{\backslash k}) = \sum_{i\neq k} r(x_i) + r(\bar{x}_k) $, we can rewrite Eqn.~\ref{eqn:DTD} as,
% \begin{equation}
%     \sum_i r(x_i) \approx \sum_i r(\bar{x}_i) + \sum_i \frac{\partial f}{\partial x_i}\left(x_i - \bar{x}_i \right)
%     \label{eqn:DTD-2}
% \end{equation}
\begin{equation}
    \sum_i r(x_i) \approx \sum_{i\neq k} r(x_i) + r(\bar{x}_k) +  \frac{\partial f}{\partial x_k}\left(x_k - \bar{x}_k \right)
    \label{eqn:DTD-2}
\end{equation}
We define $\delta r_k = f(\vec{x}) - f(\vec{{x}}_{\backslash k}) \approx \frac{\partial f}{\partial x_k}\left(x_k - \bar{x}_k \right)$ as the \textit{differential relevance score} attributed to the corresponding feature. 
When the features are loosely correlated, collecting terms with equivalent indices in Eqn.~\ref{eqn:DTD-2} and ignoring higher order effects, we can write $r(x_k) \approx  r(\bar{x}_k) + \delta r_k$ where we denote $r(\bar{x}_k)$ as the \textit{mean-behavior relevance score}. Figure~\ref{fig:avgLRPTopoDNN} show the absolute mean behavior relevance scores of different features and the relative size and  distribution of the relevance scores are very similar to what we can see in Figure~\ref{fig:featrankTOPODNN-2}. This explains that a large contribution of the LRP scores actually comes from the mean behavior relevances, and has very little to do with the network's ability to distinguish different jet types. In fact, many of the angular variables that are regarded as highly important by \dAUC and SHAP methods also show large differential relevance, as shown in the distributions of Mean Absolute Differential Relevance (MAD Relevance) scores (normalized with respect to relevance scores from the baseline model) in Figures~\ref{fig:bkgdLRPTopoDNN}~and~\ref{fig:sigdLRPTopoDNN}.

\begin{figure}[!h]
\centering
\includegraphics[width=0.5\textwidth]{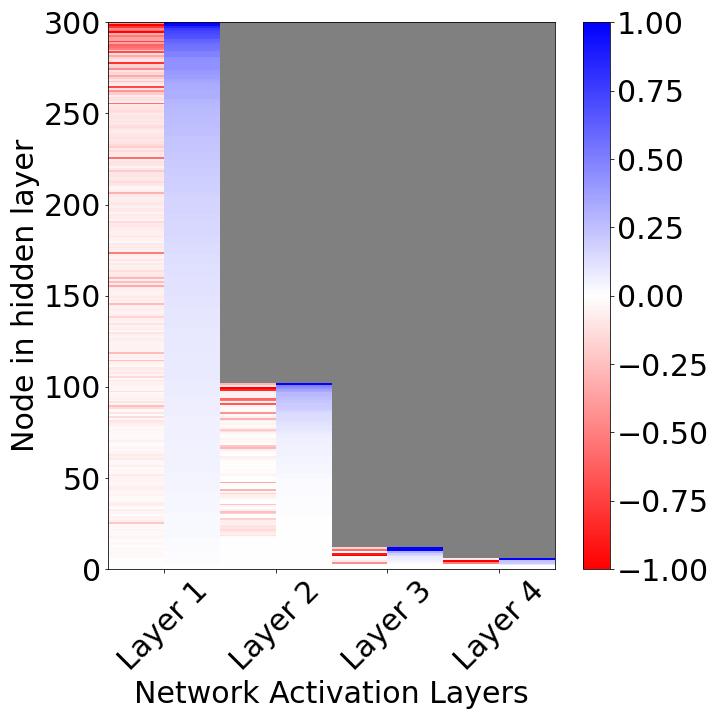}
\caption{NAP diagram for TopoDNN model visualizing a 2D map of RNA scores for different nodes of the activation layers. To simultaneously visualize the scores for QCD and top jets, we project the RNA scores of the former as negative values. The nodes in each layer are ordered according to their RNA scores for the top jets.}
\label{fig:NAPTOPODNN}
\end{figure}

Now we turn our focus to examine the behavior of the internal architecture of the model with NAP diagrams. As discussed in Section~\ref{sec:review-interp}, NAP diagrams plot the RNA scores of different nodes of the activation layers within the network. Figure~\ref{fig:NAPTOPODNN} shows the 2D map of RNA scores for QCD and top jets where the RNA scores of the former are plotted as negative values to allow simultaneous visualization. It can be readily understood that the network in Figure~\ref{fig:NAPTOPODNN} is quite \textit{sparse}, i.e. most nodes show relatively smaller activations. We can heuristically quantify \textit{sparsity} of the network by the fractional number of hidden activation nodes with an RNA score less than a given threshold. We arbitrarily choose this threshold to be 0.2 and find that the network's sparsity measures for background and signal jet categories are 0.86 and 0.76 respectively, giving an overall sparsity measure of 0.70. This implies that about 70\% of the nodes show a cumulative activity level of less than 20\% compared to that of the most active node in the corresponding layer. We also see in Figure~\ref{fig:NAPTOPODNN} that the most active nodes for different jet categories are almost completely disentangled by the time information propagates to the third layer. Large sparsity of the network and early disentanglement of jet categories indicate the network's complexity can be reduced without any noticeable compromise in its performance. To demonstrate this, we have trained variants of the TopoDNN model with lesser complexity by simultaneously reducing the depth and width of the MLP model. As shown in Table~\ref{tab:TopoDNN-perf}, these simplified models perform almost equally well while the model complexity is significantly reduced.

\subsection{\mbns{8}}
\label{sec:multibody}

\begin{figure}[!h]
\centering
\subfloat[]{
\includegraphics[width=0.5\textwidth]{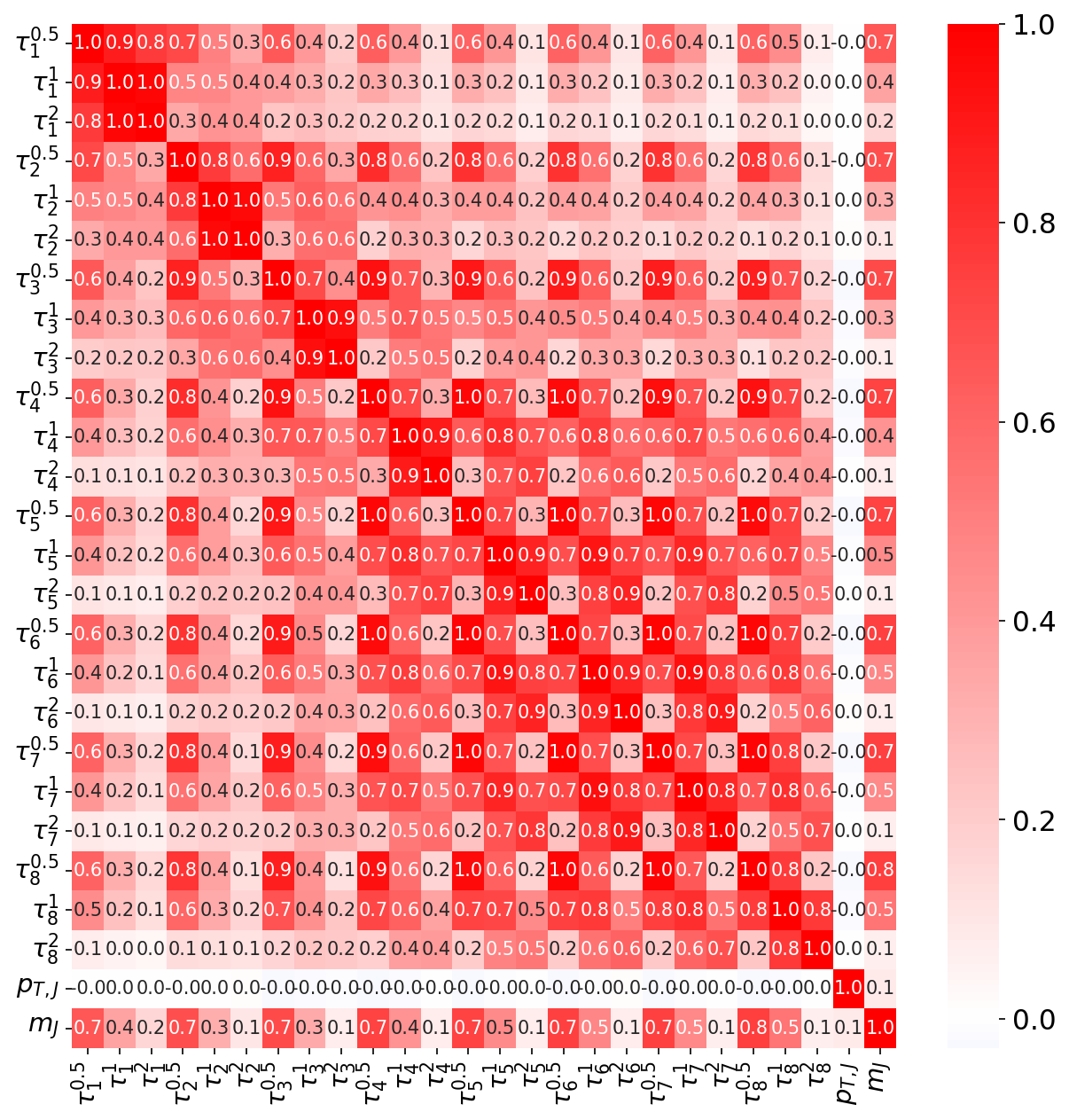}
\label{fig:featMB8Sbkg}            
}
\subfloat[]{
\includegraphics[width=0.5\textwidth]{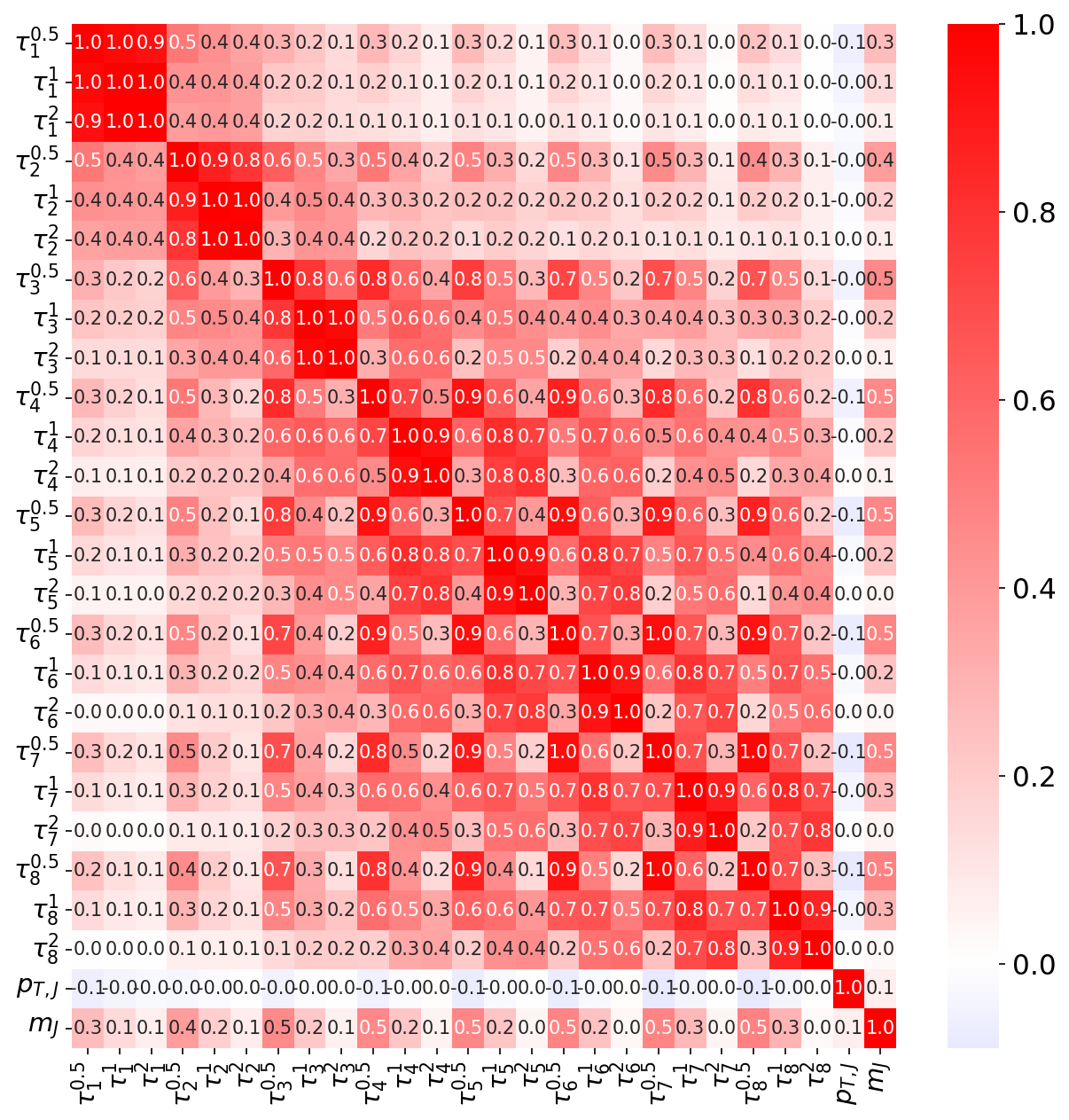}
\label{fig:featMB8Ssig}            
}
\caption{Feature correlation matrix for \protect\subref{fig:featTopoDNNbkg} background QCD jets and \protect\subref{fig:featTopoDNNbkg} signal top jets for the different input features for the \mbns{8} network}
\label{fig:MB8Sfeatcorr}
\end{figure}

Although the underlying architecture of the \mbns{8} model is an MLP, there is stark difference among the input features. Unlike the input to the TopoDNN model explored in Section~\ref{sec:topodnn}, the inputs to the \mbns{8} model are highly correlated for both jet categories (Figure~\ref{fig:MB8Sfeatcorr}). Such large correlations among features can make it hard to distinguish whether a feature truly conveys independent discriminating characteristics or if a feature is deemed important by a model simply because it is correlated with another feature. 
Training a NN with correlated feature inputs can contribute to increased model complexity~\cite{ayinde2019regularizing}, overfitting~\cite{cogswell2015reducing}, and obscure its interpretability~\cite{kaur2020interpreting}. The \mbns{8} network has been trained with \textsc{DropOut} layers~\cite{srivastava2014dropout} with a dropout rate of 0.2 (0.1) for the first (final) two hidden layers to protect it from the problem of overfitting. 
With such large correlations among input features, we want to differentiate between two aspects of a feature's importance- its independent contribution to a network's decision making process and its deemed importance in a certain instance of a trained model because of its correlations with other models.

\begin{figure}[!h]
\centering
\subfloat[]{
\includegraphics[width=0.33\textwidth]{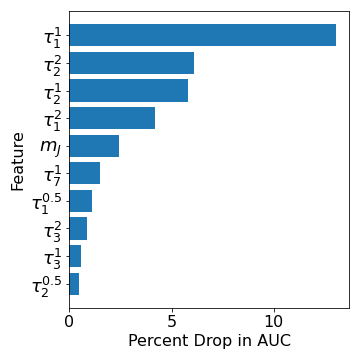}
\label{fig:dAUCMB8S}            
}
\subfloat[]{
\includegraphics[width=0.33\textwidth]{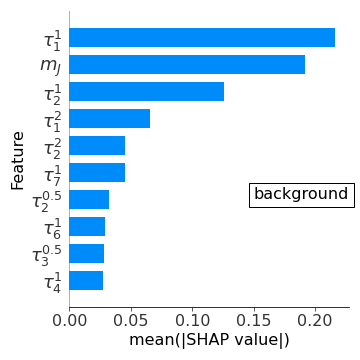}
\label{fig:bkgSHAPMB8S}            
}
\subfloat[]{
\includegraphics[width=0.33\textwidth]{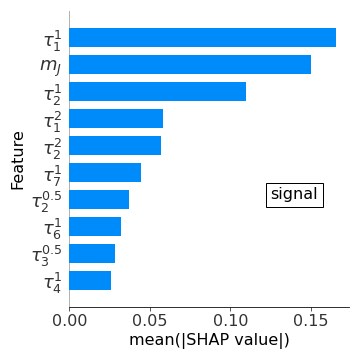}
\label{fig:sigSHAPMB8S}
}
\caption{Feature ranking obtained from the \protect\subref{fig:dAUCMB8S} \dAUC scores, \protect\subref{fig:bkgSHAPMB8S} SHAP scores for background QCD jets, and \protect\subref{fig:sigSHAPMB8S} SHAP scores for signal top jets for the \mbns{8} network. Only the 10 highest-ranked features are shown.}
\label{fig:featrankMB8S-1}
\end{figure}

Figure~\ref{fig:dAUCMB8S} shows the distribution of \dAUC score for the top 10 features. When compared with the SHAP score distributions for background and signal jets in Figures~\ref{fig:bkgSHAPMB8S} and \ref{fig:sigSHAPMB8S}, the sets of top ranking features for these different methods show significant overlap but their raking shows some noticeable difference, especially for the jet mass feature. 
As shown in Figure~\ref{fig:jet-m}, the distribution of jet mass is very different for the two types of jets and naturally, it is expected to be a strong discriminator.
This is also reflected in the distribution of SHAP scores. But the \dAUC ranking places the jet mass variable at a lower ranking compared to subjettiness variables $\tau_1^{(2)}$ and $\tau_2^{(2)}$. 
As shown in Figure~\ref{fig:MB8Sfeatcorr}, these variables demonstrate almost 100\% correlation with $\tau_1^{(1)}$ and $\tau_2^{(1)}$ respectively, both of which are identified as top ranking variables in both \dAUC and SHAP rankings and have strong discriminative distributions as shown in Figures~\ref{fig:tau_1_10} and \ref{fig:tau_2_10}. 
We do not expect a one-to-one correspondence between the feature rankings from \dAUC and SHAP scores. 
%since there is no established, trivial translation between these scores. 
Nevertheless, having a lower rank for jet mass compared to variables that display almost perfect correlations with other strong discriminators naturally intrigues the question whether it is overshadowing the importance of variables that actually adds new information to the classifier. It has been observed that the inclusion of jet mass significantly improves the performance of the \mbns{N} taggers~\cite{liam2019reports}. Hence, the jet mass distribution has the ability to better contribute to a network's decision-making process compared to other highly correlated subjettiness variables. %Hence, the jet mass distribute should rank higher.

In order to investigate whether the highly correlated variables can actually independently contribute to the network's performance, we train a variant of the \mbns{8} network where only the $\tau_x^{(1)}$ variables are included, along with the mass and transverse momentum of jets. This choice is inspired from the block-diagonal concentration of pairwise correlations in Figure~\ref{fig:MB8Sfeatcorr}. This alternate network has almost identical performance as compared to the baseline \mbns{8} network as shown in Table~\ref{tab:TopoDNN-perf}. The feature rankings via \dAUC and SHAP scores for this network also consistently identify $\tau_1^{(1)}, \tau_2^{(1)},$ and jet mass as the most important features for this classification model.
%are shown in Figure~\ref{fig:featrankMB8S_taux1}. We still see that the rank of jet mass is different between \dAUC and SHAP score distributions. However, having removed highly correlated features from the network's inputs , we can safely assume these differences are primarily methodological. 
These top ranking features display relatively weaker correlations among themselves (correlation coefficients $\leq 0.4$) and  hence, 
can contribute new information to the classifier's decision making process. 
Moreover, since the two networks demonstrate almost equivalent performance, the highly correlated subjettiness variables only marginally impact the network's performance. The relatively high \dAUC score attributed to $\tau_1^{(2)}$ and $\tau_2^{(2)}$ in Figure~\ref{fig:dAUCMB8S} must be resulting from strong feature correlations. 

This, however, does not imply that these features are unimportant for the particular instance of the  trained network we investigated. On the contrary, the large \dAUC scores associated with these variables indicate that the trained \mbns{8} model depends on these correlations for proper inference. But the large importance associated with these variables should not be interpreted as their independent contribution to jet classification. 

% \begin{figure}[!h]
% \centering
% \subfloat[]{
% \includegraphics[width=0.33\textwidth]{figures/MB8S/dAUC_MB8S_ptmass_tau_x_1.png}
% \label{fig:dAUCMB8S_taux1}            
% }
% \subfloat[]{
% \includegraphics[width=0.33\textwidth]{figures/MB8S/SHAP_MB8S_tau_x_1_bkg.png}
% \label{fig:bkgSHAPMB8S_taux1}            
% }
% \subfloat[]{
% \includegraphics[width=0.33\textwidth]{figures/MB8S/SHAP_MB8S_tau_x_1_sig.png}
% \label{fig:sigSHAPMB8S_taux1}
% }
% \caption{Feature rankings obtained from the \protect\subref{fig:dAUCMB8S_taux1} \dAUC scores, \protect\subref{fig:bkgSHAPMB8S_taux1} SHAP scores for QCD background, and \protect\subref{fig:sigSHAPMB8S_taux1} SHAP scores for top jets for the \mbns{8} network trained with $\{ \tau_x^{(1)} \} \bigcup \{ p_{T,J}, m_J \}$ variables}
% \label{fig:featrankMB8S_taux1}
% \end{figure}

\begin{figure}[!h]
\centering
\subfloat[]{
\includegraphics[width=0.25\textwidth]{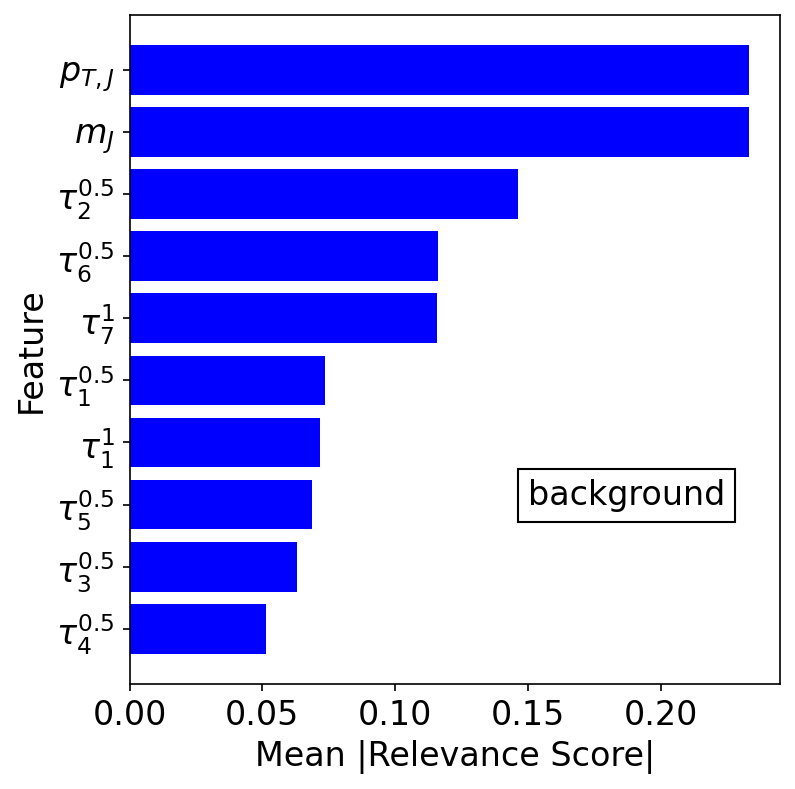}
\label{fig:LRPMB8Sbkg}            
}
\subfloat[]{
\includegraphics[width=0.25\textwidth]{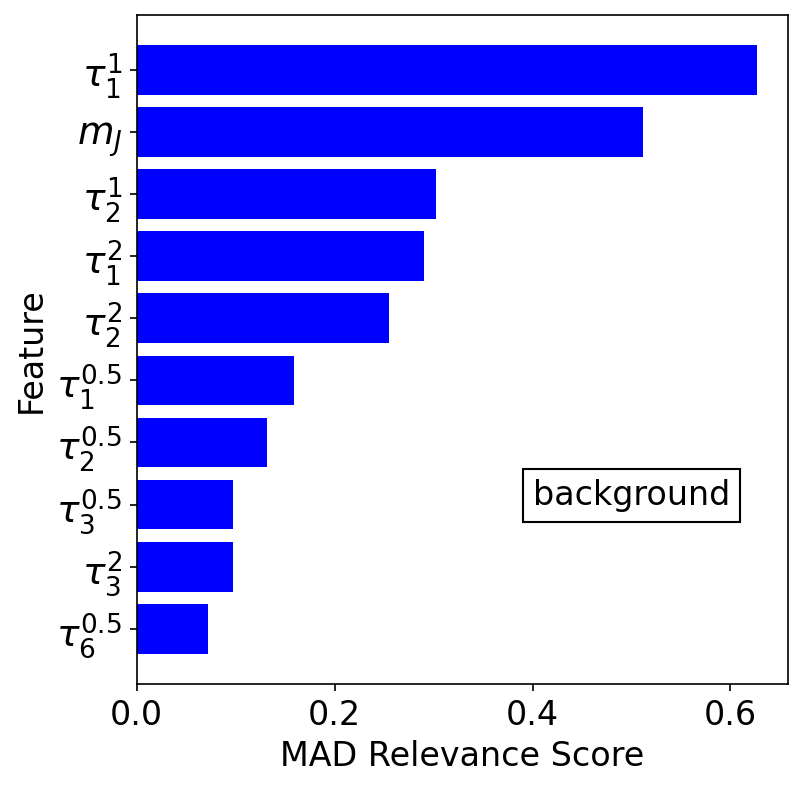}
\label{fig:dLRPMB8Sbkg}            
}
\subfloat[]{
\includegraphics[width=0.25\textwidth]{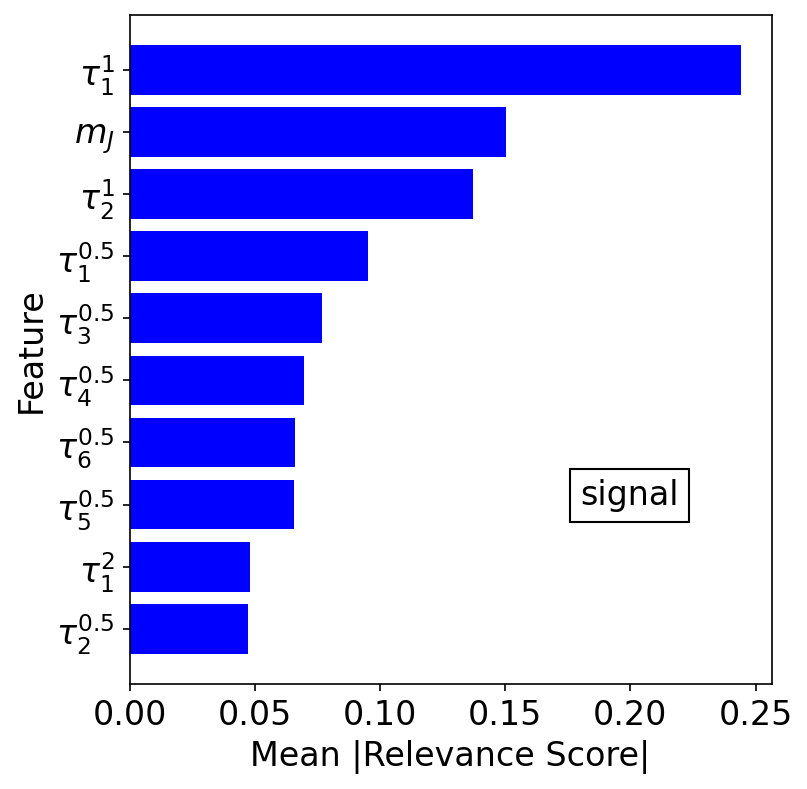}
\label{fig:LRPMB8Ssig}            
}
\subfloat[]{
\includegraphics[width=0.25\textwidth]{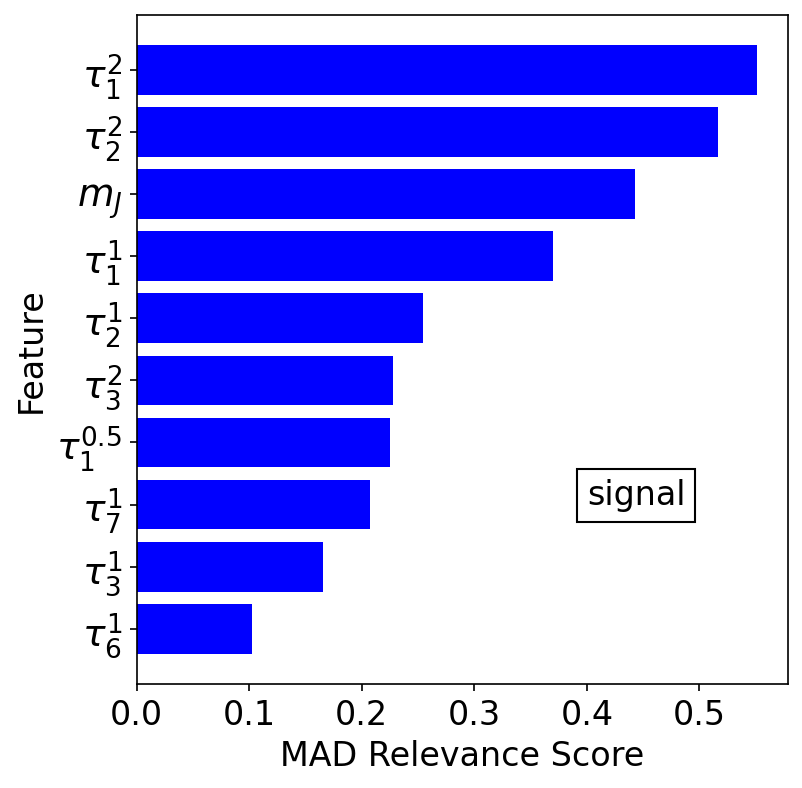}
\label{fig:dLRPMB8Ssig}            
}
\caption{Distribution of \protect\subref{fig:LRPMB8Sbkg} relevance scores for background QCD jets, \protect\subref{fig:dLRPMB8Sbkg} MAD relevance scores for background QCD jets, \protect\subref{fig:LRPMB8Ssig} relevance scores for signal top jets, and \protect\subref{fig:dLRPMB8Ssig} MAD relevance scores for signal top jets for the \mbns{8} network.  Only the 10 highest-ranked features are shown for each jet category.}
\label{fig:featrankMB8S-2}
\end{figure}

Next we turn our attention to relevance scores attributed to different input features by the LRP method. From our studies of the LRP method for the TopoDNN model, we know the relevance scores can be an unreliable measure in identifying how important a feature really is. We can see that pattern repeated for the \mbns{8} network too. Figures~\ref{fig:LRPMB8Sbkg} and \ref{fig:LRPMB8Ssig} show the mean absolute relevance scores attributed to different features for background and signal jets. LRP attributes a large relevance score to $p_{T,J}$, the transverse momentum of the jet for the background jets.
%s attributed for background jets is very different from what we find from previously explored feature ranking methods. 
%Jet $p_T$ is assigned the largest relevance score for background jets. 
However, this variable is one of the least expressive features for the network. It has almost no correlation with other features, and has a very similar distribution for both jet types. Hence, assigning this feature a very large relevance score definitely raises some concerns about the reliability of the feature ranking obtained from LRP. 
%interpretation of this network.  
The distribution of MAD relevance scores in Figure~\ref{fig:dLRPMB8Sbkg} gives a more appropriate distribution for feature importances. For the top jets, the MAD relevance distribution (Figure~\ref{fig:dLRPMB8Ssig}) identifies $\tau_1^{(2)}, \tau_2^{(2)}, m_J$ as the most important features. $\tau_1^{(2)}, \tau_2^{(2)}$ also ranked high in the \dAUC metric and we have explained how their importance primarily stems from their correlations with other expressive input features to the network. 
%, similar to what we have found from SHAP scores. 
% However, in the distribution of MAD relevance score (Figure~\ref{fig:dLRPMB8Ssig}) $\tau_1^{(2)}$ ranks higher. Also, $\tau_2^{(2)}$ and $\tau_1^{(0.5)}$ have larger MAD relevance scores compared to $\tau_2^{(1)}$. This ambiguity can be attributed to the very large correlations among these features. 

\begin{figure}[!h]
\centering
\includegraphics[width=0.5\textwidth]{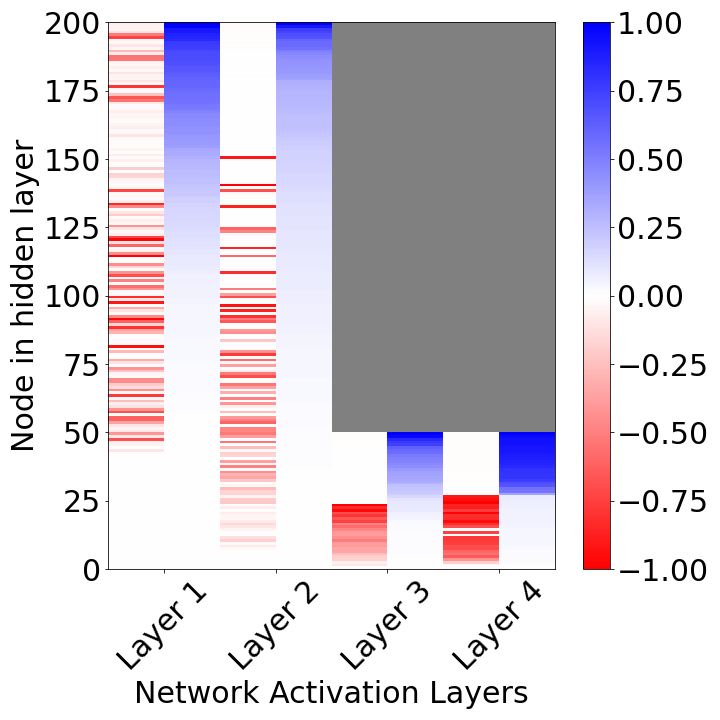}
\caption{NAP diagram for \mbns{8} model visualizing a 2D map of RNA scores for different nodes of the activation layers. To simultaneously visualize the scores for QCD and top jets, we project the RNA scores of the former as negative values. The nodes in each layer are ordered according to their RNA scores for the top jets.}
\label{fig:NAPMB8S}
\end{figure}

To demonstrate how the different hidden activation layers contribute to information propagation for the two jet categories, we show the 2D map of the RNA scores of different nodes in Figure~\ref{fig:NAPMB8S}. Although the network appears to be relatively sparse for different jet categories, with a sparsity measure of 0.74 and 0.64 for background and signal jets measured with respect to a threshold of 0.2 on RNA scores, different nodes are most strongly activated for signal and jet categories. As a result, the network's overall sparsity becomes 0.44.   However, we can already see that the nodes that are most strongly activated by the two jet categories are almost completely disentangled at Layer 2. This indicates that the network might be simplified by choosing a shallower network and indeed verified in Table~\ref{tab:TopoDNN-perf} where we see that models trained without the final two layers have almost identical performance metrics.

\subsection{PFN}
\label{sec:pfn}

While the previously analyzed TopoDNN and \mbns{8} models both employed a single MLP to perform the jet classification, PFN utilizes a deep set topology that linearly combines particle-level neural embeddings to obtain a jet level latent space, which eventually is used to train a second MLP to learn the classification task.
% Before going into details about interpretability of PFN, we would like to point out that our implementation of PFN obtained significantly better performance benchmark compared to what is reported in Ref.~\cite{kasieczka2019machine} where the different models reported accuracies between 0.89 and 0.93 with ROC-AUC score ranging between 0.955 and 0.985. We obtained an accuracy of 0.977 and an ROC-AUC score of 0.997, which also outperforms the permutation invariant symmetry-guided PELICAN~\cite{bogatskiy2022pelican}, graph-based Interaction Network (IN)~\cite{moreno2020jedi} and LorentzNet~\cite{gong2022efficient} models as well as the transformer-based Particle Transformer (ParT)~\cite{qu2022particle} model. The major difference comes from the model's inputs. While the PFN model trained in Refs~\cite{kasieczka2019machine} directly used the constituents' four momenta as input to the, we trained the network with $p_T, \eta, \phi$ variables of the constituents. The constituent $\eta$ and $\phi$ values were standardized according to the prescription in Ref.~\cite{komiske2019energy} where $(\eta_i, \phi_i, p_{T,i}) \to (\eta_i - \eta_J, \phi_i - \phi_J, \frac{p_{T,i}}{\sum_j p_{T,j}})$. 
% %However, instead of normalizing the constituent $p_T$ by the inverse of scalar sum of the constituents' $p_T$, they were scaled by the inverse of the jet's transverse momentum.

\begin{figure}[!h]
\centering
\subfloat[]{
\includegraphics[width=0.33\textwidth]{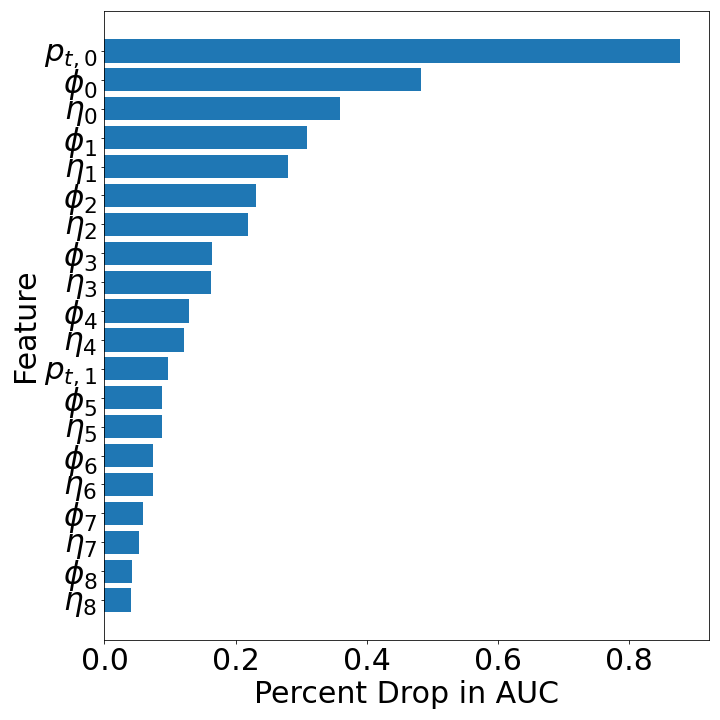}
\label{fig:dAUCptetaphiPFN}            
}
\subfloat[]{
\includegraphics[width=0.33\textwidth]{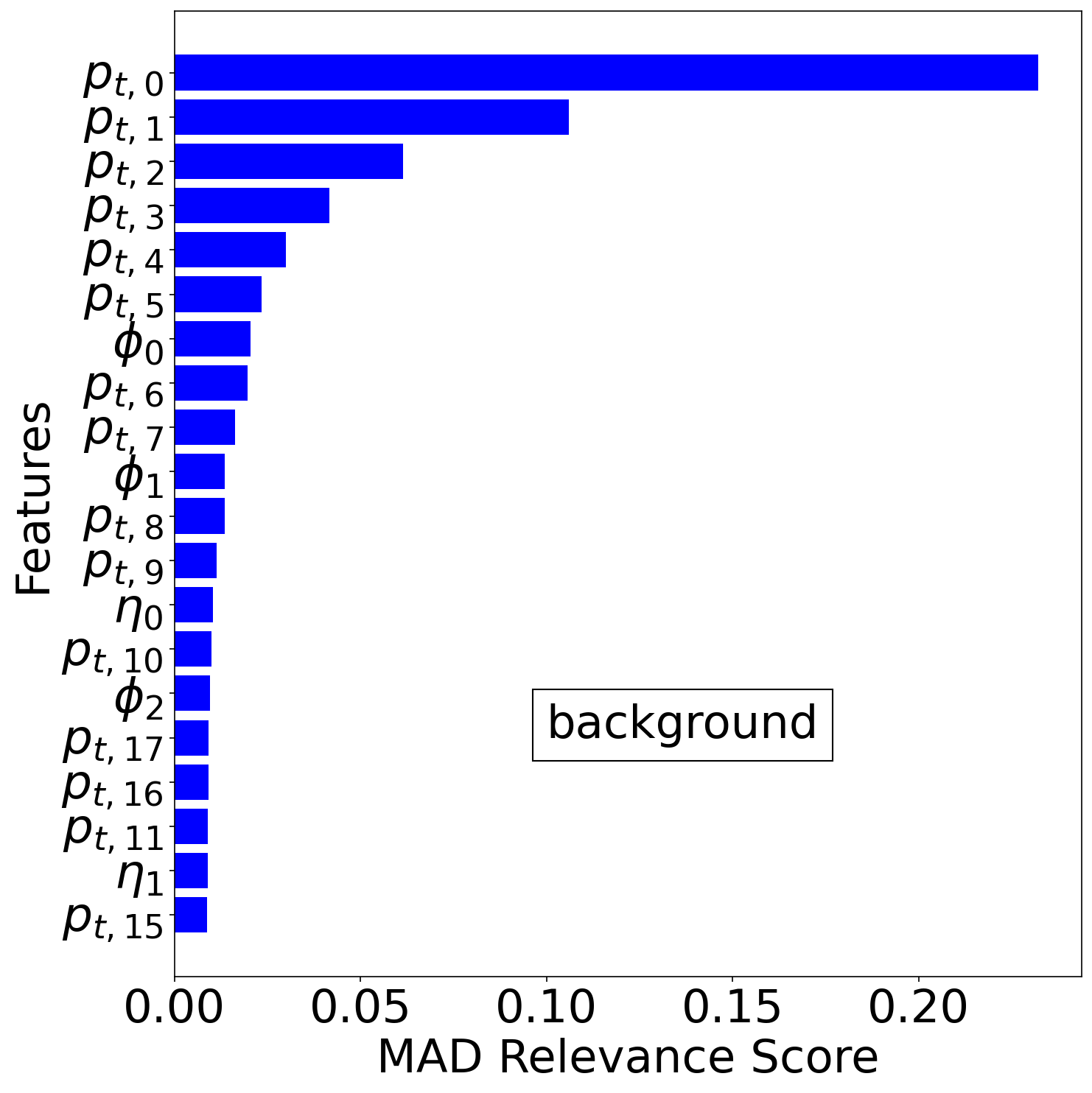}
\label{fig:bkgdLRPptetaphiPFN}            
}
\subfloat[]{
\includegraphics[width=0.33\textwidth]{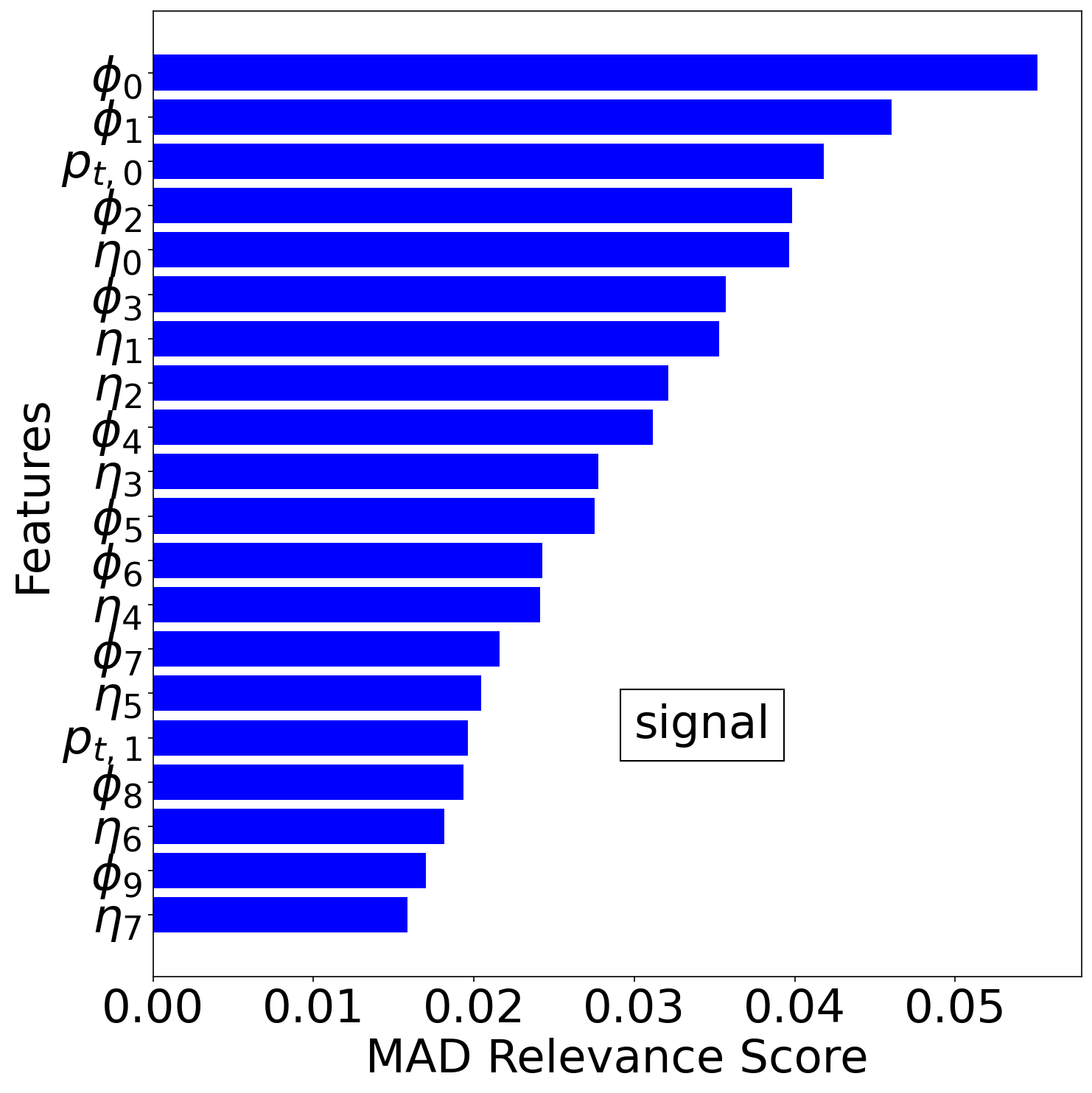}
\label{fig:sigdLRPptetaphiPFN}
}
\caption{Feature ranking obtained from the \protect\subref{fig:dAUCptetaphiPFN} \dAUC scores, \protect\subref{fig:bkgdLRPptetaphiPFN} MAD relevance scores for background QCD jets, and \protect\subref{fig:sigdLRPptetaphiPFN} MAD relevance scores for signal top jets for the PFN network.  Only the 20 highest-ranked features are shown.}
\label{fig:featrankptetaphiPFN-1}
\end{figure}

We start by examining the feature ranking from the \dAUC and MAD relevance scores in Figure~\ref{fig:featrankptetaphiPFN-1}.
We note that the ranking of features obtained from the \dAUC metric is somewhat different from the ranking of the MAD relevance scores for the two jet categories. 
For instance, the azimuthal angle of the most energetic jet constituent, $\phi_0$ appears with a low MAD relevance score for background QCD jets while appearing as the top-ranked feature for the signal jets while also reporting a relatively large \dAUC value.
On the other hand, the relative transverse momentum of that same constituent has the largest MAD relevance score for background jets as well as the largest \dAUC score while being ranked after $\phi_0, \phi_1$ for the signal jets. 
This difference in MAD relevance ranking for the two jet classes is somewhat unlike what we have seen for the classifier models previously investigated. 
These deviations in relative rankings are understood by examining the distributions of the corresponding features. 
Note that in linear order, the differential relevance $\delta r_k \propto x_k - \Bar{x}_k$. Hence, larger deviations from the mean can lead to larger differential relevance scores. 
%It is instructive to examine the distributions of deviation from mean for some of these input features to understand their relative rankings. 
The distribution of $\phi_0 - \Bar{\phi}_0$ is sharply peaked at zero for background jets. Hence, it is reasonable for this variable to have a low MAD relevance score for background jets. On the other hand, the same variable shows long-tailed distributions for the signal jets and hence, has a larger impact on its MAD relevance score. Similarly, the distribution of $p_{t,0} - \Bar{p}_{t,0}$ has long tails for both jet classes with a wider spread for the background jets. As a result, this variable shows up high in ranking for both jet classes, while being ranked higher for background jets compared to signal jets.

% The rankings obtained from both methods show a consistent distribution with the azimuthal distribution of the higher energy constituents being identified as some of the most important features, along with the transverse momenta of the two most energetic constituents. 
The TopoDNN network is also trained on a similar set of inputs and hence, it is naturally expected to see noticeable overlap in the set of important features for these two networks.
%there are two major differences that prohibit one-to-one comparison between the feature rankings for these variables. 
However, the input data preprocessing is very different for these two networks which may result into significant differences on how these variables are treated by the corresponding models. 
TopoDNN always centers the most energetic constituent along the origin in the $(\eta, \phi)$ plane while using a fixed scale normalization for the $p_t$ variables. 
Hence, the $\phi_0$ variable is trivially set to zero for all jets  and it does not appear in TopoDNN's list of top-ranked variables. 
While such differences make it difficult to obtain a one-to-one comparison between the two sets of features rankings, some common conclusions can be drawn from both models. For instance, distributions pertaining to the highest energy constituents are (trivially) more important than the lower energy constituents. We also see that angular distributions of these constituents play a more decisive role in determining the jet class for the top jets. On the other hand, the distributions of their transverse momenta have a larger impact in classifying the background jets.
%There is an additional Lorentz transformation applied to align the second most energetic 
%Additionally, particle-level information fed to PFN is directly used to obtain particle-level neural embeddings while they are directly used to perform jet classification for the TopoDNN network. Hence, the importance metric attributed to each feature in PFN is indicative of how impactful that constituent is in obtaining the jet level embeddings. A similar interpretation 
%while the relative ranking of the features of a given constituent suggests how important those features are for the network to obtain the constituent level embedding. 
% For instance, the variables $\phi$ and $p_T$ for any given constituent outrank the corresponding  $\eta$ which suggests that the $\Phi$ network assigns more importance to the former in order to obtain its neural embedding.
%
\begin{figure}[!ht]
\centering
\subfloat[]{
\includegraphics[width=0.5\textwidth]{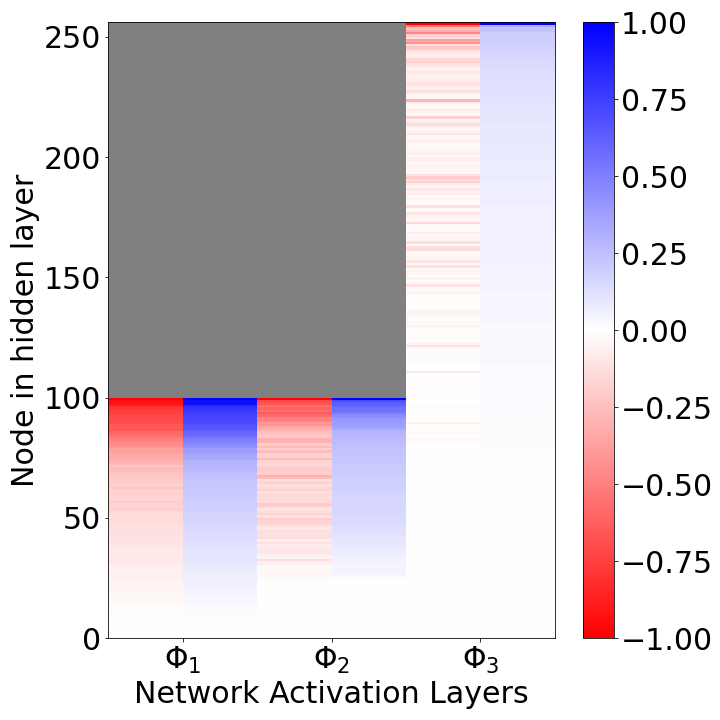}
\label{fig:NAPPHIPFN}            
}
\subfloat[]{
\includegraphics[width=0.5\textwidth]{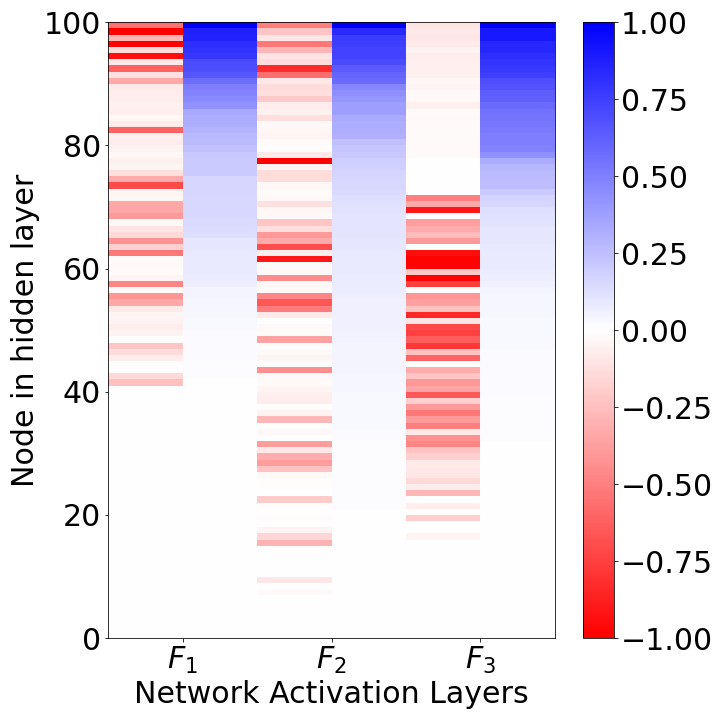}
\label{fig:NAPFCPFN}            
}
\caption{NAP diagrams for \protect\subref{fig:NAPPHIPFN} $\Phi$ and \protect\subref{fig:NAPFCPFN} $F$ networks for the PFN model showing the RNA scores of different nodes in these networks for the two jet classes. To simultaneously visualize the scores for QCD and top jets, we project the RNA scores of the former as negative values. The nodes in each layer are ordered according to their RNA scores for the top jets. }
\label{fig:NAPPFN}
\end{figure}

The particle level embeddings obtained by the $\Phi$ network are summed over to obtain a latent space representation of the jet.  Characterization of latent spaces has been a topic of general interest in many areas of machine learning application. For instance, disentangling semantic features of images via latent spaces in Variational AutoEncoders (VAEs)~\cite{kingma2013auto} and its variants~\cite{burgess2018understanding, hadjeres2017glsr,bajaj2021invariance,zhao2019variational} has been widely studied in modern machine learning literature. In the context of collider physics, how latent spaces embed information and can be used as effective candidates for anomaly detection and bump hunting has been studied~\cite{moreno2020jedi, komiske2019energy, bortolato2022bump}. The proponents of PFN performed detailed studies showing how the latent space representation forms discernible contours in the $(\eta, \phi)$ plane of jet image representations. While such studies are useful to divulge geometric features of the latent space configuration, interpreting how they actually contribute to the network's decision making process, especially for large latent spaces with $\mathcal{O}(100)$-dimensions, remains unexplored.

Since the PFN network is trained to obtain  classification scores in two disentangled dimensions in the very last layer of $F$ network, it is only expected that the information propagation pathways for the different types of jets will show some level of disentanglement within the hidden layers of the networks. This is indeed verified by the NAP diagrams shown in Figure~\ref{fig:NAPPFN}. These NAP diagrams reveal some crucial insights. 
Firstly, we clearly see that the activity level of different nodes in the final layer of the $\Phi$ network for background and signal jets is very similar. It implies that the network embeds the jet-level information in the same latent subspace. 
Secondly, the latent space appears to be very sparse and we indeed found that many of the latent space variables are identically zero for all events in both jet categories.  
A third observation is that the $F$ network effectively learns to disentangle the representation of jet classes only at the third hidden layer (Figure~\ref{fig:NAPFCPFN}). But the first layer of the $F$ network is quite sparse with RNA scores for almost 40\% of the nodes being very close to zero for both jet classes. 
With these observations in mind, we retrained variant networks with latent space dimensions of 64 and 32 while reducing the number of nodes in the first hidden layer of $F$ and still got comparable performance (Table~\ref{tab:TopoDNN-perf}).

\begin{figure}[!h]
\centering
\subfloat[]{
\includegraphics[width=0.33\textwidth]{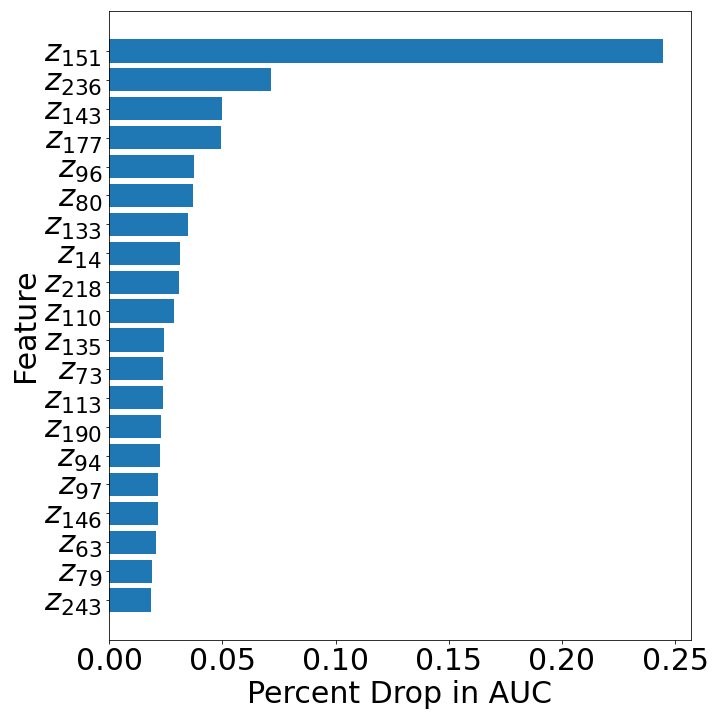}
\label{fig:dAUCPFNz}            
}
\subfloat[]{
\includegraphics[width=0.33\textwidth]{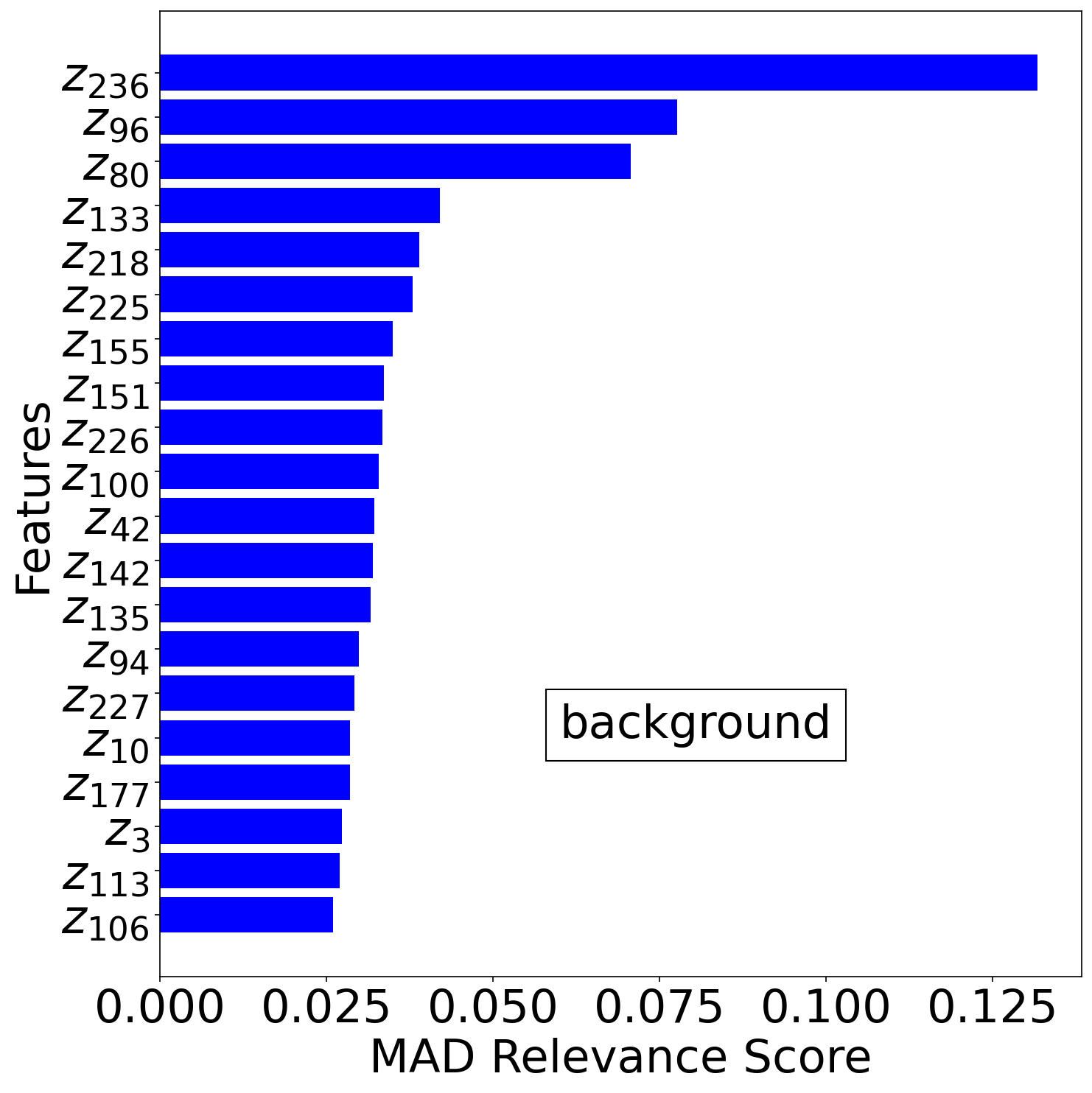}
\label{fig:bkgdLRPPFNz}            
}
\subfloat[]{
\includegraphics[width=0.33\textwidth]{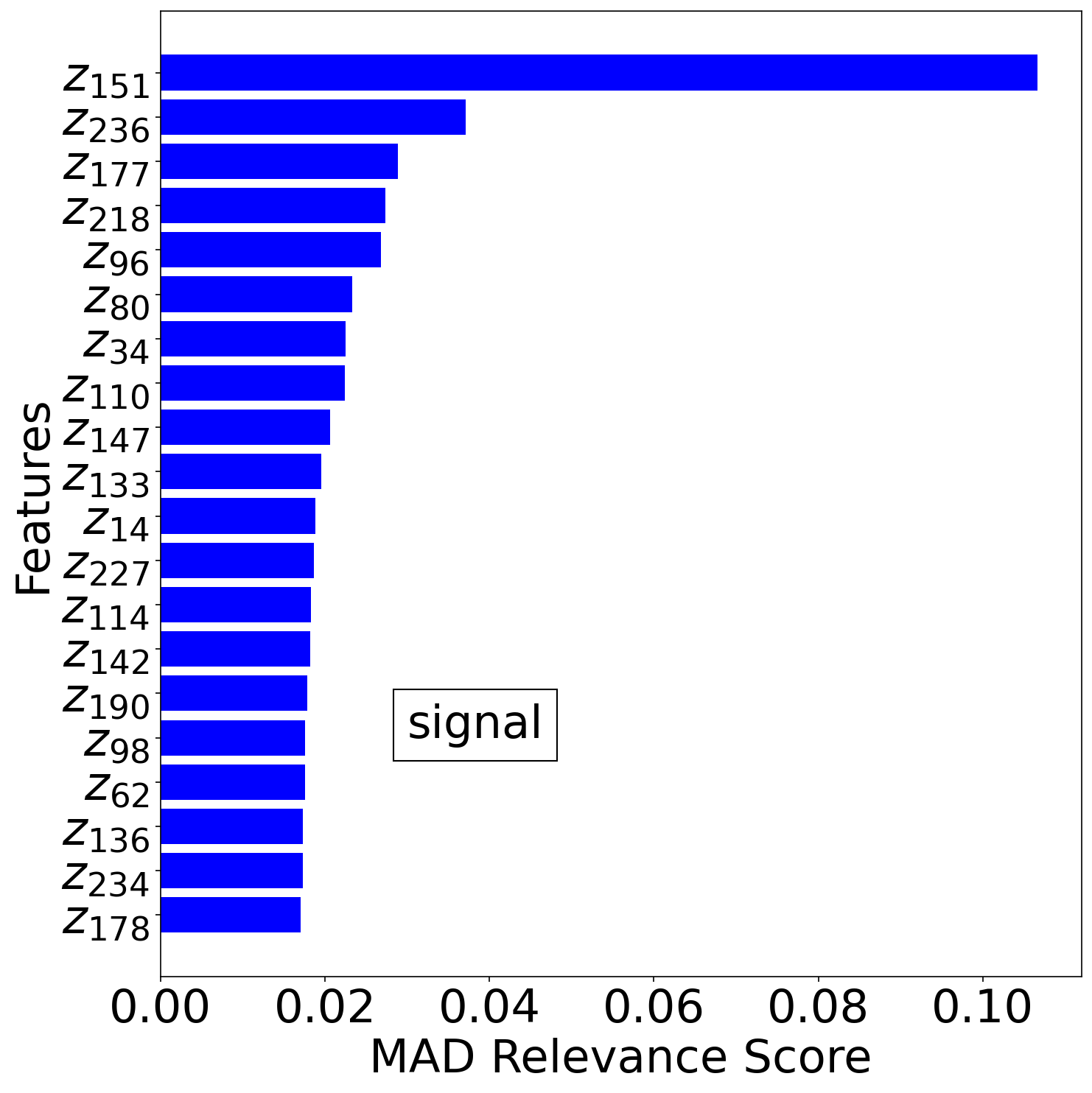}
\label{fig:sigdLRPPFNz}
}
\\
\subfloat[]{
\includegraphics[width=0.25\textwidth]{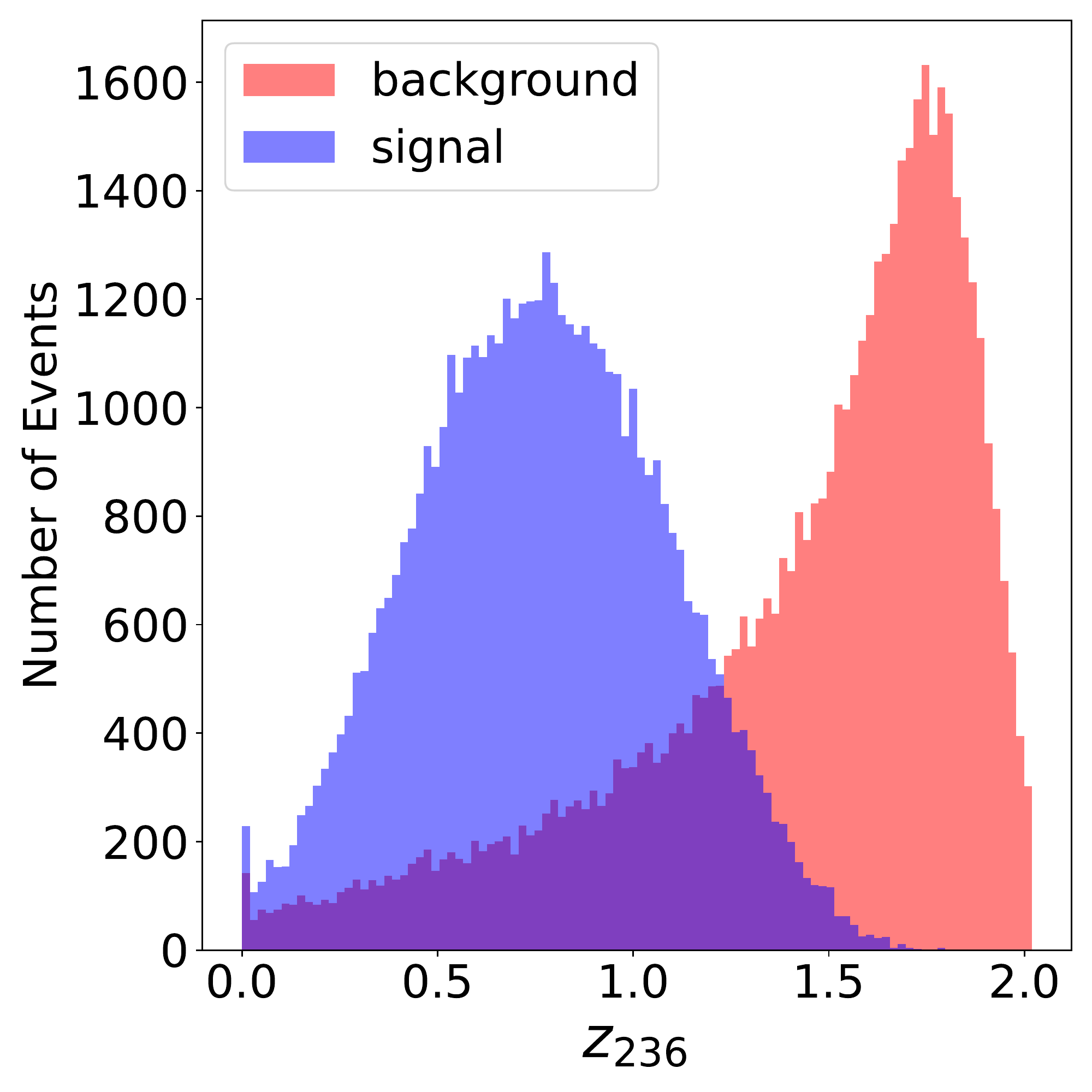}
\label{fig:z236PFN}            
}
\subfloat[]{
\includegraphics[width=0.25\textwidth]{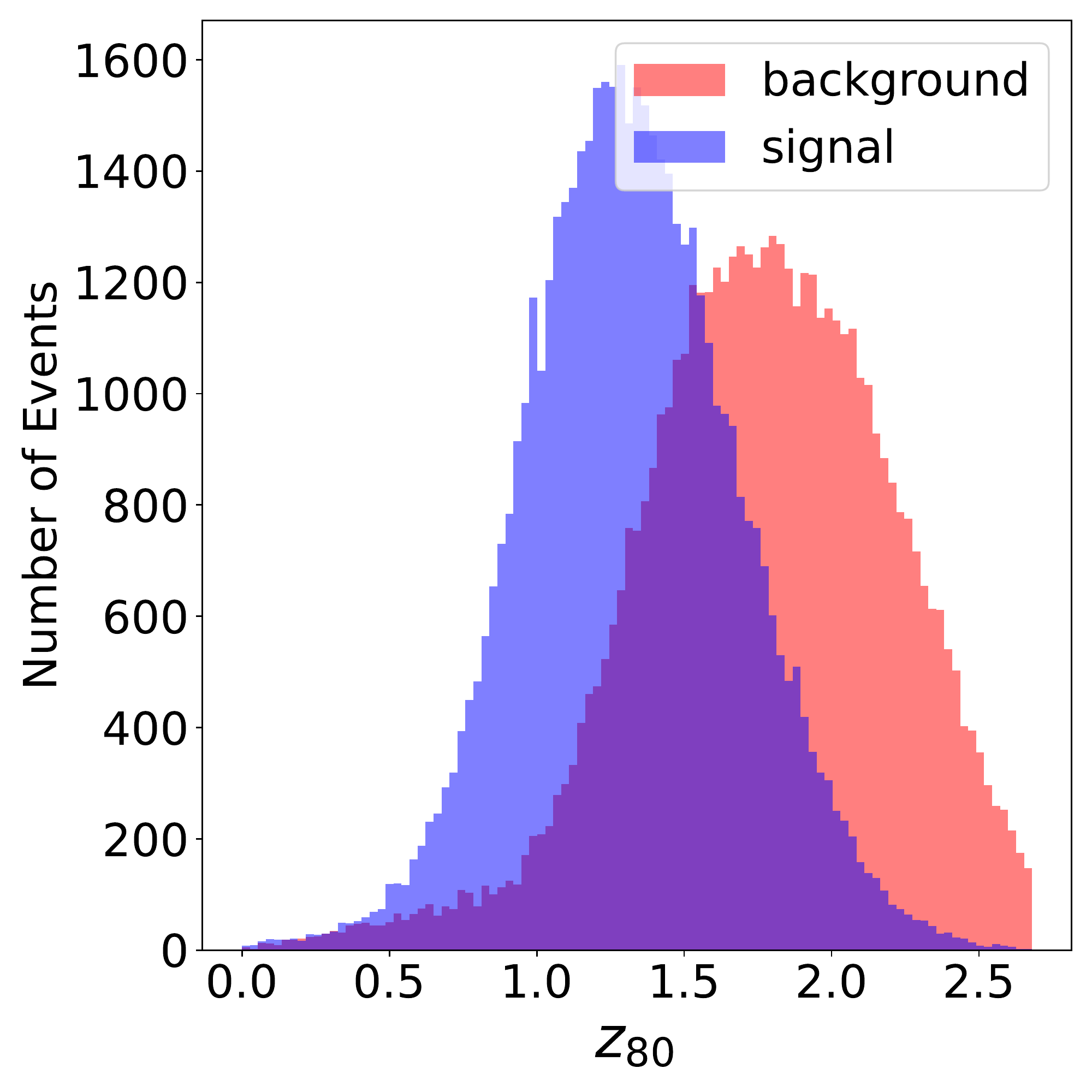}
\label{fig:z96PFN}            
}
\subfloat[]{
\includegraphics[width=0.25\textwidth]{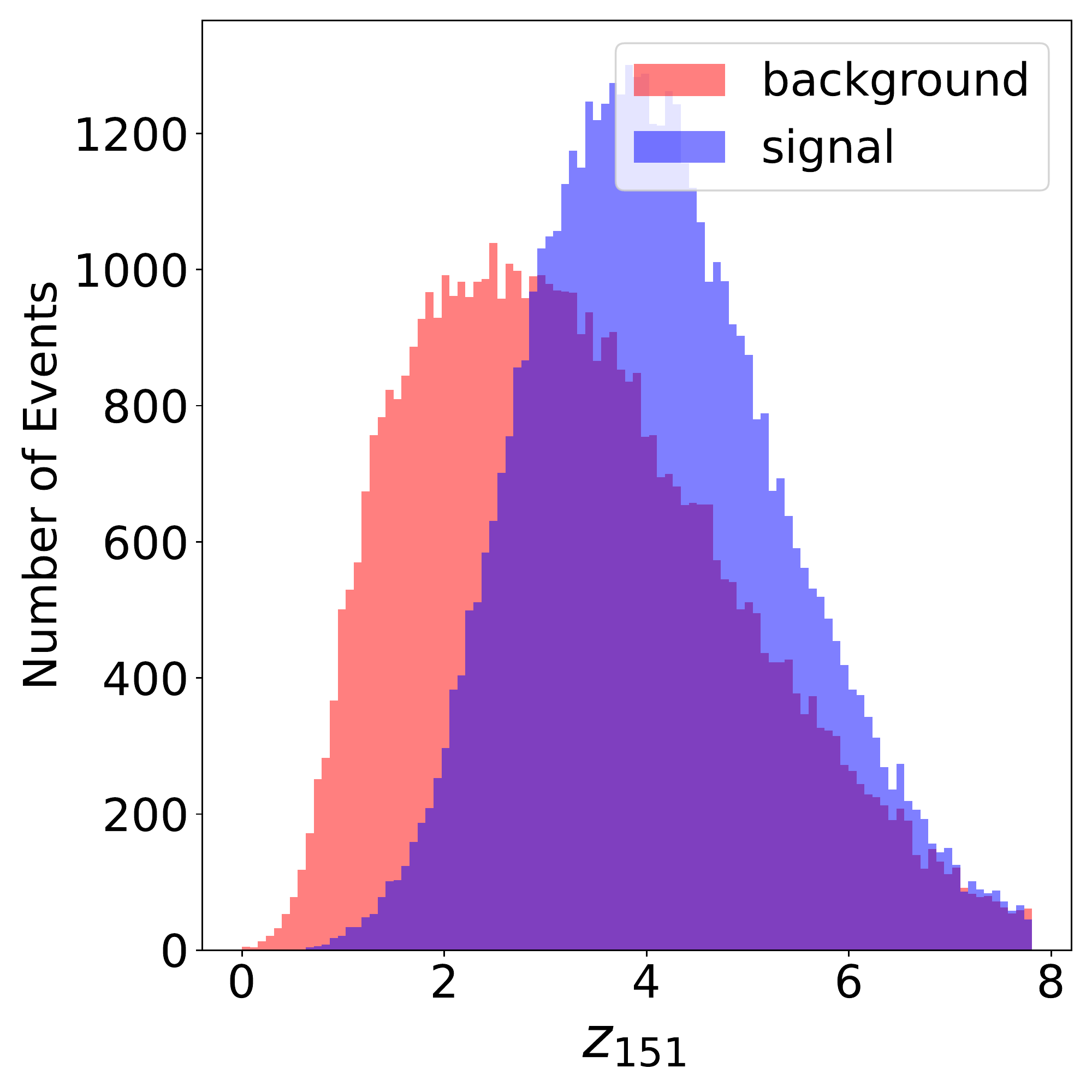}
\label{fig:z151PFN}
}
\subfloat[]{
\includegraphics[width=0.25\textwidth]{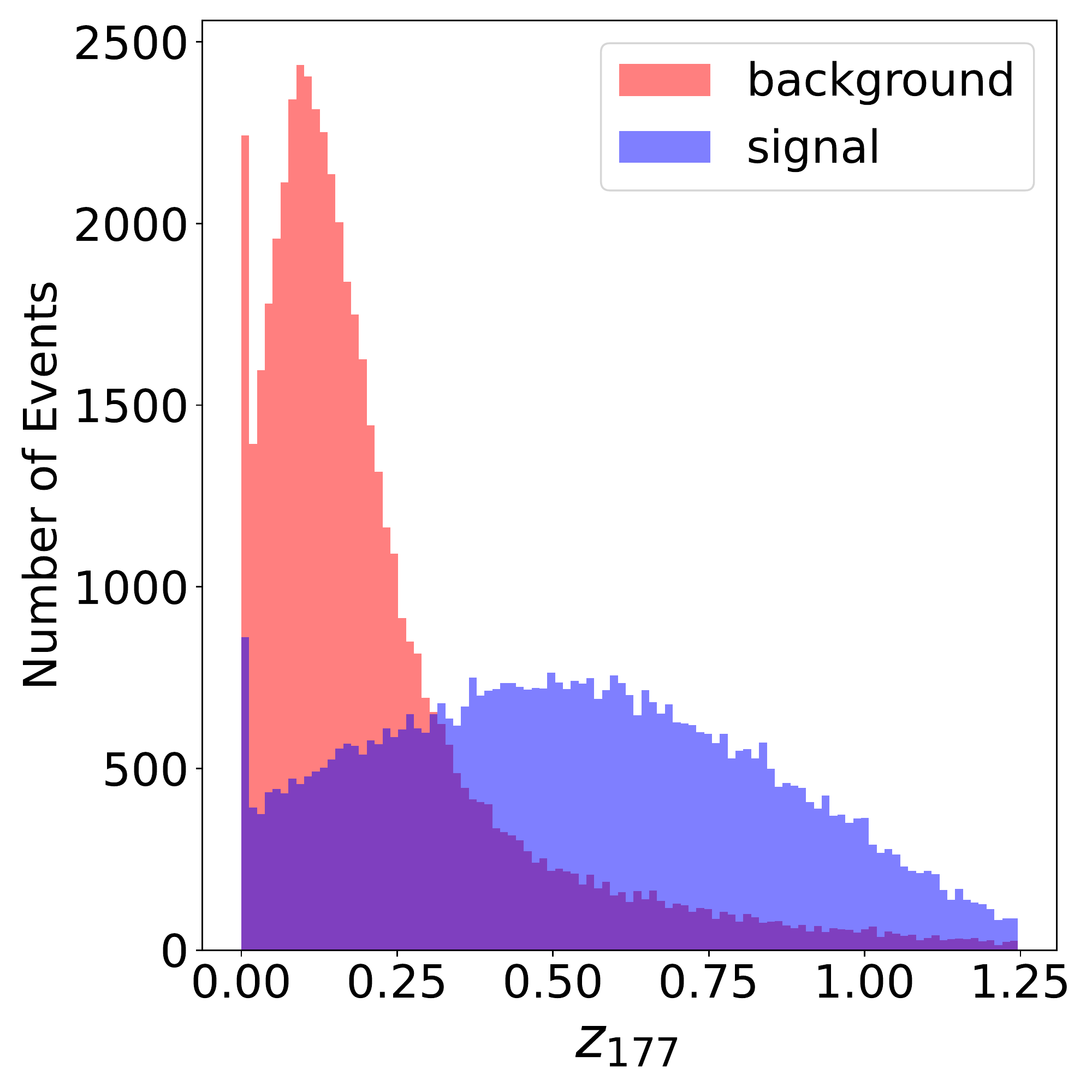}
\label{fig:z177PFN}
} 
%\\
% \subfloat[]{
% \includegraphics[width=0.25\textwidth]{figures/PFN/z65.pdf}
% \label{fig:z65PFN}            
% }
% \subfloat[]{
% \includegraphics[width=0.25\textwidth]{figures/PFN/z231.pdf}
% \label{fig:z231PFN}            
% }
% \subfloat[]{
% \includegraphics[width=0.25\textwidth]{figures/PFN/z250.pdf}
% \label{fig:z250PFN}
% }
% \subfloat[]{
% \includegraphics[width=0.25\textwidth]{figures/PFN/z222.pdf}
% \label{fig:z222PFN}
% }
\caption{Figures in the top row show the latent space feature rankings obtained from the \protect\subref{fig:dAUCPFNz} \dAUC scores, \protect\subref{fig:bkgdLRPPFNz} MAD relevance scores for background QCD jets, and \protect\subref{fig:sigdLRPPFNz} MAD relevance scores for signal top jets.  Only the 20 highest-ranked features are shown for each jet category. Figures in the bottom row show the distribution of some of the highest-ranked latent space features for both jet categories.}
\label{fig:featrankzPFN}
\end{figure}

Given the sparsity of the latent space representation, we expect that only a small subset of these latent features will actually have a strong contribution towards the decision making process of $F$ in the baseline model. The ranking of different latent features using the \dAUC score and MAD relevance scores are shown in Figures~\ref{fig:dAUCPFNz}--\ref{fig:sigdLRPPFNz}. There are noticeable overlaps among the features that rank high with these methods, though the actual sequence of latent variables, understandably, shows some differences. We show the distributions of some of these latent space embeddings in Figures~\ref{fig:z236PFN}--\ref{fig:z177PFN}.  

\begin{figure}[!ht]
\centering
\subfloat[]{
\includegraphics[width=0.5\textwidth]{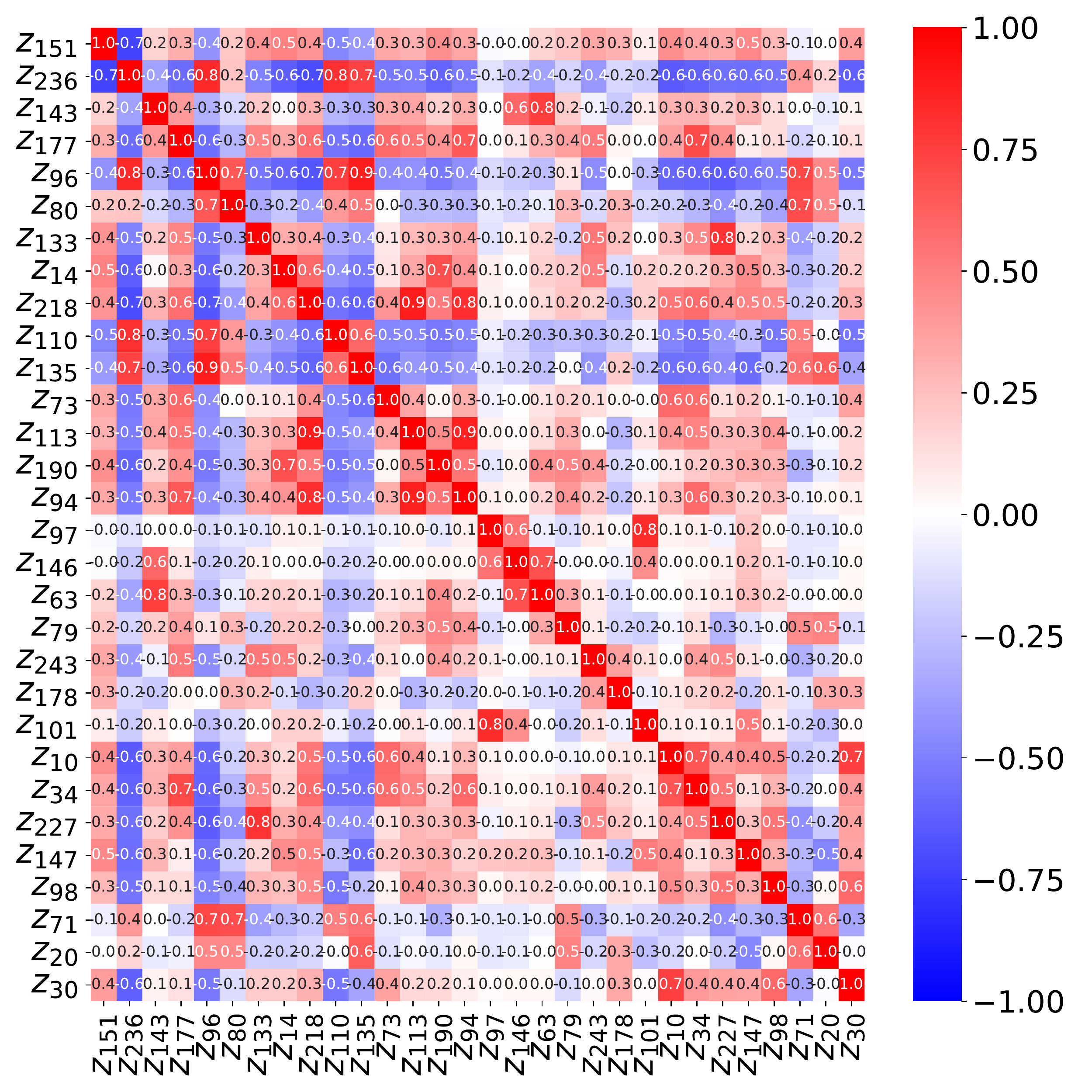}
\label{fig:FeatcorrbkgzPFN}            
}
\subfloat[]{
\includegraphics[width=0.5\textwidth]{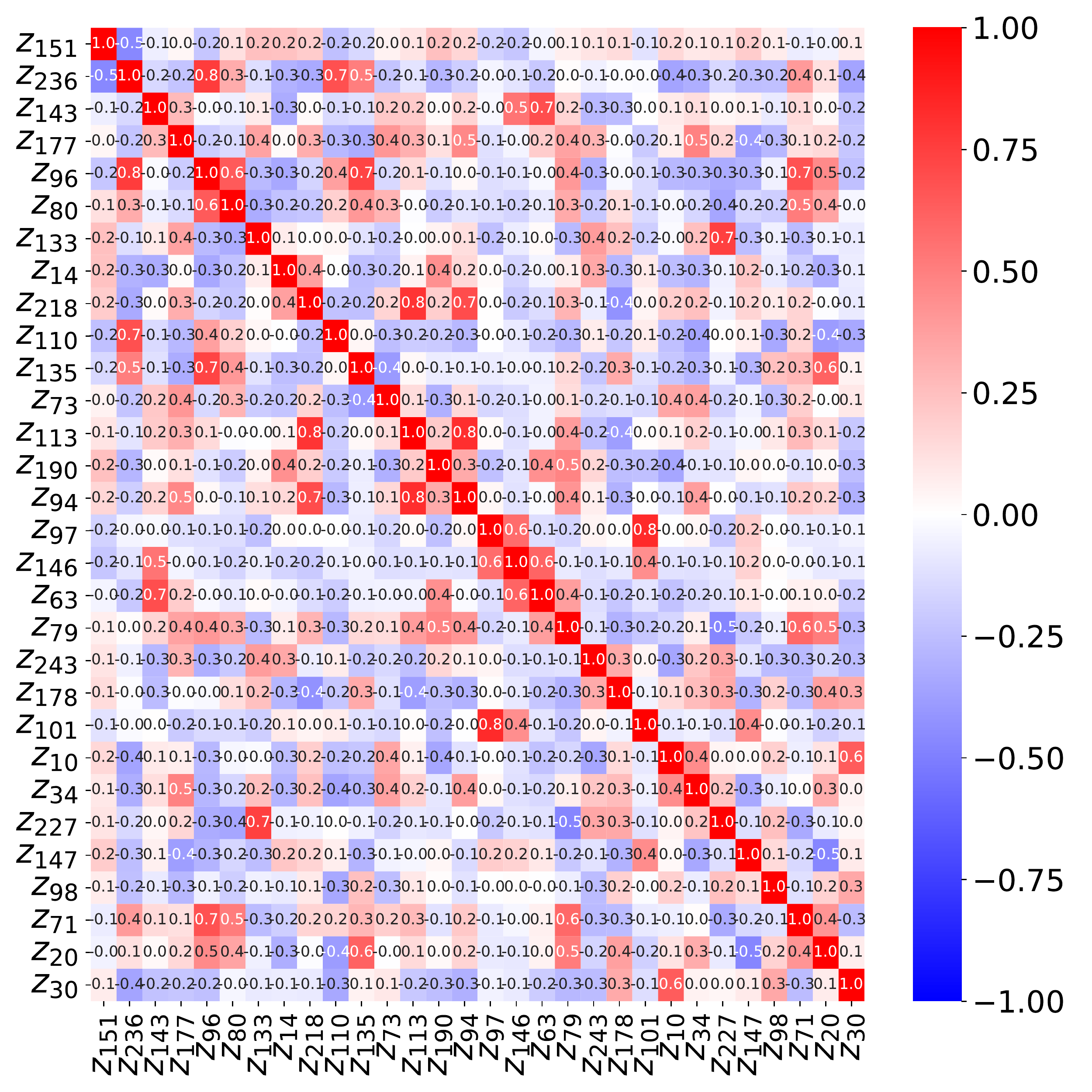}
\label{fig:FeatcorrsigzPFN}            
}
\caption{Correlation matrix for latent space features for \protect\subref{fig:FeatcorrbkgzPFN} background QCD jets and \protect\subref{fig:FeatcorrsigzPFN} signal top jets. Only the top 30 latent features obtained from the \dAUC metric are shown. }
\label{fig:FeatCorrzPFN}
\end{figure}

These distributions highlight a stark contrast between the the latent space representation for PFN when compared to those studied in the context of generative models like VAEs. 
The latter class of models are known to provide semantic disentanglement in their latent spaces creating a  clear separation in latent space dimensions that account for variabilities in input distributions. 
For instance, VAEs trained on the popular celebrity protrait dataset CelebA~\cite{liu2015faceattributes} have successfully demonstrated disentanglement of semantics like age, gender, skin tone, hairstyle etc.~\cite{burgess2018understanding} in the latent space. 
In the context of the top tagging dataset, it is the jet kinematics that is embedded in the latent space. 
As seen from the NAP diagram in Figure~\ref{fig:NAPFCPFN} and the latent space feature distributions in Figures~\ref{fig:z236PFN}--\ref{fig:z177PFN}, the PFN latent space does not provide any disentanglement between the jet classes.
%and such embedding alone is often insufficent for classification~\cite{dillon2021better}. 
In fact, for models like PFN that are trained to learn the jet classes and not the distribution of training dataset, the model has no additional incentive in disentangling the jet classes. Rather, PFN learns to embed the information regarding jet classes in correlations among latent features. This can be seen from the latent space correlation matrices of the two jet classes shown in Figure~\ref{fig:FeatCorrzPFN}. The pairwise feature correlations are quite different for the two jet classes, creating a clearer context for the classifier network $F$ to obtain the desired jet classification.

\begin{figure}[!h]
\centering
\subfloat[]{
\includegraphics[width=0.25\textwidth]{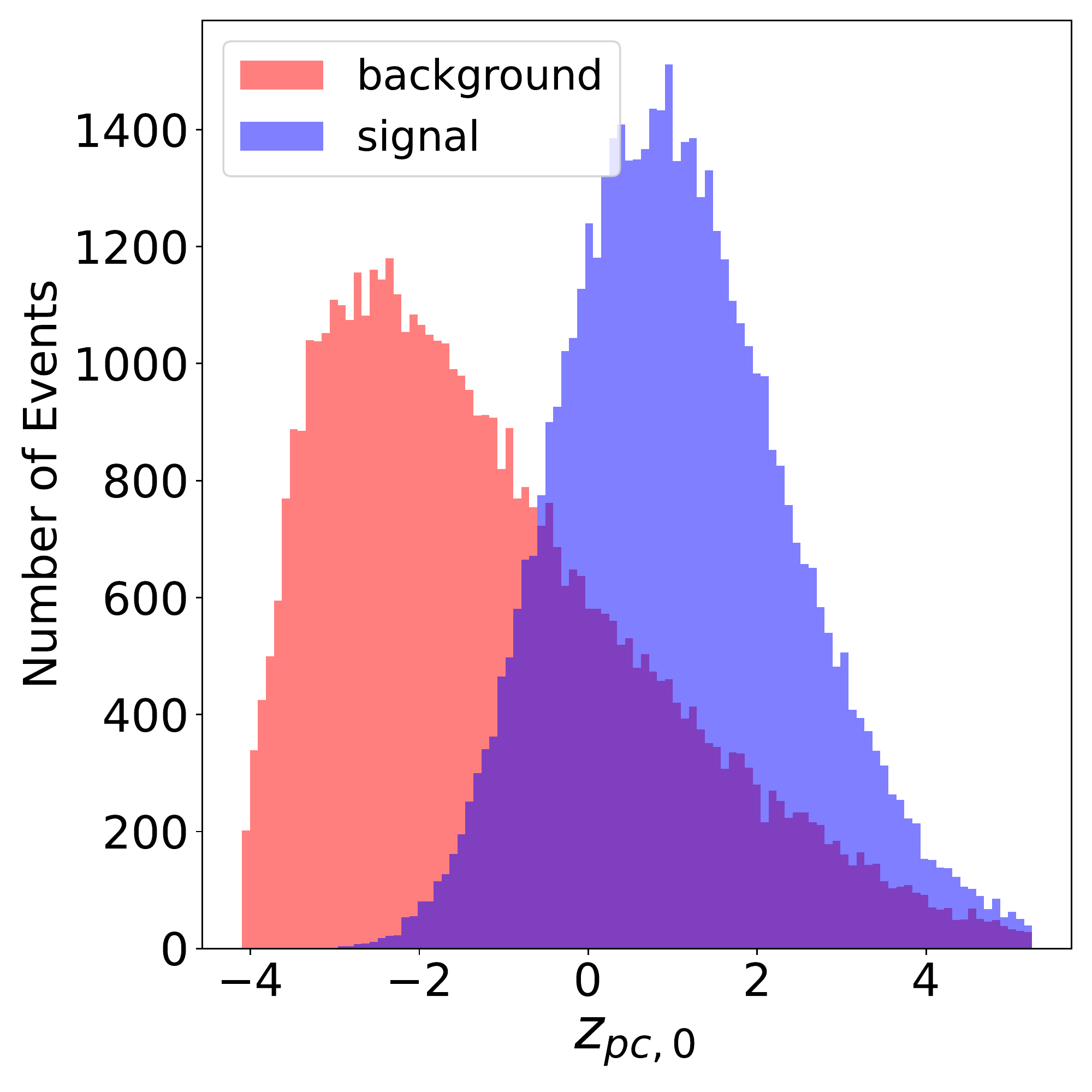}
\label{fig:zpca0PFN}            
}
\subfloat[]{
\includegraphics[width=0.25\textwidth]{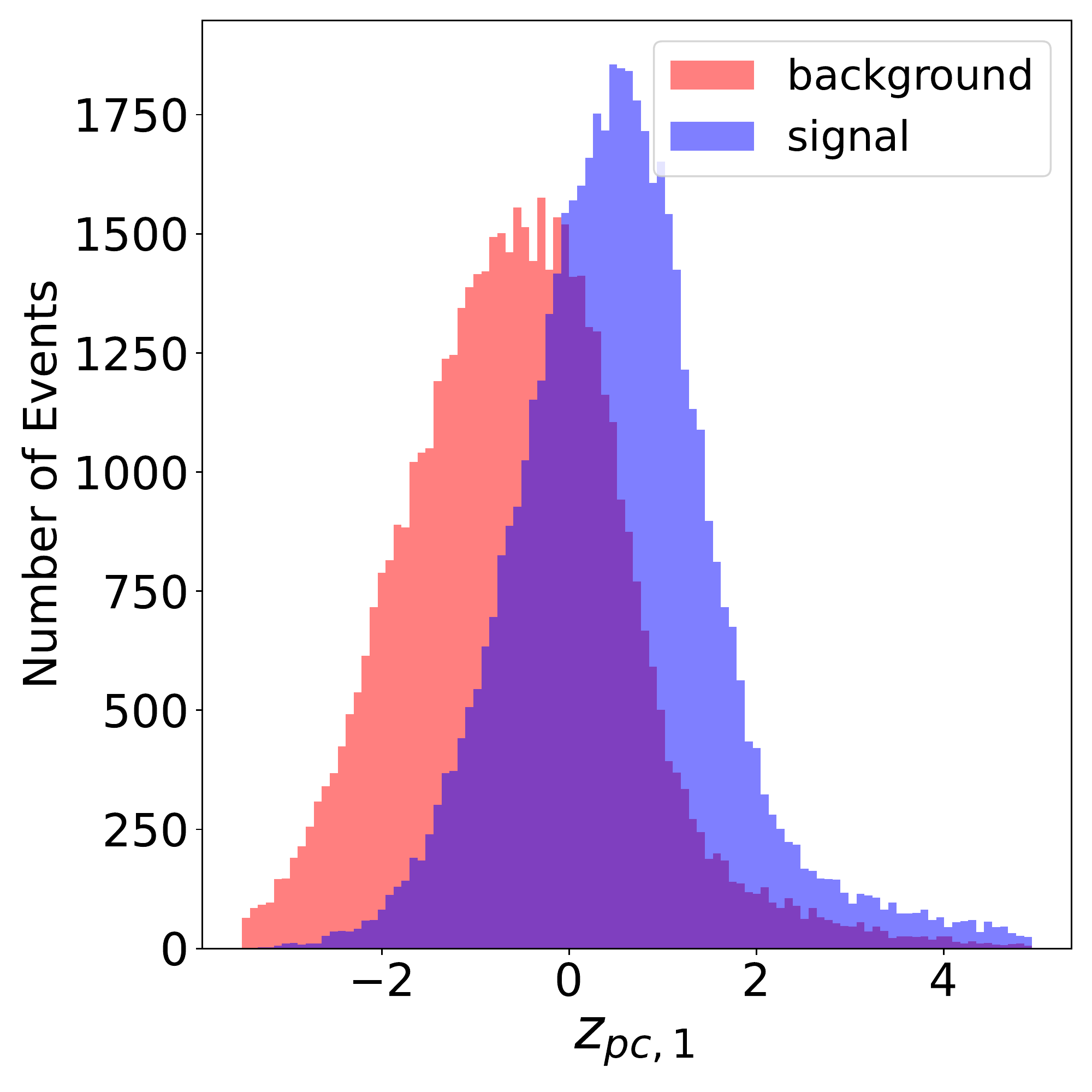}
\label{fig:zpca1PFN}            
}
\subfloat[]{
\includegraphics[width=0.25\textwidth]{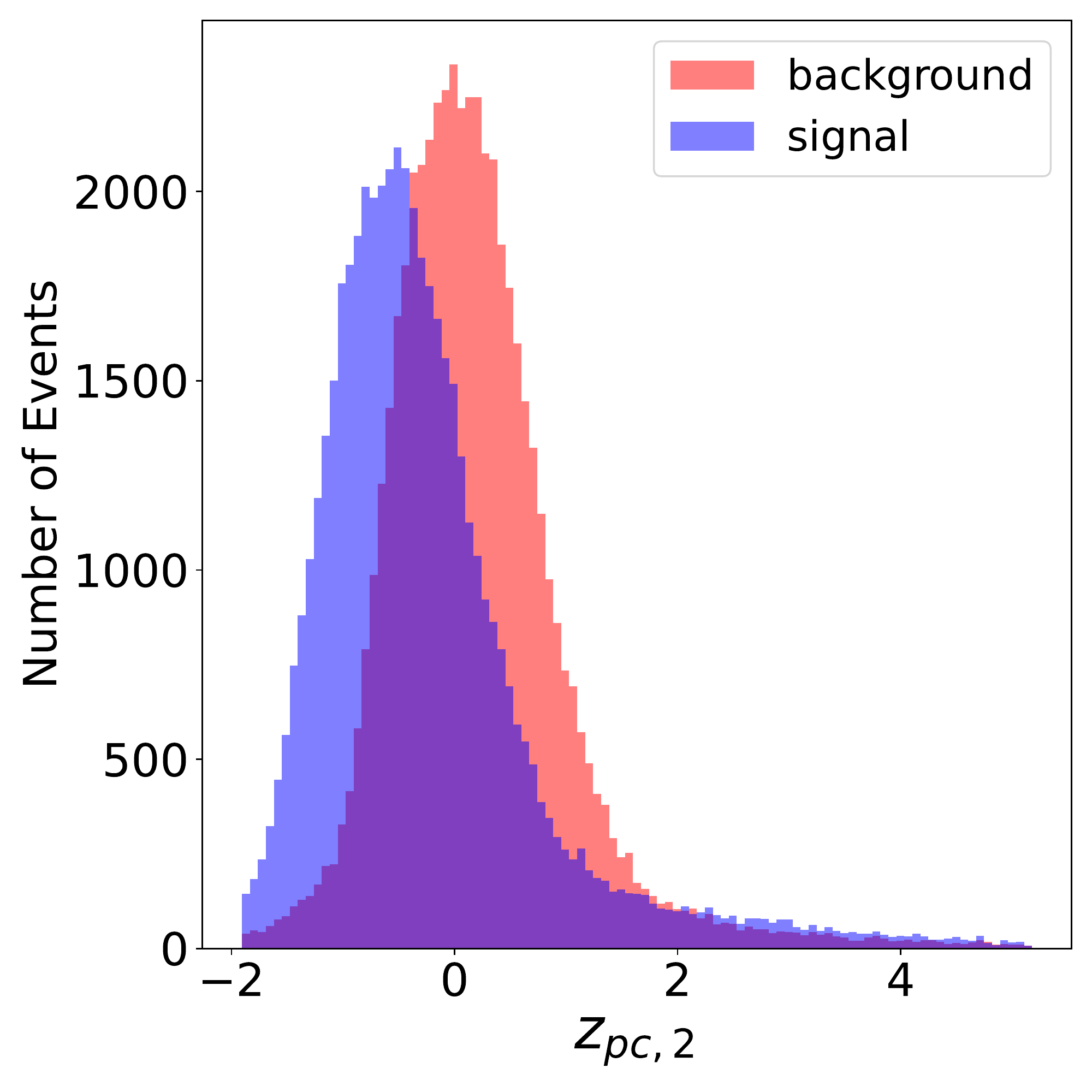}
\label{fig:zpca2PFN}
}
\subfloat[]{
\includegraphics[width=0.25\textwidth]{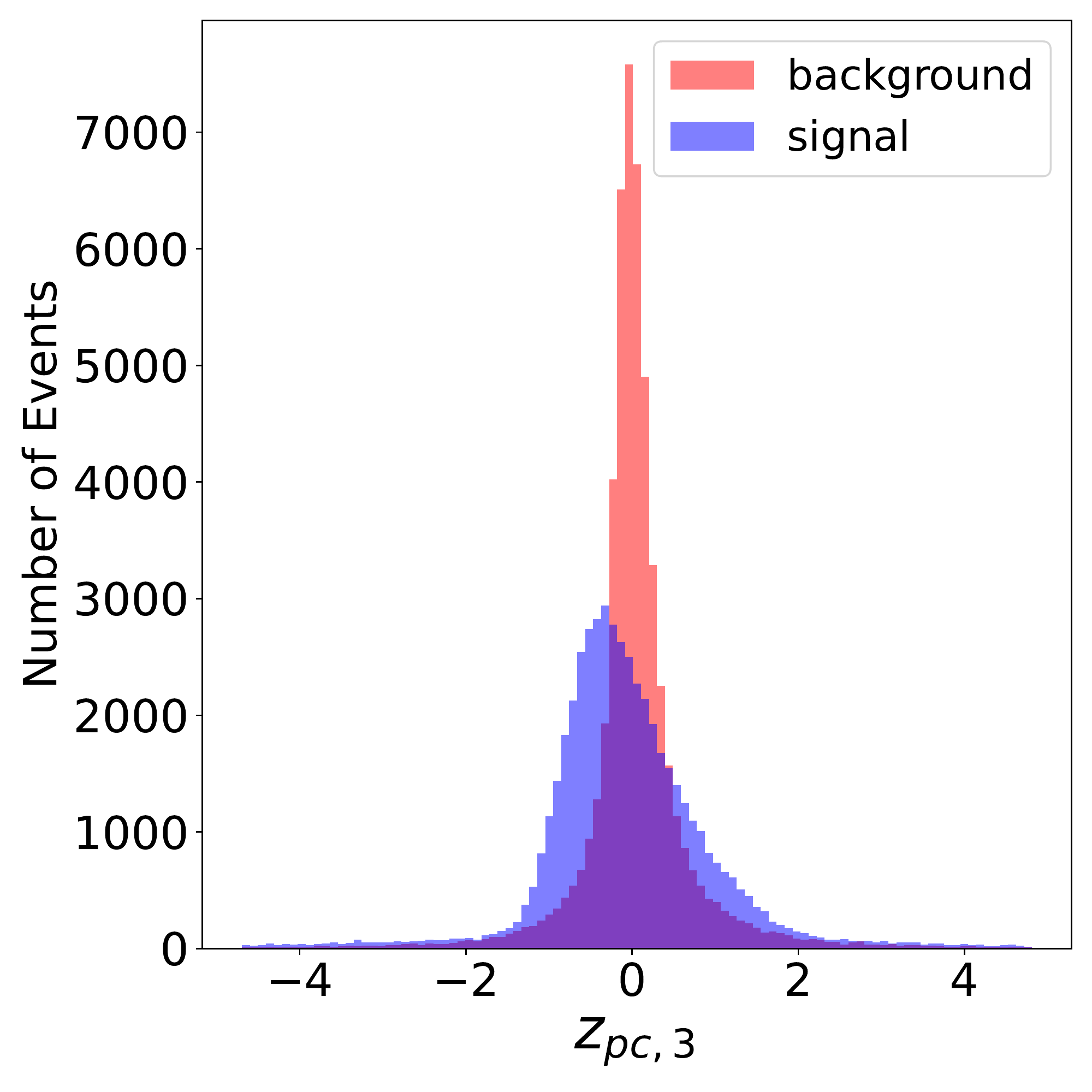}
\label{fig:zpca3PFN}
}
\\
\subfloat[]{
\includegraphics[width=0.33\textwidth]{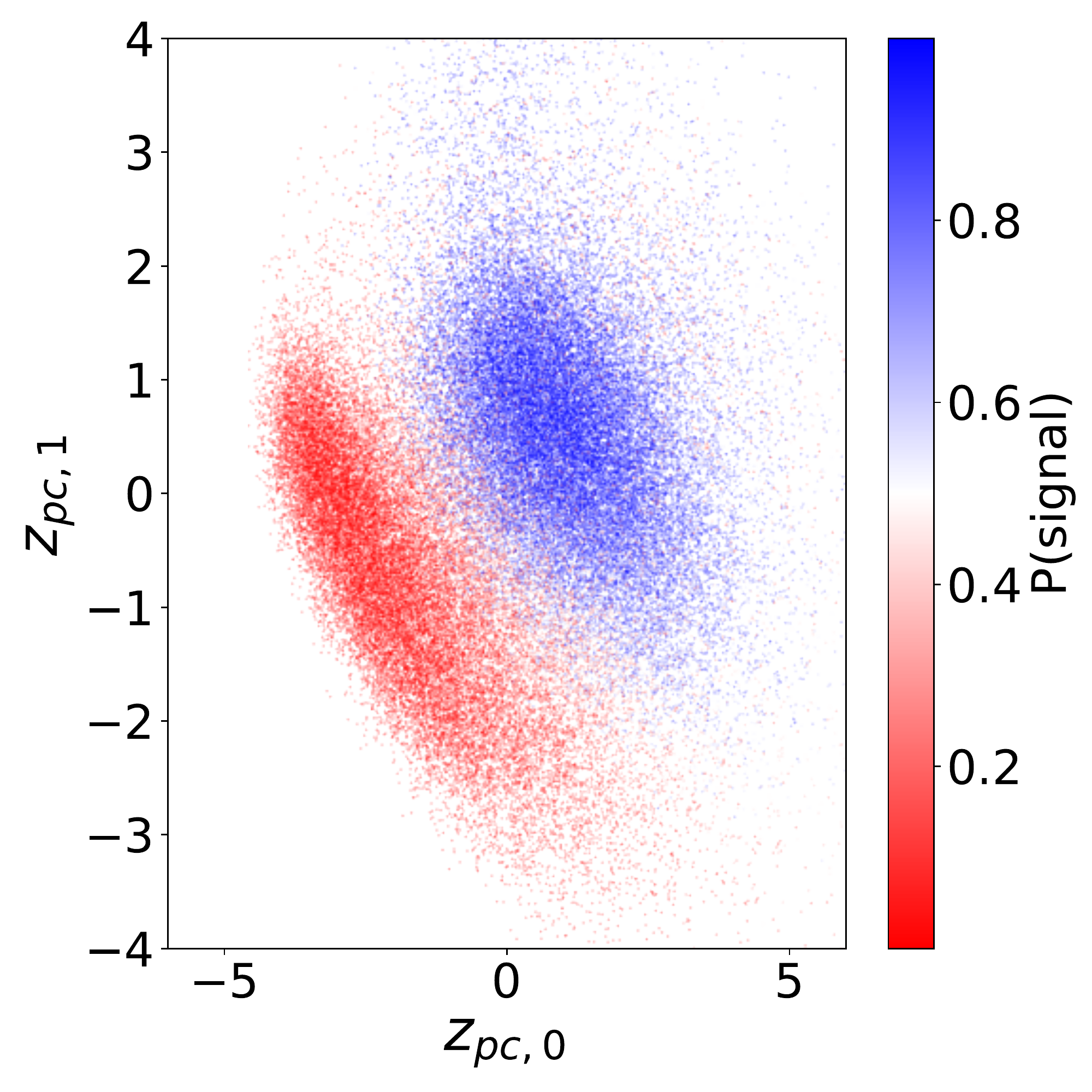}
\label{fig:zpca0zpca1PFN}            
}
\subfloat[]{
\includegraphics[width=0.33\textwidth]{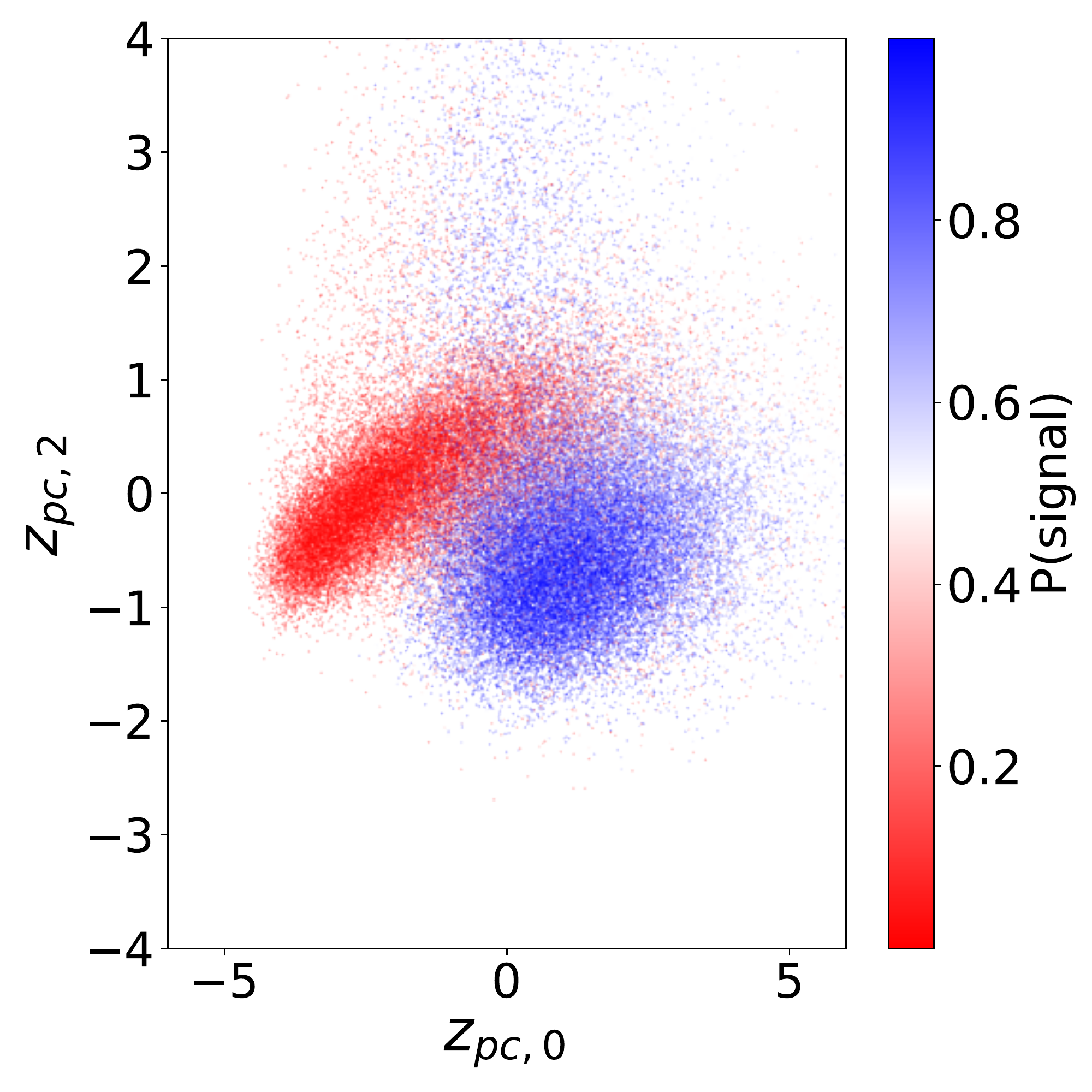}
\label{fig:zpca0zpca2PFN}            
}
% \subfloat[]{
% \includegraphics[width=0.33\textwidth]{figures/PFN/zpca0_zpca11.pdf}
% \label{fig:zpca0zpca11PFN}            
% }
\caption{Distributions of some of the principal components for  background QCD jets and signal top jets (top row) and pairwise distributions for some of these components as a function of the predicted signal class probability (bottom row)}
\label{fig:pcazPFN}
\end{figure}

To illustrate the nature of the jet class identifying characteristics in the  correlations among the latent features, we examine the distribution of variances in the datasets using Principal Component Analysis (PCA)~\cite{jolliffe2016principal}. PCA performs a linear transformation on these features to obtain a set of orthogonal feature spaces with no cross-correlation among the transformed features.
Since the size of the latent space is large but sparse, we select the top-ranking subset of latent features so that simultaneously masking each latent feature in the remaining subset causes at most  1\% drop in the AUC score from the test data. For our baseline PFN model, this requires choosing 95 of the 256 latent features.  
%The reduced feature space preserves the sample variance of the dataset but redistributes them along orthogonal basis vectors. 
We found that 99\% of the observed variance in the test data was described by the top 37 principal components. We show the distribution of the top four components in Figures~\ref{fig:zpca0PFN}-~\ref{fig:zpca3PFN}. 
We can readily see how these PCA-transformed latent features can differentiate between the two jet classes in Figures~\ref{fig:zpca0zpca1PFN}-~\ref{fig:zpca0zpca2PFN}.

Having examined the disentanglement between the jet classes by the principal components of the latent space,
%demonstrates how the jet class information is actually embedded in the correlation among different dimensions of the latent space. 
%It 
it is also instructive to investigate the physical nature of the latent space learned by the network. While it is neither trivial nor obvious for neural networks to learn about features 
%that bear resemblance with features 
that bear meaning to humans, it has been seen that latent space networks can occasionally learn about physical variables~\cite{moreno2020jedi}. 
In case of the PFN, since the latent space dimensions are highly correlated with each other, we chose to study the correlation between the principal components and  jet features like jet mass, the number of constituents, and the subjettiness variables which, as shown in Figures~\ref{fig:jet-feats} and~\ref{fig:jet-taus}, can have moderate to strong discriminative power. 
We found that the first principal component, $z_{pc,0}$ (Figure~\ref{fig:zpca0PFN}) shows a strong correlation with jet mass for both jet categories with correlation coefficients being 0.82 and 0.64 for background and signal jets respectively. 
%This is particularly impressive considering that the network was trained with only information regarding the 3-momentum of the constituents and information about the energy or jet mass was absent from the set of inputs. 
$z_{pc,0}$ also shows strong correlations with the number of jet constituents. 
$z_{pc,1}$ and $z_{pc,2}$ showed moderate to strong correlations with the subjettiness variables $\tau_1^{(1)}$ and $\tau_2^{(1)}$, implying the PFN model also learns to somewhat reconstruct distributions similar to these variables. 

\section{Interpretability Inspires: The Particle Flow Interaction Network (PFIN)}
\label{sec:new-tagger}

The performance metrics of PFN are found to be very similar to that of the \mbns{8} network. Our studies suggest that PFN learns to loosely reconsturct some of the expressive jet features in its latent space. However, what constrains PFN's performance can be understood by examining the construction of the latent space. The PFN latent space is constructed by linearly combining the individual particle-level embeddings from the $\Phi$ network. Such linear combinations constrain the latent space's ability to learn any inter-particle interaction. 
The particle-level embeddings obtained from the $\Phi$ network do not take into account the ensemble of particles constituting the jets. 
As a result, the network learns to create per-particle embeddings by emphasizing particle-level features that are known to have moderate-to-strong expressive distributions (Figure~\ref{fig:featrankptetaphiPFN-1}).
Hence, we can expect a noticeable improvement in PFN's performance if the latent space can be augmented with interaction-level representations.
In fact, modern architectures known to outperform the PFN model for top tagging take inter-particle interactions in some form into account.  

Inspired by our observations regarding feature importance and latent space distributions for PFN as well as the trend in building modern architectures for top tagging, we propose an augmentation of PFN by including an Interaction Network (IN)~\cite{IN,moreno2020jedi} to demonstrate how particle-level interactions allow for better-performing models. 

\begin{figure}
    \centering
    \includegraphics[width=1.1\textwidth]{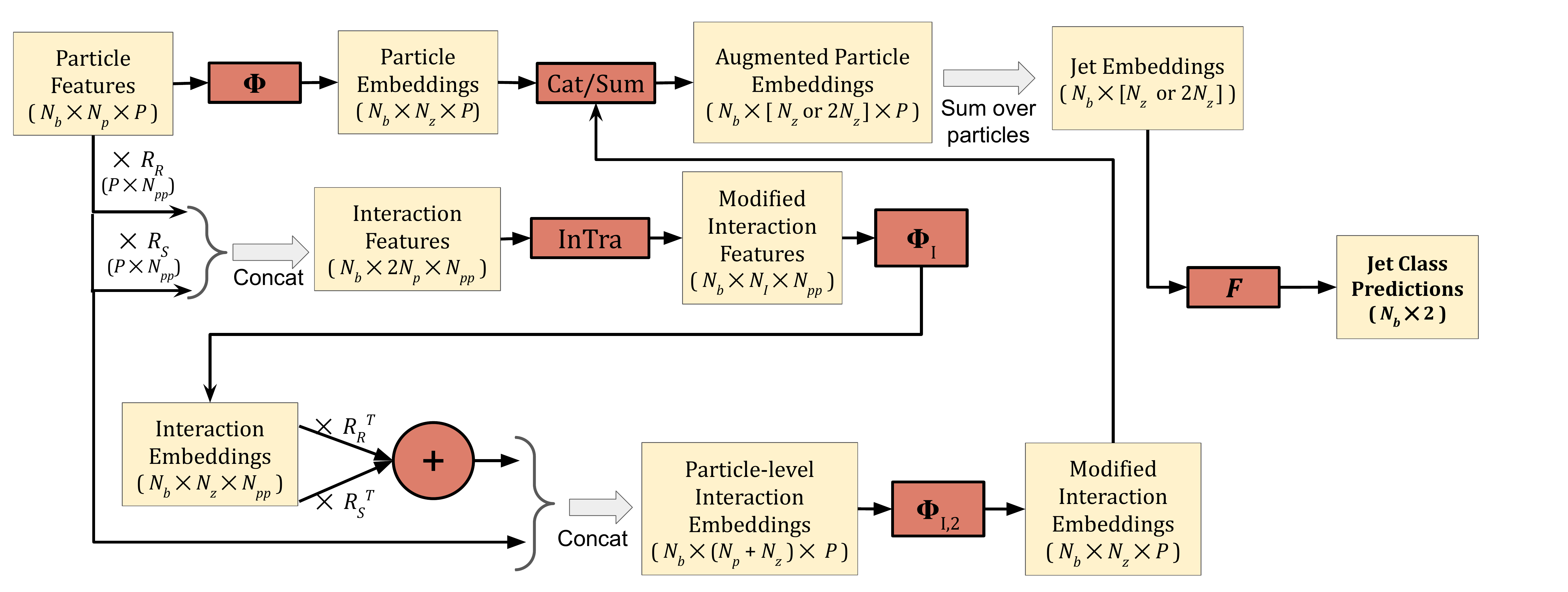}
    \caption{Model architecture and data flow for the PFIN model. $N_b$ represents the batch size. The \texttt{InTra} block computes the pairwise particle interaction features given in Eqn.~\ref{eqn:phys-feats}. The \texttt{Cat/Sum} block creates the augmented particle embeddings by either concatenating or summing the outputs of $\Phi_{I,2}$ with $\Phi$. $\Phi, \Phi_I, \Phi_{I,2}, F$ networks are implemented as fully connected MLPs with ReLU activation.}
    \label{fig:PFIN-flow}
\end{figure}

The dataflow for the proposed Particle Flow Interaction Network (PFIN) model is shown in Figure~\ref{fig:PFIN-flow}. The interactions are modeled in PFIN by constructing a fully connected undirected graph with $N_{pp} = \frac{P(P -1)}{2}$ edges where $P$ is the maximum number of constituent particles the network is trained with. Each particle is represented with $N_p$ features. For our purpose, we use $N_p=3$ with $(p_t, \eta, \phi)$ for each particle with the same preprocessing used for PFN. Each edge is initially represented with $2N_P$ features by concatenating the individual particle-level features. This node-to-edge level feature construction is facilitated by a couple of interaction matrices of size $P \times N_{pp}$ called $R_R$ and $R_S$. For $P = 4$, these matrices are constructed in the following manner:
\begin{equation}
    \begin{pmatrix}
R_R\\
\hline
R_S 
\end{pmatrix}
= 
\begin{pNiceMatrix}[first-row,first-col]
    & (0,1) & (0,2) & (0,3) & (1,2) & (1,3) & (2,3)  \\
P0  & 1     & 1     & 1     & 0     & 0     & 0  \\
P1  & 0     & 0     & 0     & 1     & 1     & 0 \\
P2  & 0     & 0     & 0     & 0     & 0     & 1 \\
P3  & 0     & 0     & 0     & 0     & 0     & 0 \\
\hline 
P0 & 0      & 0     & 0     & 0     & 0     & 0  \\
P1 & 1      & 0     & 0     & 0     & 0     & 0 \\
P2 & 0      & 1     & 0     & 1     & 0     & 0 \\
P3 & 0      & 0     & 1     & 0     & 1     & 1 \\
\end{pNiceMatrix}
\label{eqn:intmat}
\end{equation}
where we have labeled the rows with the particle ID and each column label $(i,j)$ represents which particles are connected by this edge.
The edge-level features are transformed by the Interaction Transformation (\texttt{InTra}) block to calculate a $N_I = 4$ dimensional representation for each edge by calculating the physics-inspired quantities~\cite{qu2022particle, erdmann2019lorentz}: $\ln\Delta, \ln k_T, \ln z, \ln m^2$ where
\begin{align}    \label{eqn:phys-feats}
    \Delta &= \sqrt{(\eta_1 - \eta_2)^2 + (\phi_1 - \phi_2)^2} \nonumber \\
    k_T &= \min\left(p_{t,1}, p_{t,2}\right)\Delta \nonumber \\
    z &= \frac{\min\left(p_{t,1}, p_{t,2}\right)}{p_{t,1} + p_{t,2}} \\
    m^2 &= (E_1 + E_2)^2 - || \vec{p}_1 + \vec{p}_2 ||^2 . \nonumber 
\end{align}

The subscripts $1$ and $2$ represent the two particles associated with the edge and each quantity in the aforementioned relations represents its unpreprocessed value. Given these quantities are symmetric with respect to the particles, the actual ordering of the particles does not impact PFIN's dataflow, maintaining the permutation-invariant property of PFN. 
These interaction features are transformed into $N_z$ dimensional interaction embeddings by the trainable $\Phi_I$ network. These embeddings are propagated back to particle level using the interaction 
matrices and only those interactions are considered where both constituents are present. These particle-level interaction embeddings are concatenated with the original particle features and further transformed into $N_z$ dimensional modified per-particle interaction embeddings via a trainable $\Phi_{I,2}$ network. These embeddings are concatenated or summed with per-particle embedding from PFN's $\Phi$  network to obtain augmented particle embeddings. These augmented features are then summed over its constituents to  obtain the jet-level latent space. Finally, the $F$ network obtains jet class probabilities for each of the jet class based on these jet-level latent space features.

The model hyperparameters and performance metrics for our implementation of PFIN are given in Table~\ref{tab:PFIN-results}. The choices of numbers of nodes in the hidden layers of $\Phi$ and $F$ as well as the size of the latent space were inspired from the study of RNA scores and NAP diagrams for PFN. This allows us to keep the increase in number of trainable parameters manageably small. Both versions of PFIN show notable improvement in performance outperforming both PFN and IN models. PFIN  outperforms LGN~\cite{bogatskiy2020lorentz} and its performance is comparable to those of ParticleNet and ResNeXt models while requiring a significantly smaller number of parameters, providing much faster training and model convergence.

\begin{table}[!h]
    \centering
    {
    \begin{tabular}{|c|c|}
    \hline
    \multicolumn{2}{|c|}{Model Hyperparameters} \\
    \hline
    \hline 
    Number of constituents, $N_p$ & 60 \\
    Nodes in $\Phi$ Network    & (100,100,64) \\
    Nodes in $\Phi_I$ Network  & (128,128,64) \\
    Nodes in $\Phi_{I,2}$ Network  & (128,128,64) \\
    Nodes in $F$ Network & (64,100,100) \\
    Latent space dimension  & 64 (s), 128 (c) \\
    Number of Parameters & 97k (s), 101k (c) \\
    \hline 
    \hline
    \multicolumn{2}{|c|}{Performance Metrics} \\
    \hline 
    \hline
    ROC-AUC & 0.9839 (s), 0.9838 (c) \\
    Accuracy & 0.937 (s), 0.937 (c) \\
    Background Rejection Rate $(1/\epsilon_B)$ & 1041 (s), 1030 (c) \\
    \hline
    \end{tabular}
    }
    \caption{Model hyperparameters and performance metrics for the PFIN model. (s) and (c) respectively represent architectures where per-particle interaction embeddings from $\Phi_{I,2}$ are summed and concatenated with the particle embeddings from $\Phi$. The background rejection rate $1/\epsilon_B$ is evaluated at a signal efficiency of 30\%. }
    \label{tab:PFIN-results}
\end{table}

\begin{figure}[!ht]
\centering
\subfloat[]{
\includegraphics[width=0.33\textwidth]{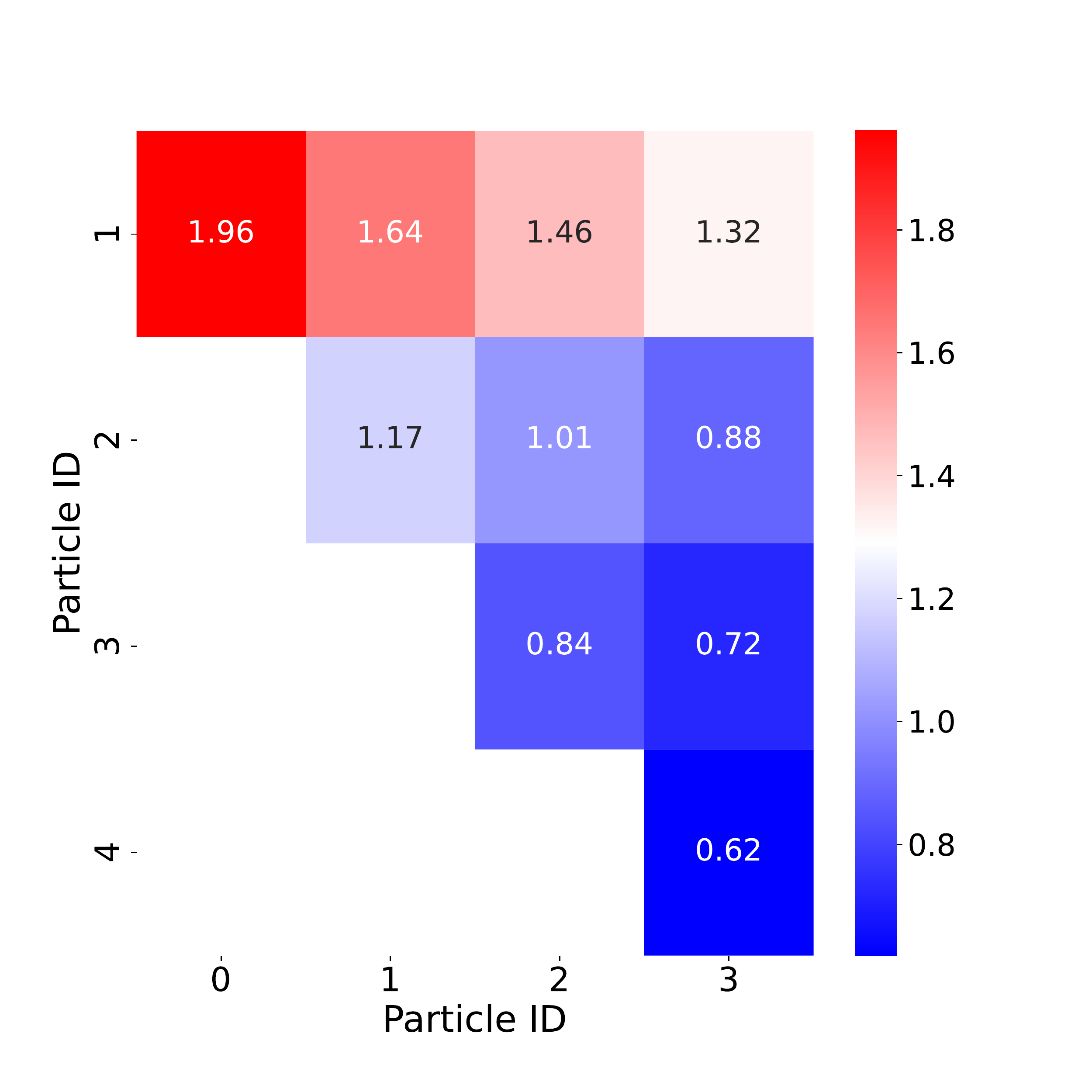}
\label{fig:dAUC-PFIN}            
}
\subfloat[]{
\includegraphics[width=0.33\textwidth]{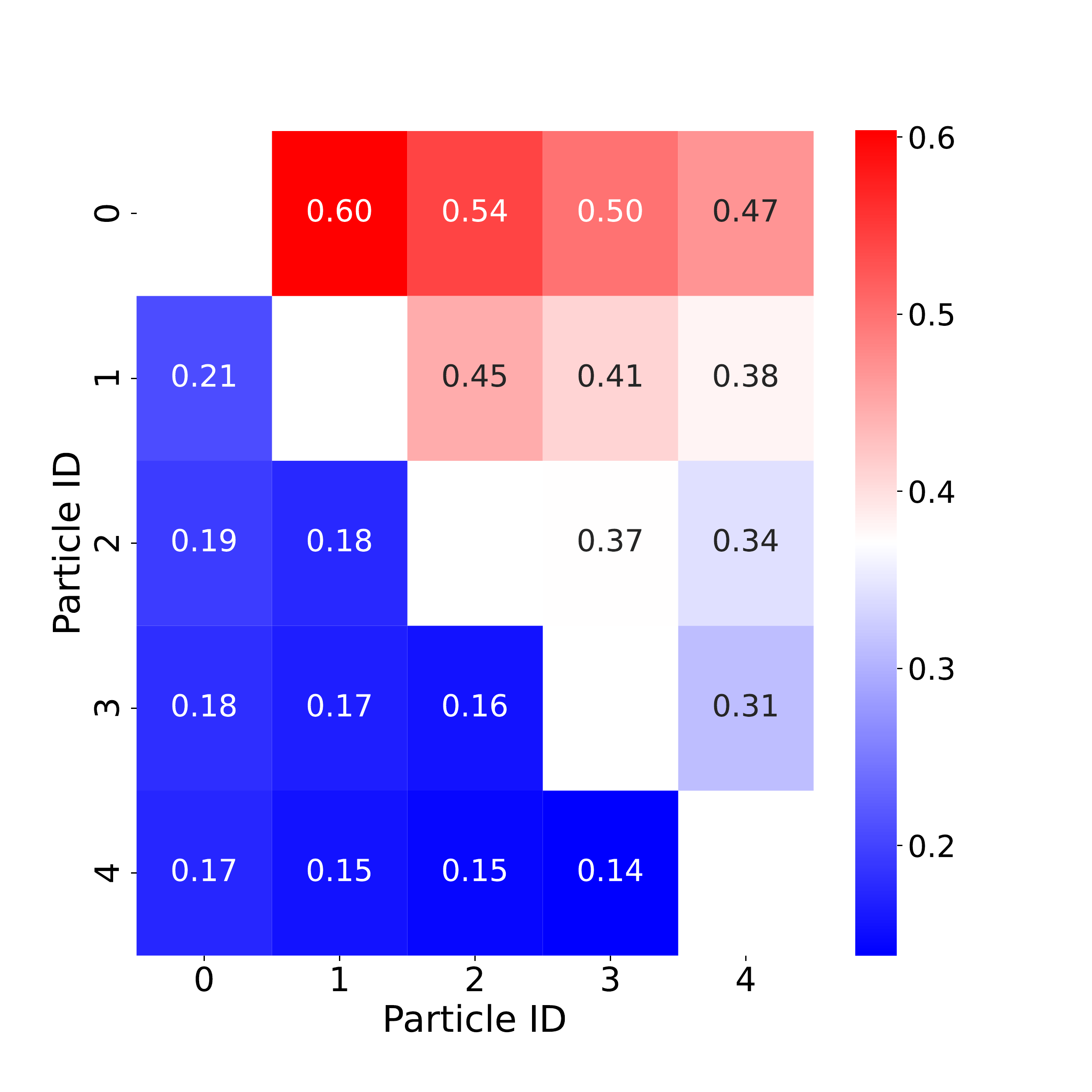}
\label{fig:MAD-PFIN}            
}
\subfloat[]{
\includegraphics[width=0.33\textwidth]{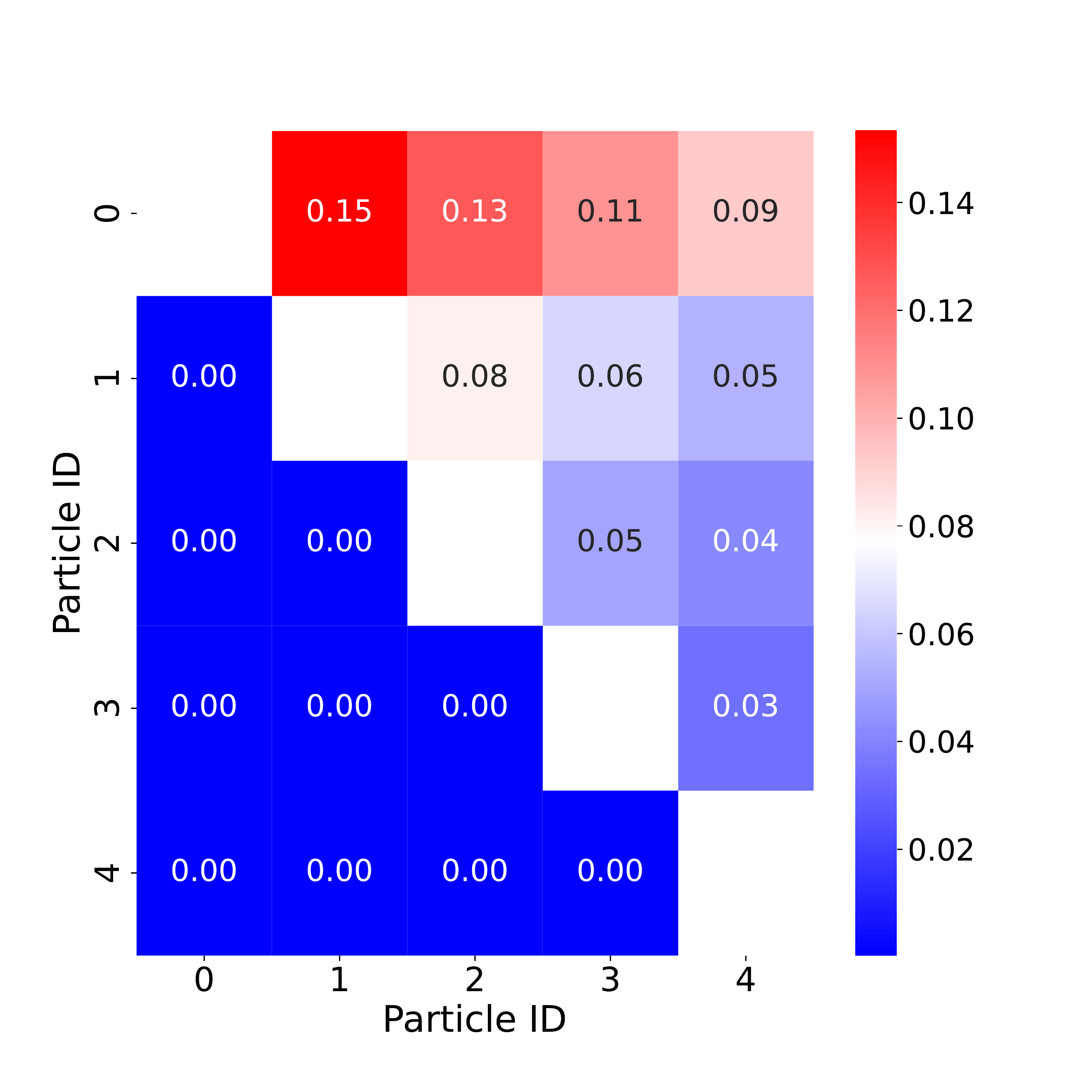}
\label{fig:dpred-PFIN}            
}
\caption{
\protect\subref{fig:dAUC-PFIN} \dAUC,
\protect\subref{fig:MAD-PFIN} normalized MAD relevance score, and
\protect\subref{fig:dpred-PFIN} mean deviation in jet class prediction probabilities for the PFIN model (with summation being used for particle level feature augmentation) for pairwise masking of particle interaction features for the five most energetic constituents. Particle constituents are arranged in decreasing order of their energies. For \protect\subref{fig:MAD-PFIN} and \protect\subref{fig:dpred-PFIN}, the upper and lower diagonal entries represent the corresponding scores for the signal and background jet classes respectively.
}
\label{fig:PFIN-XAI}
\end{figure}

PFIN allows us to explore the impact of pairwise particle interaction on jet classification. We calculate the \dAUC and MAD relevance score for each pair of particles by masking the corresponding input to the $\Phi_I$ network and  calculating the deviation in model prediction with respect to the baseline model's result. We additionally calculate the deviation in the model's jet class prediction probabilities.  The results for the PFIN network with summation used for augmented particle embeddings are shown in Figure~\ref{fig:PFIN-XAI}. The pairwise particle interactions play a particularly important role in identifying the signal jets. The mean deviations in the background jet class probabilities are barely impacted by masking interaction features. However, for the signal jets, this impact is found to be rather large, the mean prediction probability is reduced by almost 20\% when the interaction between the two most energetic  jets is masked. This clearly outlines the importance of particle interactions in detecting signal jets and consequently, explains the improvement observed in  model performance.

Similar to what we observed for the PFN model, the jet class information is found to be embodied in the distribution of correlations among the latent space features. Hence, we further investigated PFIN's latent space by performing PCA and one crucial observation is a stronger correlation between  jet mass and the top principal component of the PFIN latent space. These correlation coefficients were found to be close to 90\% for both jet classes for both variants of the PFIN model. The other principal components also showed moderately improved correlations with the subjettiness variables and the number of jet constituents. As a result, the latent jet features allow  construction of notably more expressive distributions, contributing to the observed improvement in jet tagging performance.

\section{Conclusion}
\label{sec:conclusion}

This paper presents a comprehensive study of the interpretability of DNN based top tagger models. Our work has unveiled a number of important aspects regarding how these models connect with the corresponding datasets. We have observed intriguing inconsistencies in feature ranking from different ranking methods, especially how the LRP method can produce a relevance distribution that can lead to a misleading interpretation of feature importance. Modifying LRP to obtain \textit{differential} relevance scores has been found to be more consistent with other approaches in XAI. Furthermore, explainability metrics need to be carefully studied and understood for models trained with highly correlated input features since, as we show with the \mbns{8} model, interpretability of AI models can be obscured in such cases. Our investigation suggests that models learn to embed jet class information in correlations among latent space embeddings and can learn to mimic distributions that closely resemble physical jet features. On the other hand, RNA scores and NAP diagrams can lead to an effective understanding of how information is propagated through different layers of a network and can lead to efficient model reoptimization strategies. 
Using NAP diagrams to reduce network complexity can be especially beneficial when multiple variants of the same network are required to be trained, e.g. to quantify uncertainties on event classification scores from systematic variations or adapt an architecture to learn jet classification in different phase spaces.
RNA scores and NAP diagrams also open the possibility of incorporating these methods to obtain \textit{in-situ} model optimization during training. 

Observations regarding feature importance and model sparsity can also lead to better model building, as we demonstrate with the PFIN model which learns to take advantage of individual particle-level feature embeddings as well as pairwise particle interactions to noticeably improve over the PFN model. PFIN's performance is better than or comparable to a number of other novel models while its implementation needs a smaller number of parameters, ensuring faster training and quicker convergence. Studying the impact of pairwise particle interactions on PFIN using \dAUC and MAD relevance scores reveals how these interactions can play an important role to identify top jets. 

This work establishes a methodological paradigm to demonstrate the usage of XAI tools for day-to-day applications of DNNs and MLPs in HEP. Many modern HEP analyses heavily rely on DNNs not only for  jet tagging and object reconstruction, but also for event classification and identification of signal events in search for physics beyond the standard model. Our work lays the foundation of exploring quantitative and robust explanations for such models. Compared to many of the existing tools to find feature importance, calculating \dAUC and MAD relevance scores is much faster and can produce equally reliable explanations for the model performance. While these methods may require further sophistication to be applicable with other data structures, we expect them to provide satisfactory performance for most classification problems relying on tabular data. 
%understanding how a model works can provide valuable insights into what a model learns and how it can be improved. While our work has rediscovered the potential of the PFN model by correctly implementing the prescribed preprocessing, thoroughly investigating its latent space distributions allowed us to discover its failure to take into account the expressive subjettiness features. As a result, the proposed AugPFN model has learned to largely outperform rather novel deep learning architectures by simply augmenting the PFN latent space with the subjettiness variables. 
Building on the results and observations presented in this work, our future work will take a deeper look into adapting these novel tools and methods for more general data types and interpreting novel architectures like graph nets and transformers in the context of top tagging and more general jet classification scenarios. 

\FloatBarrier

\setcounter{table}{0}  \renewcommand{\thetable}{\Alph{section}\arabic{table}}
\appendix

\section{Performance of Baseline and variant models}

\begin{table}[!ht]
    \centering
    \resizebox{0.9\textwidth}{!}{
    \begin{tabular}{|c|c|c|c|c|c|c|}
    \hline
   Architecture & Description & Params & AUC & Acc & $1/\epsilon_B$ & Sparsity \\
    \hline
    \multirow{4}{*}{TopoDNN} & 
    \textit{Baseline} &
    59k &
    0.971 & 
    0.916 & 
    278 &
    0.705 \\
    %\hline
    & 
    Trained without $p_{T,0}$ & 
    59k &
    0.970 & 
    0.914 &
    267 &
    0.852 \\
    %\hline
    & 
    Hidden Layers: $(240,80,10)$ & 
    42k &
    0.972 & 
    0.916 & 
    309 &
    0.818  \\
    %\hline
    & 
    Hidden Layers: $(120,40,6)$ & 
    16k &
    0.972 & 
    0.916 & 
    305 &
    0.584 \\
    \hline
    \multirow{5}{*}{\mbns{8}} & 
    \textit{Baseline} & 
    57k &
    0.980 & 
    0.928 & 
    796 & 
    0.426 \\
    %\hline
    & 
    Trained without jet $p_T$ & 
    57k &
    0.980 & 
    0.928 & 
    775 &
    0.444 \\
    %\hline
    & 
    Trained with $\{\tau_x^{(1)}\} \bigcup \{p_{T,J}, m_J\}$ & 
    55k &
    0.976 & 
    0.921 & 
    516 &
    0.404 \\
    %\hline
    & 
    Hidden Layers: $(200,200,50)$ & 
    55k &
    0.980 & 
    0.928 & 
    816 &
    0.416 \\
    & 
    Hidden Layers: $(200,200)$ & 
    45k &
    0.980 & 
    0.928 & 
    775 &
    0.452 \\
    \hline
    \multirow{3}{*}{PFN} &  
    \textit{Baseline} & 
    82k &
    0.980 & 
    0.928 & 
    699 &
    0.811 $(\Phi)$, 0.530 $(F)$ \\
    & 
    $\Phi : (100,100,64), F:(64,100,100)$ & %v9
    38k &
    0.978 & 
    0.925 &  
    653 &
    0.617 $(\Phi)$, 0.439 $(F)$ \\
    & 
    $\Phi : (100,64,32), F:(64,100,100)$ & %v7
    28k &
    0.978 & 
    0.924 & 
    603 &
    0.576 $(\Phi)$, 0.436 $(F)$ \\
    \hline
    \multirow{4}{*}{PFIN} &  
    \multirow{2}{*}{\textit{Baseline (concatenation)}} & 
    \multirow{2}{*}{101k} &
    \multirow{2}{*}{0.984} & 
    \multirow{2}{*}{0.937} & 
    \multirow{2}{*}{1030} &
    0.178 $(\Phi)$, 0.584 $(\Phi_I)$ \\
    & & & & & &
    0.675 $(\Phi_{I,2})$, 0.705 $(F)$ \\
    &
    \multirow{2}{*}{\textit{Baseline (summation)}} & 
    \multirow{2}{*}{97k} &
    \multirow{2}{*}{0.984} & 
    \multirow{2}{*}{0.937} & 
    \multirow{2}{*}{1041} &
    0.208 $(\Phi)$, 0.625 $(\Phi_I)$ \\
    & & & & & &
    0.70 $(\Phi_{I,2})$, 0.712 $(F)$ \\
    \hline
    \end{tabular}
    }
    \caption{Performance and sparsity of the baseline and model variants for different model architectures.  The background rejection rate $1/\epsilon_B$ is evaluated at a signal efficiency of 30\%.}
    \label{tab:TopoDNN-perf}
\end{table}

\FloatBarrier

\section*{Acknowledgements}
We would like to thank Volodymyr Kindratenko and the Center for Artificial Intelligence Innovation at the NCSA for support through our affiliation. 
We would like to thank Javier Duarte for useful discussions, comments, and suggestions. 
We are grateful to Gregor Kasieczka and Huilin Qu for pointing us to important resources and code repositories. 
Comments and suggestions from Sung Hak Lim, Ranit Das, Gregor Kasieczka, and David Shih helped us identify and rectify some inconsistencies in the previous version of our code.
This research is part of the Delta research computing project, which is supported by the National Science Foundation (award OCI 2005572), and the State of Illinois. Delta is a joint effort of the University of Illinois at Urbana-Champaign and its National Center for Supercomputing Applications. This work utilizes resources supported by the National Science Foundation’s Major Research Instrumentation program, grant 1725729, as well as the University of Illinois at Urbana-Champaign. This work was supported by the FAIR Data program of the U.S. Department of Energy, Office of Science, Advanced Scientific Computing Research, under contract number DE-SC0021258 and by the U.S. Department of Energy, Office of Science, High Energy Physics, under contract number DE-SC0023365. 

%\section*{References}
\newcommand{\newblock}{}
\bibliography{main2}
\bibliographystyle{JHEP}

\end{document}